\documentclass[twocolumn,UTF8]{aastex63}

\usepackage{graphicx}			
\usepackage{xcolor}
\usepackage{CJKutf8}
\usepackage{bm}
\usepackage{multirow}
\usepackage{enumitem}
\usepackage{tabularx}
\usepackage{gensymb}
\usepackage[hyphens]{url}

\newcolumntype{Z}{>{\centering\arraybackslash}X}   
\newcolumntype{Y}{>{\hsize=1\hsize\raggedright\arraybackslash}X}

\turnoffedit
\shorttitle{Energy evolution from the chromosphere to the heliosphere} 
\shortauthors{Reeves et al.} 

\begin{document}

\begin{CJK*}{UTF8}{gbsn}

\title{Energy Evolution from the Chromosphere to the Heliosphere in the 2021 October 28 Solar Eruption}

\author[0000-0002-6903-6832]{Katharine K. Reeves}
\affiliation{Center for Astrophysics \textbar\ Harvard \& Smithsonian, Cambridge, MA, 02138, USA}

\author[0000-0002-0494-2025]{Daniel B. Seaton}
\affiliation{Southwest Research Institute, Boulder, CO, 80302, USA}

\author [0000-0002-3805-320X]{Cynthia Cattell}
\affiliation{School of Physics and Astronomy, Minnesota Institute for Astrophysics, University of Minnesota, 116 Church St. SE,  Minneapolis, MN 55455 USA}

\author[0000-0002-0660-3350]{Bin Chen (陈彬)}
\affiliation{Center for Solar-Terrestrial Research, New Jersey Institute of Technology, Newark, NJ 07102, USA}

\author[0000-0003-1713-9466]{Liam David}
\affiliation{Princeton Plasma Physics Laboratory, Princeton, NJ, 08540, USA}

\author[0000-0002-5456-4771]{Federico Fraschetti}
\affiliation{Center for Astrophysics \textbar\ Harvard \& Smithsonian, Cambridge, MA, 02138, USA}
\affiliation{Lunar \& Planetary Laboratory, University of Arizona, Tucson, AZ 85721, USA}

\author[0000-0002-0850-4233]{Joe Giacalone}
\affiliation{Lunar \& Planetary Laboratory, University of Arizona, Tucson, AZ 85721, USA}

\author[0000-0003-1377-6353]{Phillip Hess}
\affiliation{U.S. Naval Research Laboratory, Washington, DC 20375, USA}

\author[0000-0003-1398-1752]{Andryi Koval}
\affiliation{University of Maryland, Baltimore County, Baltimore, MD 21250, USA}
\affiliation{NASA Goddard Space Flight Center, Greenbelt, MD 20771, USA}

\author[0000-0003-2102-0070]{Dana W.\ Longcope}
\affiliation{Dept.\ of Physics, Montana State University, Bozeman, MT 59717}

\author[0000-0002-2325-5298]{Surajit Mondal}
\affiliation{Center for Solar-Terrestrial Research, New Jersey Institute of Technology, Newark, NJ 07102, USA}

\author[0000-0002-4103-6101]{Christopher S. Moore}
\affiliation{Center for Astrophysics \textbar\ Harvard \& Smithsonian, Cambridge, MA, 02138, USA}

\author[0000-0002-0945-8996]{Sophie Musset}
\affiliation{European Space Research and Technology Centre: Noordwijk, South Holland, NL}
\affiliation{Johns Hopkins University Applied Physics Laboratory: North Laurel, Maryland, US}

\author[0000-0001-6692-9187]{Tatiana Niembro}
\affiliation{Center for Astrophysics \textbar\ Harvard \& Smithsonian, Cambridge, MA, 02138, USA}

\author[0000-0002-6176-4077]{Daniel Pacheco}
\affiliation{Deep Space Exploration Laboratory/School of Earth and Space Sciences, University of Science and Technology of China, Hefei, 230026, China}

\author[0000-0002-8748-2123]{Yeimy J. Rivera}
\affiliation{Center for Astrophysics \textbar\ Harvard \& Smithsonian, Cambridge, MA, 02138, USA}

\author[0000-0003-2215-7810]{Soumya Roy}
\affiliation{Center for Astrophysics \textbar\ Harvard \& Smithsonian, Cambridge, MA, 02138, USA}
\affiliation{Inter-University Centre for Astronomy and Astrophysics, Pune, Maharashtra, 411007, India}
\affiliation{Manipal Centre for Natural Sciences, Manipal Academy of Higher Education, Karnataka, Manipal, 576104, India}

\author[0000-0003-4043-616X]{Xudong Sun (孙旭东)}
\affiliation{Institute for Astronomy, University of Hawai`i at Manoa, Pukalani, HI 96768, USA}

\author[0000-0003-1689-6254]{Durgesh Tripathi}
\affiliation{Inter-University Centre for Astronomy and Astrophysics, Pune, Maharashtra, 411007, India}

\author[0000-0002-0608-8897]{Domenico Trotta}
\affiliation{European Space Agency (ESA), European Space Astronomy Centre (ESAC), 28692, Madrid, Spain}

\author[0000-0002-0631-2393]{Matthew J. West}
\affiliation{European Space Research and Technology Centre: Noordwijk, South Holland, NL}

\author[0000-0003-2872-2614]{Sijie Yu (余思捷)}
\affiliation{Center for Solar-Terrestrial Research, New Jersey Institute of Technology, Newark, NJ 07102, USA}

\author[0000-0003-3218-5487]{Chunming Zhu}
\affiliation{Dept.\ of Physics, Montana State University, Bozeman, MT 59717}

\begin{abstract}
We perform a detailed study of the energetics for a well-observed solar eruption and flare that occurred on 28 October 2021. This event included a GOES class X1.0 flare, a global EUV wave, and a coronal mass ejection that reached speeds of $>$2000\,km\,s$^{-1}$.  The event was observed from a variety of spacecraft in NASA's Heliophysics System Observatory, including multiple missions near Earth, STEREO-A off the Sun-Earth line, and Solar Orbiter, near the Sun-Earth line at about 0.8~au. Using remote sensing, in situ observations, and in some cases scaling laws based on previous observations, we characterize the following quantities: \edit2{free} magnetic energy, energy in non-thermal electrons, energy in non-thermal ions, bolometric energy, energy deposited in the chromosphere, thermal energy radiated in the flare loops, energy dissipated by the EUV wave, CME kinetic and gravitational potential energy, CME energy flux in the heliosphere, and the energy partition in the CME shock. We find that the total energy released during the event is consistent with estimates of the pre-event stored magnetic energy, and the CME kinetic + potential energy dominates the energy partition.
\end{abstract}

\section{Introduction}
\label{section:intro}
Solar eruptions serve as a pathway to convert the Sun's stored magnetic energy into other forms, including thermal energy, kinetic energy, and gravitational potential energy. The release of stored energy manifests in a variety of ways, including heating in the chromosphere, heated flare loops in the corona, accelerated electrons and ions, EUV waves, and accelerated ejecta in coronal mass ejections.

Several previous attempts have been made to characterize and understand the energy partitioning in solar eruptive events.  \citet{Emslie04} and \citet{Emslie05} examined two well-observed solar eruptions associated with X-class solar flares that occurred on 21 April 2002 and 23 July 2002.  For these events, the studies found that the CME contains the largest fraction of the released energy, though the uncertainties in the energy budget calculations are large.  A more comprehensive study by \citet{emslie12} evaluated 38 solar eruptive events, mostly associated with M- and X-class flares, and similarly concluded that most of the energy released during solar eruptions goes into the kinetic energy of the CME. On the other hand, in a study of 399 M and X flares, \citet{Aschwanden2017} found that \edit2{only approximately 7\% of the released magnetic energy is partitioned into the launch of the CME}.

The process of accounting for the energy partitioned into different manifestations in a solar eruption is complex because the energy released can transform through multiple states \citep[see][and references therein]{Benz2017}. For example, electrons that are accelerated in the corona by the released magnetic energy may encounter the chromosphere and deposit energy there, generating hard X-rays and heating the chromosphere.  Flare loops, which appear subsequently, are also a result of this energy deposited in the chromosphere via a heat induced pressure differential that drives hot plasma up into the corona.  Similarly, magnetic energy that is transformed into the kinetic energy of the CME can then be further converted into energetic particles through s shock formed as the CME propagates into space. Thus a thorough and systematic accounting of all of this energy, spanning multiple observational regimes, is required to completely address the energy partition during these eruptive events. 

The eruption that we study in this paper occurred on 28\,October\,2021, and it was associated with an X1.0 flare in AR\,12887, located in the southern hemisphere near the central meridian of the Sun.  A CME and an EUV wave were associated with this event, and the CME caused the first ground level enhancement (GLE) of Solar Cycle 25 \citep{Klein2022,Papaioannou2022}, as well as disturbances to the ionosphere \citep{Habarulema2022,Fagundes2024} when it passed the Earth.  

A unique feature of this event is that it was observed by Solar Orbiter in a part of its orbit that was roughly aligned with the Earth-Sun line.  Solar Orbiter was at about 0.8\,au, allowing for a thorough diagnosis of the evolution of the in situ energy fluxes as a function of time.

In this work, we perform a thorough accounting of the energy partitioning of the 28 October 2021 event from the corona to the heliosphere to derive an energy budget for this eruption.  In \S\ref{section:obs} we present the observations of this event.  In \S\ref{section:energy} we catalog the energetics of the eruption, and \S\ref{section:discussion} presents the discussion and conclusions.
\section{Observations}
\label{section:obs}

The event we study is an eruption that occurred on 2021\,October\,28, originating near the central meridian of the Sun in the southern hemisphere. The eruption produced an X-class flare (SOL2021-10-28T15:17), an EUV wave, and a fast CME.  Below we summarize the observations of this event used in our exploration of the energetics.

\begin{figure}
    \centering
    \includegraphics[scale=0.47, trim={0.5in 0.5in 0in 0in}]{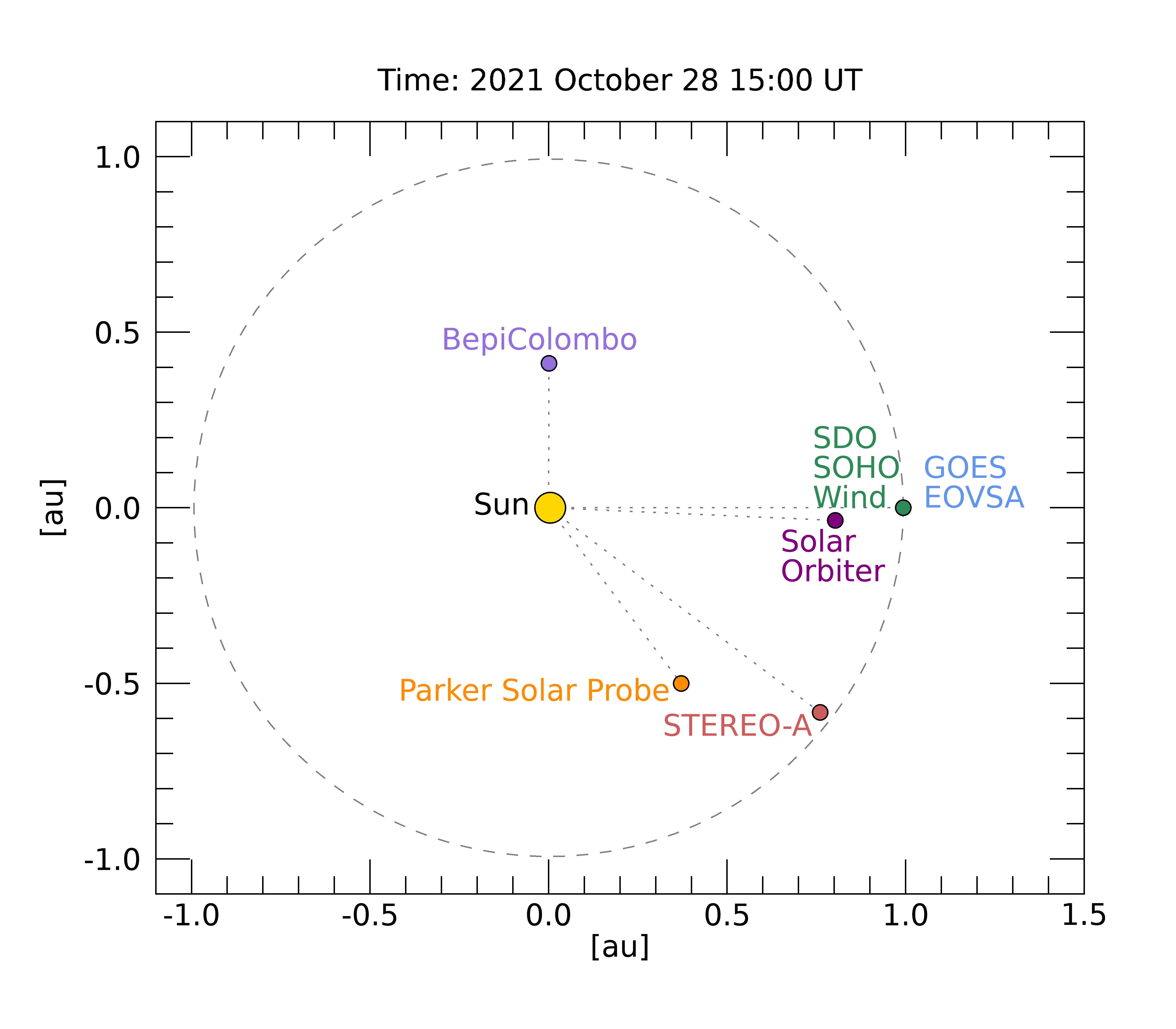}
    \caption{Locations of \textit{BepiColombo}, \textit{Parker\,Solar\,Probe}, \textit{STEREO-A}, \textit{Solar\,Orbiter} and Earth (\textit{SDO}, \textit{SoHO}, \textit{GOES} and \textit{Wind} spacecraft and EOVSA observatory) at the eruption onset time of 2021\,October\,28\,15:00\,UT.}
    \label{fig:orbit}
\end{figure}

We combine the remote sensing and in situ measurements of multiple instruments from ground- and space-based observatories. In Figure\,\ref{fig:orbit} we show, in the Earth Ecliptic (HEE) coordinate system, the locations of the spacecraft: \textit{BepiColombo} \citep{2021SSRv..217...90B}, \textit{Parker\,Solar\,Probe} \citep[\textit{PSP}, ][]{2016SSRv..204....7F}, \textit{Solar TErrestrial RElations Observatory Ahead} \citep[\textit{STEREO-A}, ][]{2005AdSpR..36.1483K}, \textit{Solar\,Orbiter} \citep[\textit{SolO}, ][]{Mueller_2020}, \textit{Solar Dynamics Observatory} \citep[\textit{SDO}, ][]{2012SoPh..275....3P}, \textit{SOlar and Heliospheric Observatory} \citep[\textit{SoHO}, ][]{1995SSRv...72...81D}, the \textit{Wind} spacecraft \citep{1997AdSpR..20..559O} and the \textit{Geostationary Operational Environmental Satellite} \citep[\textit{GOES}, ][]{Krimchansky2004}.
On Earth, we use the radio observations from the Expanded Owens Valley Solar Array \citep[EOVSA, ][]{Gary2018eovsa}. \textit{BepiColombo} and \textit{PSP} were located conveniently for multi-spacecraft analysis, however, datasets contain large data gaps during the event preventing the eruption identification and were excluded within this investigation. 

In Table \ref{tab:orbit}, we summarize the radial distance ($R$), longitude ($lon$) and latitude ($lat$) of the multiple observatories. The locations are given for the approximate eruption onset time (2021\,October\,28\,15:00\,UT) as well as when the in~situ signatures are measured by \textit{Solar\,Orbiter} (2021\,October\,30\,22:00\,UT) and \textit{Wind} (2021\,November\,01\,09:21\,UT) spacecraft. Earth locates the \textit{SDO}, \textit{SoHO}, \textit{GOES} and \textit{Wind} spacecraft, as well as the EOVSA observatory.

\begin{table}
\centering

\caption{Locations (R- radial distance, lat - latitude and lon- longitude) in HEE coordinates of \textit{BepiColombo}, \textit{Parker\,Solar\,Probe}, \textit{STEREO-A}, \textit{Solar\,Orbiter} and Earth (\textit{SDO}, \textit{SoHO}, \textit{GOES} and \textit{Wind} spacecraft and EOVSA observatory) during the eruption onset time (2021\,October\,28\,15:00\,UT) and when the in~situ signatures are measured by \textit{Solar\,Orbiter} (2021\,October\,30\,22:00\,UT) and \textit{Wind} (2021\,November\,01\,09:21\,UT) spacecraft.}
\label{tab:orbit}

\begin{tabular*}{\linewidth}{@{\extracolsep{\fill}} l r r r}
\tableline\tableline
\multicolumn{1}{c}{Eruption location} & \multicolumn{1}{c}{R [au]} & \multicolumn{1}{c}{lat [$^{\circ}$]} & \multicolumn{1}{c}{lon [$^{\circ}$]}\\
\tableline
{\it BepiColombo} &  &  & \\
\hspace{1em} at onset & 0.411 & 89.843 & 6.414 \\
\hspace{1em} at 0.81\,au & 0.432 & 96.012 & 6.665 \\
\hspace{1em} at 0.99\,au & 0.447 & 99.542 & 6.746 \\
{\it Parker\,Solar\,Probe} & & & \\
\hspace{1em} at onset & 0.623 & $-53.419$ & $-3.380$ \\
\hspace{1em} at 0.81\,au & 0.594 & $-53.646$ & $-3.388$ \\
\hspace{1em} at 0.99\,au & 0.573 & $-53.678$ & -3.391 \\
{\it STEREO-A} & & & \\
\hspace{1em} at onset & 0.958 & $-37.465$ & 0.075 \\
\hspace{1em} at 0.81\,au & 0.958 & $-37.343$ & 0.07 \\
\hspace{1em} at 0.99\,au & 0.958 & $-37.267$ & 0.067 \\
{\it Solar\,Orbiter} &  &  & \\
\hspace{1em} at onset & 0.803 & $-2.569$ & $-2.78$ \\
\hspace{1em} at 0.81\,au & 0.819 & $-1.817$ & $-2.543$ \\
\hspace{1em} at 0.99\,au & 0.828 & $-1.395$ & $-2.392$ \\
{\it Earth (Wind)} & & & \\
\hspace{1em} at onset & 0.993 & 0.0 & 0.0 \\
\hspace{1em} at 0.81\,au & 0.992 & 0.0 & 0.0 \\
\hspace{1em} at 0.99\,au & 0.992 & 0.0 & 0.0 \\
\tableline\tableline
\end{tabular*}
\end{table}

\subsection{Pre-eruptive Active Region}

\begin{figure}
\includegraphics[scale=1]{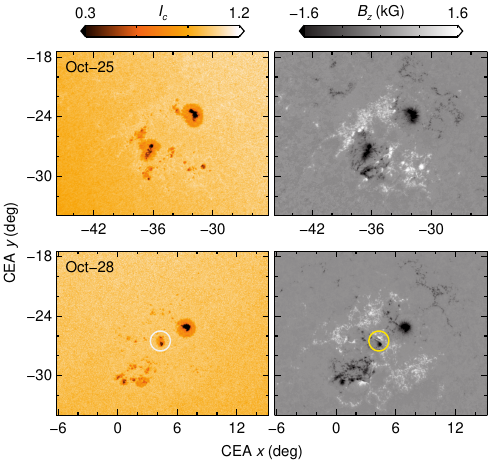}
\caption{\label{HMI.fig}  \edit1{HMI normalized continuum $I_c$ (left) and vertical magnetic field $B_z$ for 2021\,October\,25\,15:00\,UT (top) and 2021\,October\,28\,15:00\,UT (right), just before the flare. Circles in the bottom row highlight a small bipole that undergoes intense shearing motion. The animation shows the continuous evolution of the continuum and magnetic field observations of the region from October\,24\,23:59\,UT to October\,29\,22:59\,UT.}}
\end{figure}

The source region for the event, AR\,12887, was already well formed when rotated onto the disk, showing a quadrupolar configuration (Figure\,\ref{HMI.fig}, top) as observed by the Helioseismic and Magnetic Imager \citep[HMI;][]{scherrer2012} on \textit{SDO}. The subsequent evolution featured significant flux cancellation between the south-eastern, negative sunspot and the surrounding positive magnetic flux, which led to the disintegration of the spot (Figure\,\ref{HMI.fig}, bottom). Part of its negative flux moved westward and collided with a small patch of positive flux at the center (circle, see also the accompanying animation). There was strong shearing motion and apparent flux cancellation between the newly formed bipole. 

\subsection{Flare Observations}

\begin{figure}
\includegraphics[width=0.5\textwidth,trim={0.2cm 0.2cm 0.5cm 1cm},clip]{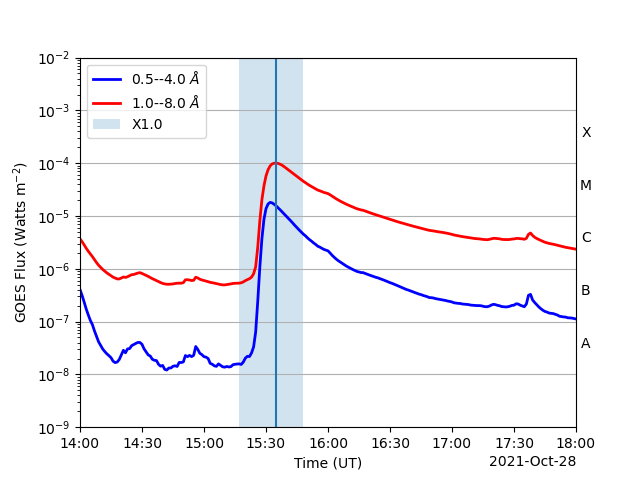}
\caption{\label{goes_overview.fig} \textit{GOES} XRS observations of the flare.  The peak time is marked with a vertical blue line, and the flare duration is indicated by the blue shading.
}
\end{figure}

The flare SOL2021-10-28T15:17 was an X1.0 class flare as observed by the X-Ray Sensor (XRS) on \textit{GOES}, shown in Figure\,\ref{goes_overview.fig}.  It started at 15:17\,UT and reached its peak at 15:35\,UT. The flare was a relatively long duration event -- even by two hours after the event peak, the {\it GOES} flux had still not returned to pre-flare values. 

\begin{figure*}
\centering
\includegraphics[width=1\textwidth]{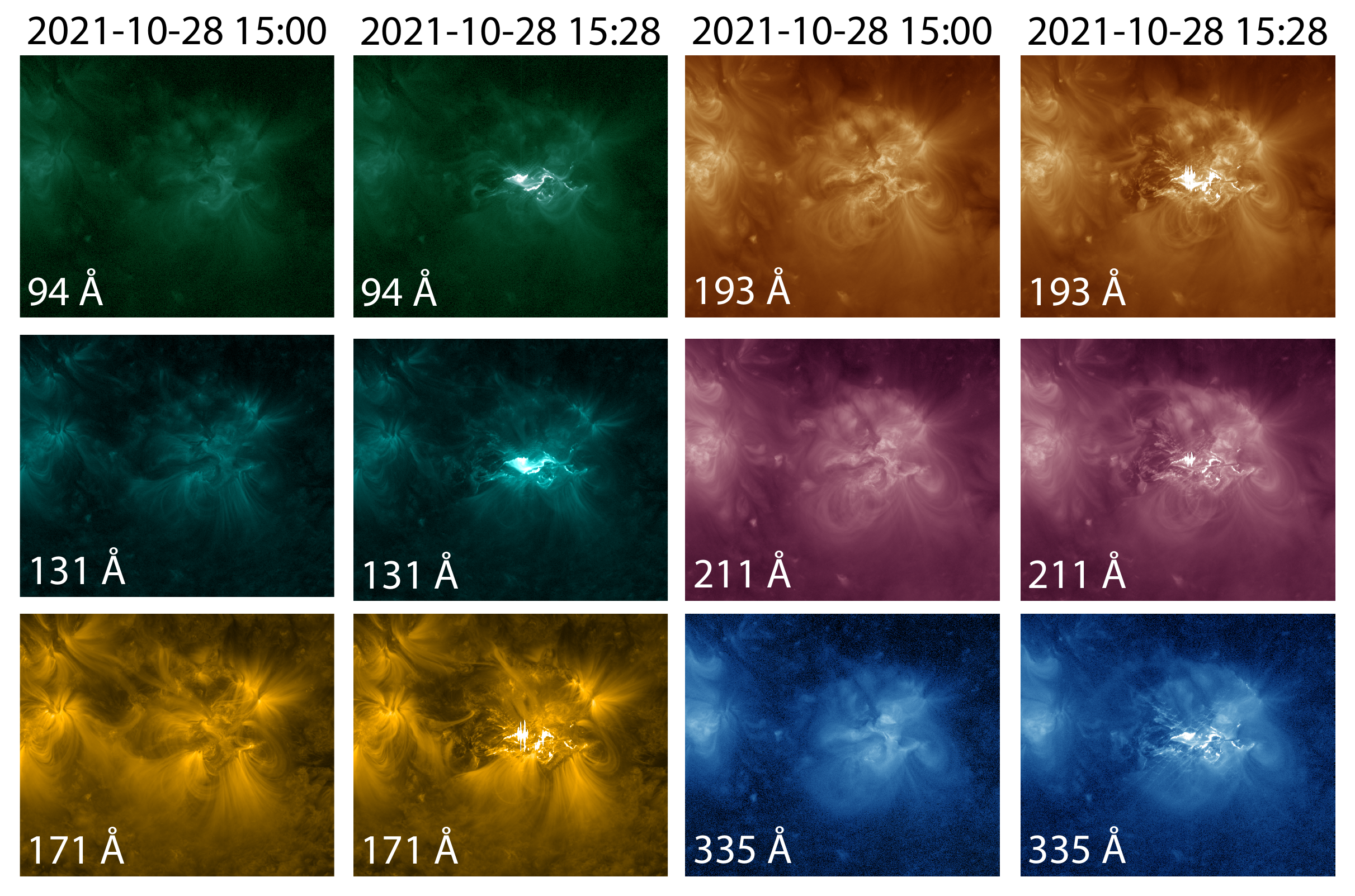}
\caption{\label{fig:aia_overview} AIA images of the corona in several EUV passbands before the onset of the flare (15:00\,UT) and during the flare (15:28\,UT).
}
\end{figure*}

The flare was observed in the EUV by the Atmospheric Imaging Assembly \citep[AIA, ][]{lemenet12} on \textit{SDO} and by the Solar Ultraviolet Imager \citep[SUVI;][]{Darnel2022} on \textit{GOES}.  AIA provides images in seven EUV passbands (94, 131, 171, 193, 211, 304, and 335\,\AA) as well as several UV channels (1550, 1600, 1700\,\AA).  Figure \ref{fig:aia_overview} shows AIA images of the flaring region before and during the flare in several EUV passbands.   SUVI imaged the flare and associated eruption with similar EUV channels to AIA (94, 131, 171, 195, 284, 304\,\AA).  {\em Hinode}/X-Ray Telescope \citep[XRT;][]{Golub2007} data were saturated during the flare peak, but there was some usable data during the decay phase in the Be-thick and Be-med filters.

This flare was also observed by \edit1{the Expanded Owens Valley Solar Array (EOVSA)}, which is a 13 element microwave interferometer that makes quasi-simultaneous observations over the 1--18\,GHz band. The event occurred when the Sun was just rising at EOVSA and only 9 of the 13 antennas were tracking the Sun. Hence the data quality and spatial resolution are somewhat degraded.

\begin{figure*}
    \centering
    \includegraphics[trim={0 6cm 0 9cm},clip,scale=0.5]{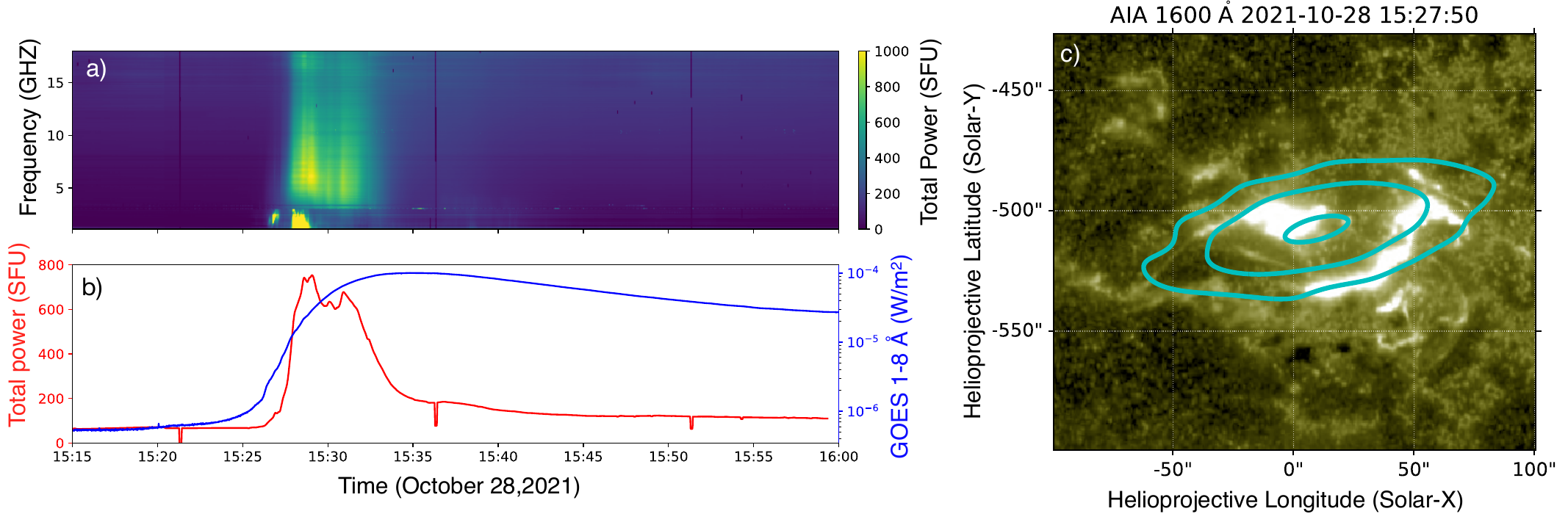}
    \caption{Panel a): EOVSA dynamic spectrum. The colorbar has been saturated at 1000\,SFU. Panel b): The red line shows the smoothed EOVSA lightcurve between 3--15\,GHz. The soft X-ray lightcurve obtained from \textit{GOES} is also shown in the same plot in blue. Panel c): EOVSA image at 13\,GHz at 15:28:05\,UT, shown with cyan contours over an AIA\,1600\,\AA$\,$ image taken close in time.}
    \label{fig:eovsa_ds}
\end{figure*}

The EOVSA observations are shown in Figures \ref{fig:eovsa_ds}. In panel a), we show the dynamic spectrum from EOVSA and panel b) shows the microwave lightcurve between 3--15\,GHz with the soft X-ray lightcurve from \textit{GOES} overlaid. We have saturated the colorscale to 1000\,SFU. 
In panel c), we overlay an example image at 13\,GHz over a nearby AIA\,1600\,\AA$\,$ image. While it is evident that the microwave source location matches that of the flaring region, the microwave source is practically unresolved. Despite the limitations of the data, we are able to obtain meaningful constraints on the nonthermal energetics of the flare as is described in Section \ref{sec:microwave}. 

\begin{figure}
    \includegraphics[width=0.5\textwidth]{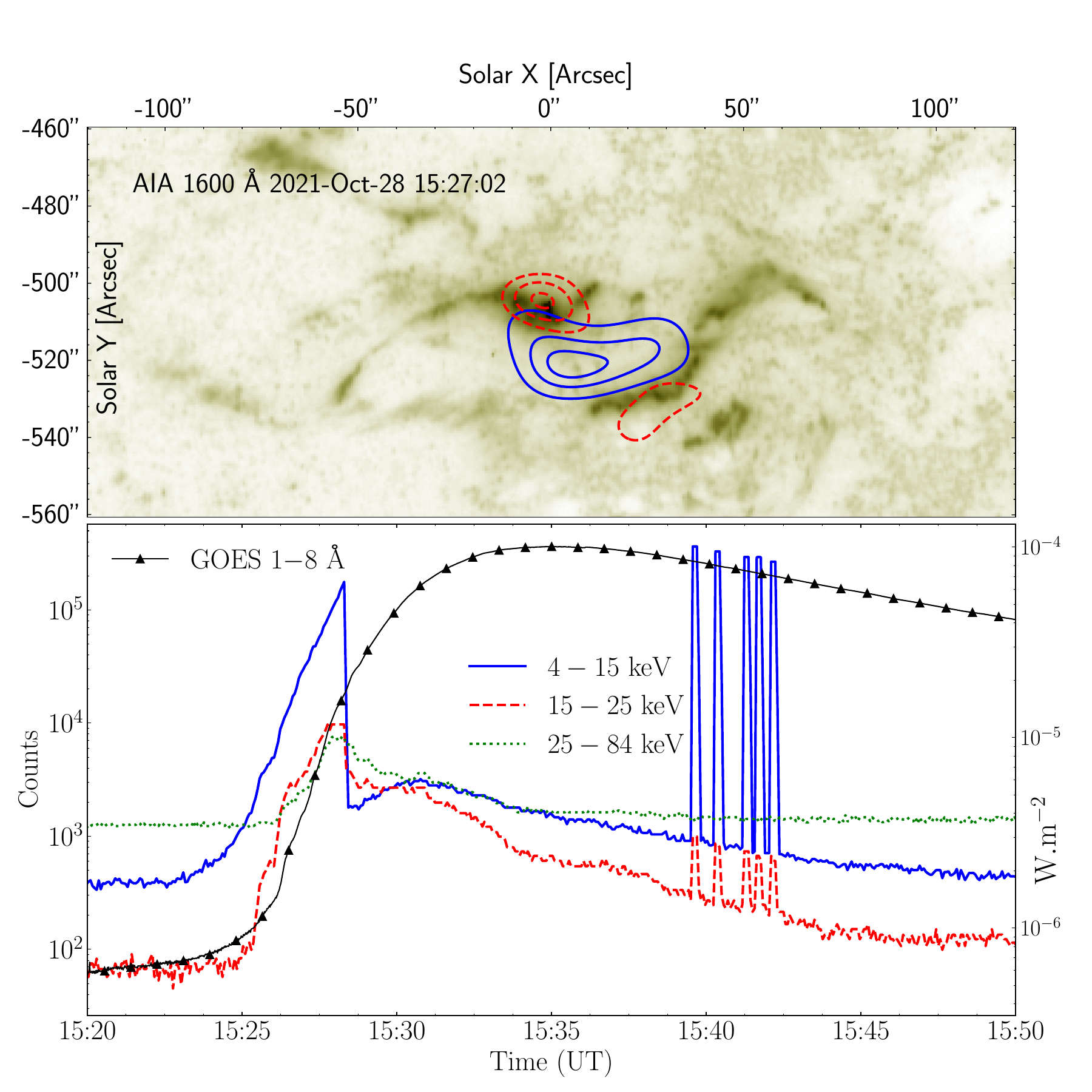}
    \caption{STIX soft and hard X-ray contours overplotted on an AIA 1600 {\AA} image with an inverted colorscale. STIX light curves from various channels are shown in the bottom panel in comparison to the {\it GOES} soft X-ray flux in 1{--}8 {\AA}. \edit2{The STIX light curves are made with 4 s time bins and the light travel time is taken into account}. The contours in the top panel are marked with same colors as the light curves in bottom channel.}
    \label{fig.stix_lc}
\end{figure}

The flare was observed by the \edit1{Spectrometer/Telescope for Imaging X-rays \citep[STIX,][]{2020A&A...642A..15K}} instrument on \textit{Solar\, Orbiter}.  In the upper panel of Figure \,\ref{fig.stix_lc} the STIX soft X-ray (4{--15} keV, solid blue) and hard X-ray (15{--}25 keV, red dashed) contours are over-plotted on the AIA~1600~{\AA} observation. The AIA~1600~{\AA} observation is in negative colormap, i.e. the brighter features appear darker. The 15{--}25 hard X-ray contours are cospatial with the AIA~1600~{\AA} ribbons.  In the lower panel of Figure\,\ref{fig.stix_lc} we plot the light curve from various STIX channels in comparison to {\it GOES} 1{--}8 {\AA} light curve. The attenuator on STIX kicks in before the soft X-ray peak of the flare to avoid saturation. The attenuator affects the softer X-ray flux considerably more than hard X-ray flux. We use the STIX spectra to obtain the constraints on the non-thermal energy deposited at the foot-points by the accelerated electrons in Section \ref{section:STIX}.  \edit1{The \textit{Fermi} Gamma-ray Burst Monitor \citep[GBM,][]{Meegan2009} observed the flare in similar energy bands as STIX during the decay phase, from 15:50 - 16:14 UT.}

\subsection{EUV Wave}
\begin{figure*}
\centering
\includegraphics[width=1.0\textwidth]{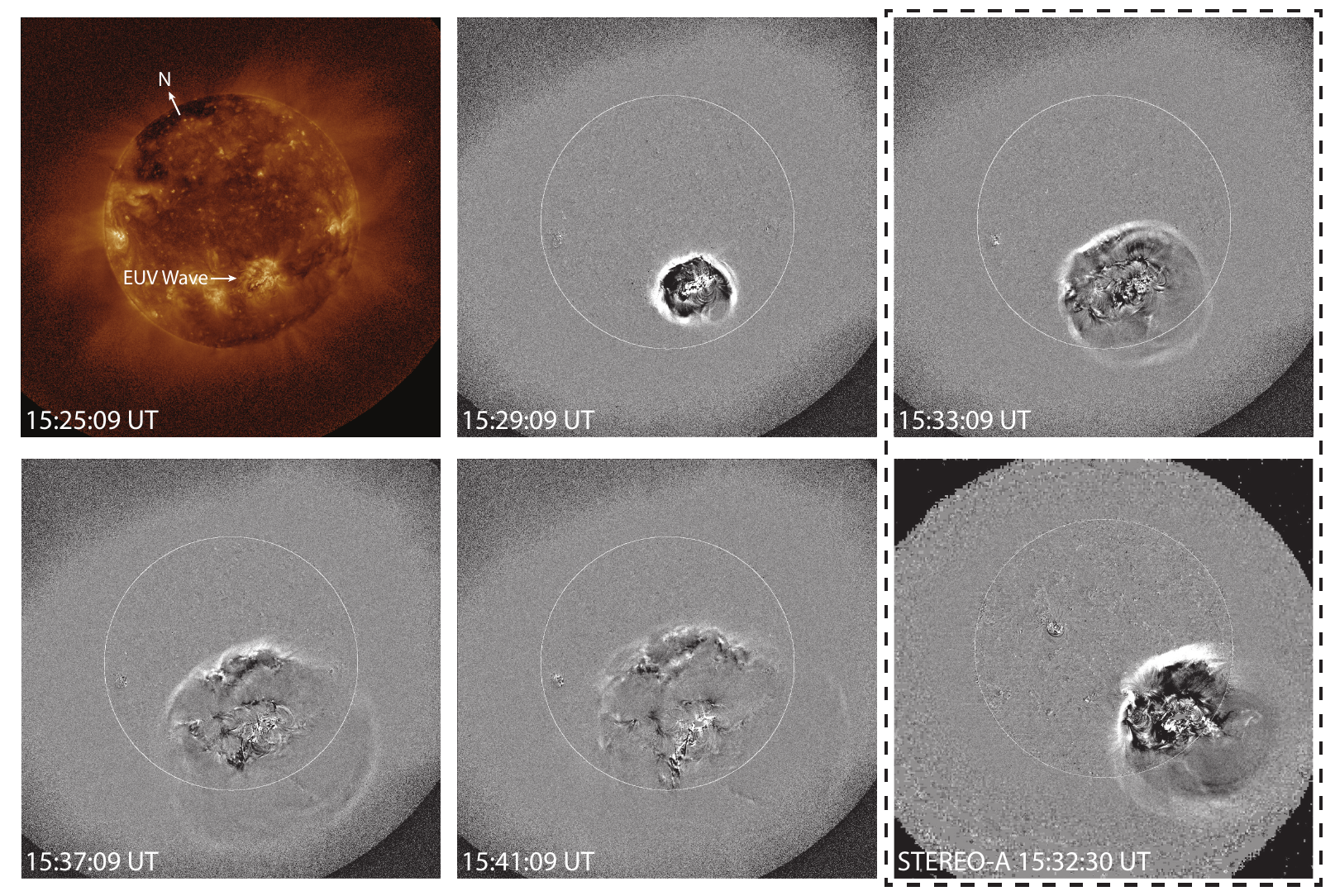}
\caption{\label{fig:wave_overview} 195\,\AA\ images of the EUV wave from SUVI and EUVI-A. The upper left image shows a normal view of the wave, enhanced with a radial filter \citep[see ][]{Seaton2023} and temporal unsharp masking technique, while the remaining panels are running difference. The two rightmost panels enclosed in the dashed line show the wave from both \textit{GOES} and \textit{STEREO-A} perspectives, separated by about $37.5^\circ$ \edit1{in longitude and 39~s in time}.
}
\end{figure*}

One of the most prominent features of the eruption was a large, circular global-scale EUV Wave \citep[sometimes referred to an ``EIT Wave'';][]{Long2017}. Although such waves can usually be seen over a wide range of EUV passbands, corresponding to a wide range of temperatures, they are often most easily seen in warm coronal bands, such as the SUVI 195\,\AA\ band (which features strong contributions from Fe~\textsc{xi} and \textsc{xii} \edit2{at} about 1.5~MK; see \citealt{Darnel2022}, Table~4). Because these waves are diffuse and highly dynamic, they are generally observed most clearly in running-difference images. Figure\,\ref{fig:wave_overview} shows an overview of the wave's evolution as seen by SUVI and the \textit{STEREO-A} Extreme Ultraviolet Imager \citep[EUVI, ][]{2008SSRv..136...67H}. Early in the event (upper left panel, in normal imaging) the wave is essentially circular in appearance, while later (subsequent panels) part of the wave that travels off of the solar disk retains its circular appearance and travels at essentially constant velocity, while the part of the wave observed on-disk begins to interact with the local magnetic structures and both slows and dissipates.

\subsection{Coronal Mass Ejection}

\begin{figure*}
\centering
\begin{interactive}{animation}{fig8_animation.mp4}
\includegraphics[width=0.95\textwidth]{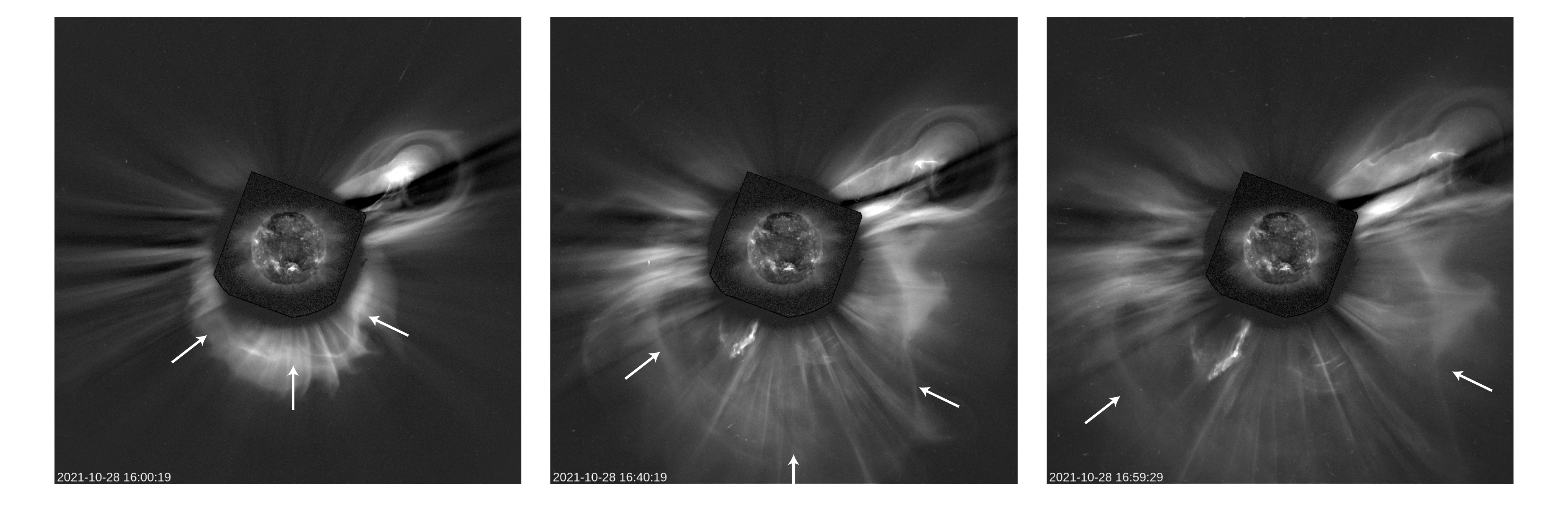}
\end{interactive}
\caption{\label{fig:lasco} Composite view of the CME using SUVI and C2 data. Arrows indicate the location of the expanding CME as it passes through the LASCO field of view. The CME in the northwest began before the onset of the halo event we study here. Note that the overlaid timestamps correspond to the SUVI image at the center of the frame, which is sampled at higher cadence than the LASCO data. \edit1{The animation shows the complete observation of the eruption from 15:00\.UT to 16:59\,UT on October\,28.}}
\end{figure*}

Figure\,\ref{fig:lasco} shows the CME's early evolution observed by SUVI and \textit{SoHO} Large Angle and Spectrometric Coronagraph \cite[LASCO][]{Brueckner1995} C2 instrument. The CME is a strong halo event, directed southward, with the dense, bright prominence core trailing the main CME bubble (visible especially in the second and third panels of the figure).

\subsection{Interplanetary Shocks: multi-point in-situ measurement}

An interplanetary shock (IPs) associated with this eruption crossed Solar\,Orbiter located at an heliocentric distance $\sim 0.8$\,au on 2021\,October\,30 at 22:02:07\,UT  and subsequently \textit{Wind} on 2021\,October\,31 at 9:33:16\,UT. Table\,\ref{tab:IPs} lists the shock parameters used in the present analysis. The IPs crossed both s/c in an oblique configuration ($\theta_{Bn} \sim 45 -60^\circ$ at both s/c). The two spacecraft were close to radial alignment (see Fig.\ref{fig:orbit}) at the time of the eruption. Thus, this event offers one of the first opportunities to measure the evolution with heliocentric distance of the energy partition at IPs, including not only the ram, thermal, and magnetic energies but also the contribution from energized particles (both ions and electrons). \textit{SolO} in-situ measurements revealed no energetic ($\gtrsim 100$\,keV) ions at the shock \citep{Trotta2023c,Dimmock2023}, which are generally expected at such a Mach number shock (see Table \ref{tab:IPs}). In the present paper we make use of multi-spacecraft in-situ measurements to provide insight on the low and high energy cutoff for energetic particles probing the shock at a different spatial locations.

\begin{table}
\begin{center}
\caption{\label{tab:IPs}Parameters for the 2021\,October\,30 and 31 shock at both \textit{SolO} and \textit{Wind}, determined with \citet{Szabo:94} and \citet{Vinas.Scudder:86}  technique for \textit{SolO} and \citet{Vinas.Scudder:86} technique for \textit{Wind}. 
}
\begin{tabular}{l l l}
\tableline\tableline
Parameter & \textit{SolO} & \textit{Wind} \\
\tableline
Passage Time (UT) & 22:02:07 Oct~30 & 09:33:16 Oct~31 \\ 
Speed [km\,s$^{-1}$] & $413.1 \pm 2.8$ & $437.7 \pm 2.2$ \\
Normal (RTN) & [0.743, $-0.003$, & [0.801, 0.299,\\
&  $-0.669$] & $-0.518$] \\
Normal $1\sigma$ error & [0.009, 0.017, & [0.009, 0.015, \\
& 0.010] & 0.014] \\
$\theta_{Bn}$ [$^\circ$] & $43.7 \pm 8.8$ & $58.7 \pm 8.5$ \\
$M_{fms}$ & $3.6 \pm 0.3$ & $2.59 \pm 0.09$ \\
\tableline\tableline 
\end{tabular}
\end{center}
\end{table}

\subsubsection{\textit{SolO} and \textit{Wind} Magnetic Field and Plasma Data}
\begin{figure}
\includegraphics[width=0.48\textwidth]{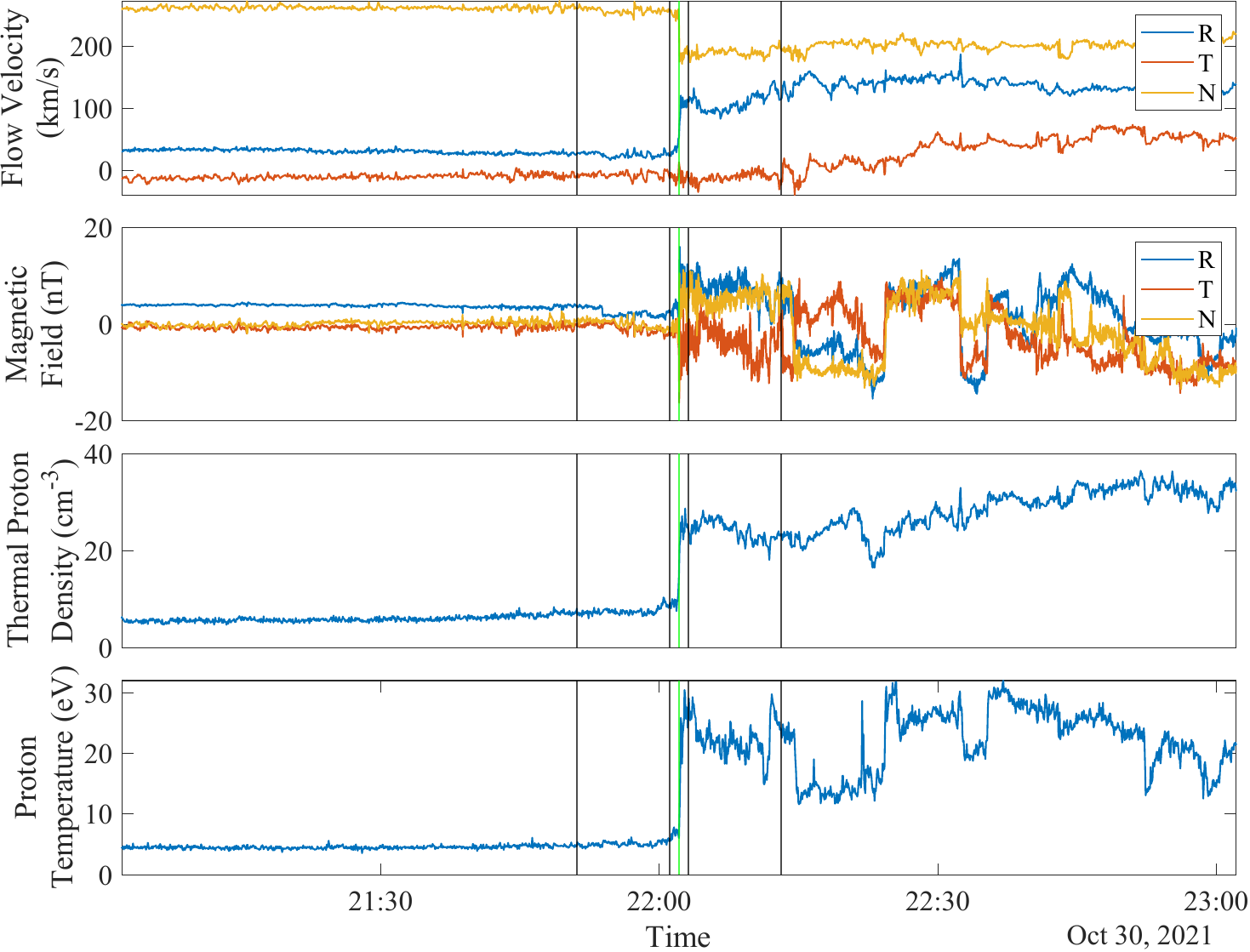}
\caption{\textit{SolO} time-series of flow velocity components in the shock frame (not in the spacecraft frame), magnetic field components in RTN coordinates, and proton density and temperature for the 2021\,October\,30 shock. The vertical green line in each panel denotes the time of the shock passage and the four vertical lines indicate the start and end times of the upstream and downstream averaging intervals for the IPs analysis in Section \ref{sec:shocks}.}
\label{SolO_Oct30_time_series.fig}
\end{figure}

\begin{figure}
\includegraphics[width=0.48\textwidth]{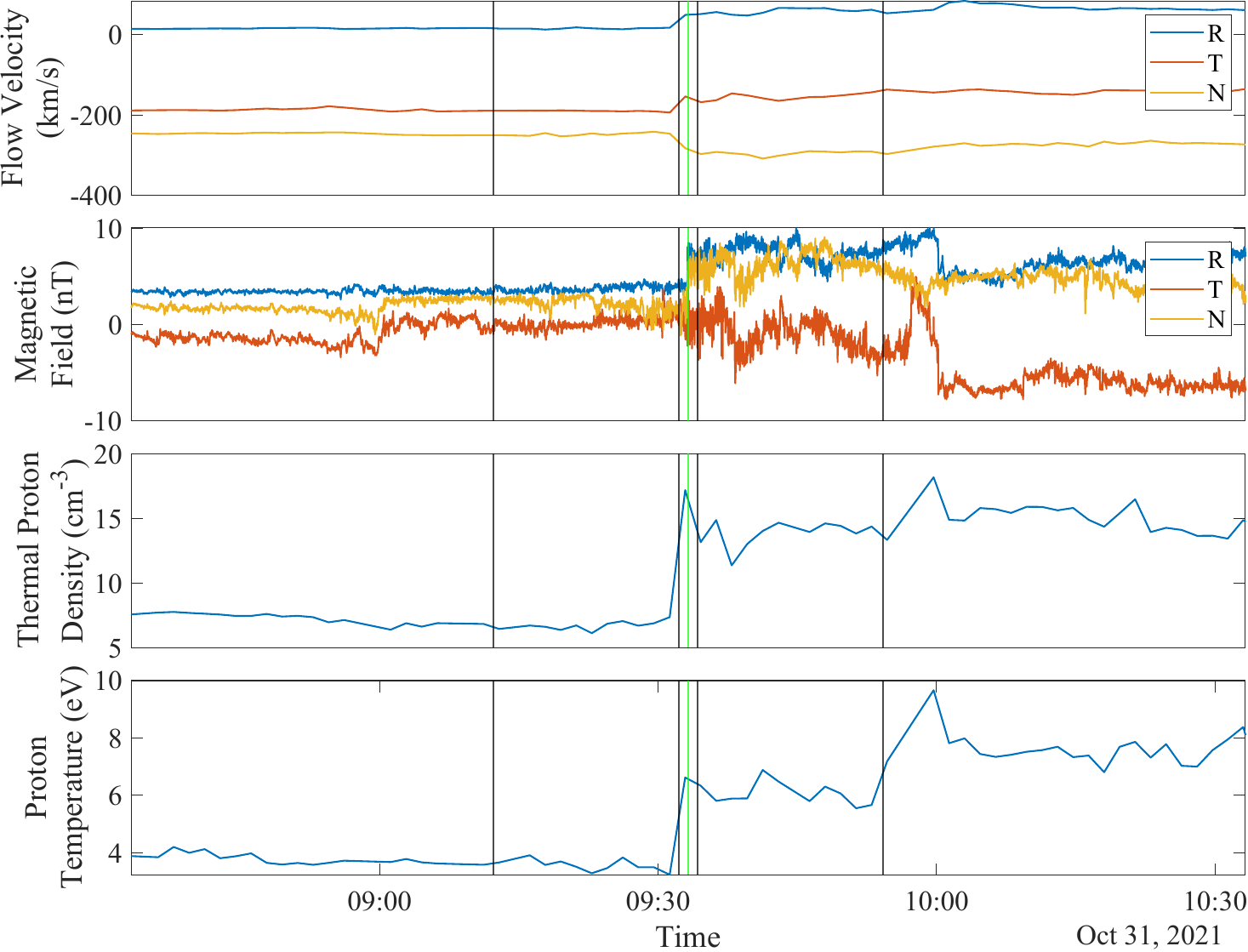}
\caption{\textit{Wind} time-series of flow velocity components in the shock frame (not in the spacecraft frame), magnetic field components in RTN coordinates, and proton density and temperature for the 2021\,October\,31 shock. The vertical green line in each panel denotes the time shock of shock passage and the four vertical lines indicate the upstream and downstream averaging intervals for the IPs analysis in Section \ref{sec:shocks}.}
\label{Wind_Oct30_time_series.fig}
\end{figure}

{For the plasma environment at \textit{SolO} (see Fig.\ref{SolO_Oct30_time_series.fig}), we use the magnetic field measured with a resolution of 64 vectors$/$s by MAG \citep{Horbury.etal:20}. Ion bulk flow, density and temperature are the
plasma moments measured by the Proton Alpha Sensor of the Solar Wind Analyser suite~\citep[SWA-PAS,][]{Owen.etal:20}, with a 3 second time resolution.} For \textit{Wind} in-situ measurements (see Fig.\ref{Wind_Oct30_time_series.fig}), we use the Magnetic Field Instrument \citep[MFI,][]{Lepping.etal:95} 11 vectors$/$s magnetic field measurements in RTN coordinates \edit1{\citep{Koval2023}} and 92-second proton density, bulk velocity and temperature data\footnote{\edit1{These data are determined with non-linear fitting of the current distribution function from the Solar Wind Experiment \citep[SWE,][]{Ogilvie.etal:95, Ogilvie2021} suite.}}. 
These quantities are also plotted in Figures\,\ref{SolO_Oct30_time_series.fig} and \ref{Wind_Oct30_time_series.fig}. 

Shock parameters at \textit{SolO} are computed using a least-squares fit of the full set of Rankine-Hugoniot conservation equations to the magnetic field and thermal plasma measurements \citep{Szabo:94, Vinas.Scudder:86}, assuming thermal equilibrium of electrons and protons. For the shock parameters at \textit{Wind}, we use the publicly available solution from the CfA shock database\footnote{https://lweb.cfa.harvard.edu/shocks}. Among different methods for shock parameter determination included in the CfA database, for this shock the selected method is RH08, which is a least-squares fit of the reduced set of Rankine-Hugoniot conservation equations devoid of temperature terms\footnote{Since the shock normal direction in the CfA database is in GSE coordinates, we convert it to RTN coordinates.}.

\subsubsection{\textit{SolO} and \textit{Wind} Energetic Particle Data}
\begin{figure}
\includegraphics[width=0.48\textwidth]{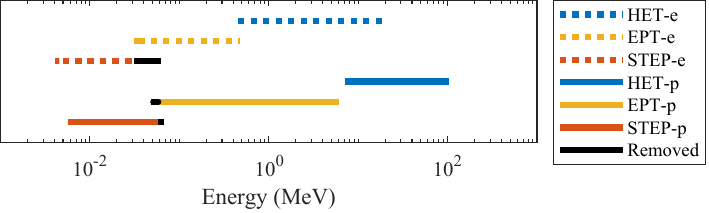}
\includegraphics[width=0.48\textwidth]{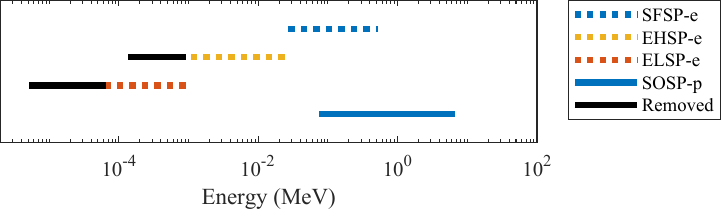}
\caption{\label{solo energy ranges.fig} Top panel: Energy ranges for the \textit{SolO}/STEP, EPT, and HET instruments used in this study, where solid lines correspond to protons and dashed lines to electrons. The black lines denote the energy ranges that were removed due to overlap.Bottom panel: Energy ranges for the \textit{Wind} data products used in this study, where solid lines correspond to protons and dashed lines to electrons. The black line over EHSP-e denotes the energy range that was removed 
due to overlap with ELSP-e. The black line over ELSP-e indicates the energies that correspond to the thermal electron spectrum and were also removed.
}
\end{figure}

For \textit{SolO}, we use data from the SupraThermal Electron Proton (STEP) instrument, the Electron Proton Telescope (EPT), and the High-Energy Telescope (HET).  The energy ranges for all instruments are plotted in the top panel of Figure\,\ref{solo energy ranges.fig}.  We average the signal across all 15 SSDs present in each channel.
Details on the instruments can be found in \citet{RodriguezPacheco2020} and \citet{Wimmer2021}. From STEP, we use the data product\footnote{The STEP electron differential intensities for each energy channel were calculated from the available data products as Electron\_Flux\_01\_Mult$\times$ (Integral\_01\_Flux - Magnet\_Flux\_01).} containing proton energies $5.73-68.0\,\text{keV}$ in 24\,bins (STEP-p) and $4.08-62.7\,\text{keV}$ electrons in 32\,bins (STEP-e) both with a one-second cadence. 

Among the 4 apertures, the flux is not isotropic: the EPT partial pressures, integrated over the entire EPT energy range \citep[see][and Sect.\ref{section:shocks} below for details]{David2022} after the shock, normalized to the Sun-facing aperture, are $1$, $0.40$, $0.55$, and $0.65$ in the {\it Sun}, {\it Asun}, {\it North}, and {\it South} apertures, respectively.
The greater particle pressures (or fluxes) of the two apertures {\it Sun} (facing the Sun) and {\it South} (facing southward) can be explained with the source location with respect to \textit{SolO} at the time of shock passage: {\it Sun} indicates a source (shock) located inside \textit{SolO} orbit and {\it South} a source mostly coming from southern heliolatitudes; consistently with this geometry, remote observations show a CME directed southward. Thus, we discard fluxes from the Asun and North apertures, for both EPT and HET, and sum the fluxes from the Sun and South apertures only, over a $2\pi$ solid angle. We use the data product containing $49.5\,\text{keV}-6.13\,\text{MeV}$ protons 
(EPT-p) and $31.2-471\,\text{keV}$ electrons 
both with a 5-second cadence. We use the HET data product containing $7.05-105\,\text{MeV}$ protons 
(HET-p) and $453\,\text{keV}-18.8\,\text{MeV}$ electrons 
(HET-e) both with a 5-second cadence. As with the EPT data, for HET we use only the fluxes from the Sun and South apertures.

There exists overlap in the energy ranges of STEP-p and EPT-p which we removed by excluding the upper two and lower three bins, respectively. This splitting was chosen to minimize the remaining overlap of $470\,\text{eV}$ ($57.8\,\text{keV}-58.3\,\text{keV}$) which is small enough to not affect the energetic proton contribution to the shock. Data are unavailable between the EPT-p and HET-p ranges from $6.13-7.05\,\text{MeV}$, but due to the small fluxes and thus energy contributions of particles at such energies, this gap is likely to have no effect on our analysis. In addition, the overlap between STEP-e and EPT-e was removed by excluding the upper eight bins of STEP-e. 

For the \textit{Wind} spacecraft, the omnidirectional flux of  energetic protons (9\,channels in the range $70\,\text{keV}$ to $6.8\,\text{MeV}$) is retrieved from Three-Dimensional Plasma and Energetic Particle Investigation (3DP) instrument (SOSP-p data product) with a 12.5-second cadence. \textit{Wind} omnidirectional flux of energetic electrons data from 3DP is retrieved from three data products: ELSP-e ($5.18\,\text{eV}-1.11\,\text{keV}$
), EHSP-e ($136\,\text{eV}-27.6\,\text{keV}$) and SFSP-e ($27.0\,\text{keV}-517\,\text{keV}$ 
) with 100-second, 100-second, and 12.5-second cadences, respectively \citep{Lin.etal:95}. The overlap between ELSP-e and EHSP-e was eliminated by removing the first six EHSP-e bins, leaving a $226\,\text{eV}$ gap ($1.11\,\text{keV}-1.34\,\text{keV}$) in the flux data. We did not compensate for the small overlap ($27.0\,\text{keV}-27.7\,\text{keV}$) between SFSP-e and EHSP-e since removing either the highest EHSP-e or lowest SFSP-e energy bin would leave a large energy range ($18.9\,\text{keV}-27.0\,\text{keV}$ or $27.7\,\text{keV}-40.1\,\text{keV}$, respectively) unaccounted for. The energy ranges used and removed are summarized in the bottom panel of Figure \ref{solo energy ranges.fig}. All data were downloaded from the NASA CDAWeb database\footnote{\url{https://cdaweb.gsfc.nasa.gov/}}.

\

\section{Results: Energetics}
\label{section:energy}
    \begin{figure*}[!ht]
\centering
\includegraphics[width=0.96\textwidth]{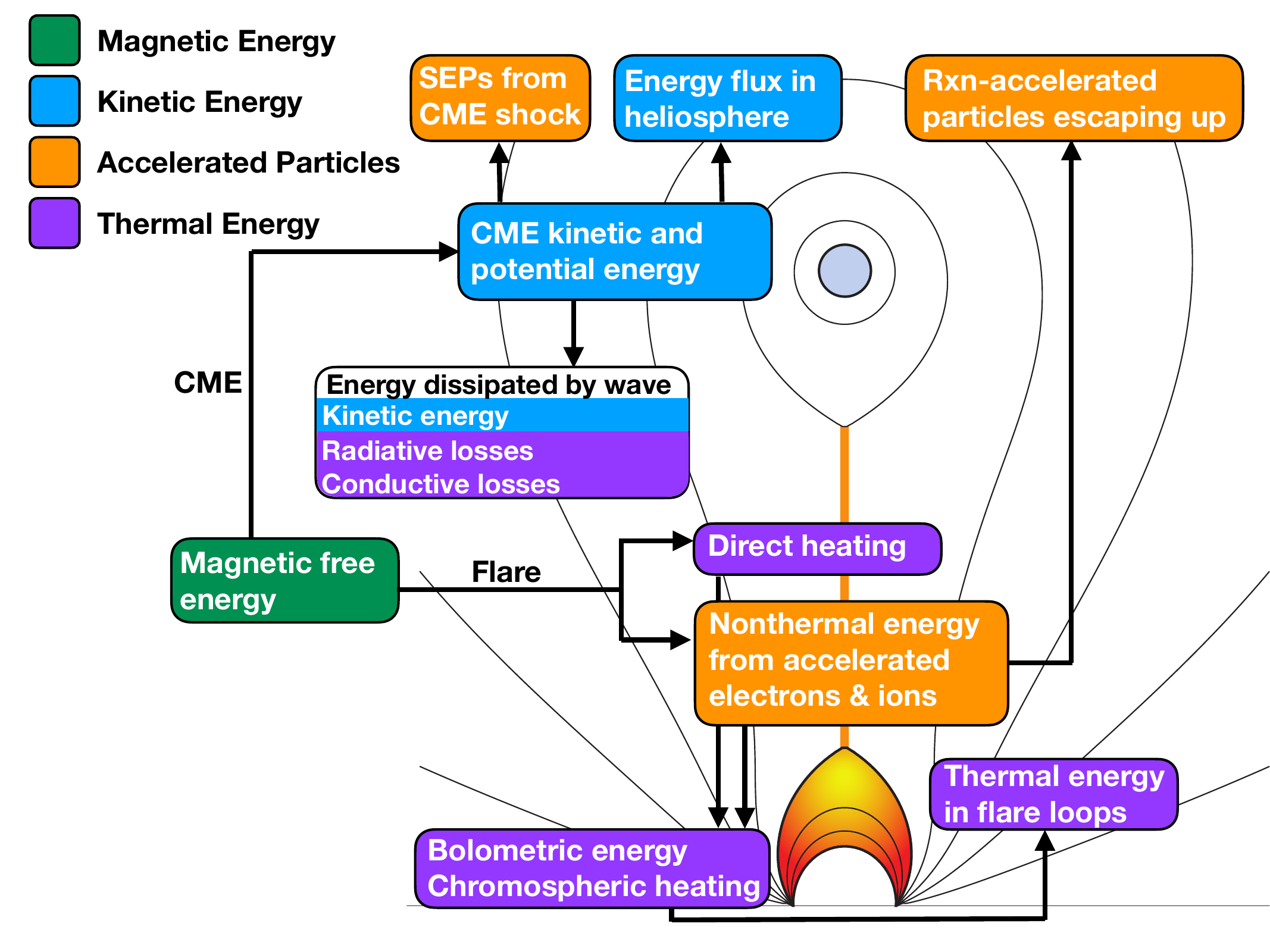}
\caption{\label{fig:energy_partition} Energy partitioning and flow in the 28 October 2021 event.  Underlying CME cartoon modified from \citet{Reeves2005}.}
\end{figure*}

In this section, we undertake the ambitious task of cataloging the energy build-up and release for the well-observed eruption that occurred on 28 October 2021, originating from AR\,12887. Figure~\ref{fig:energy_partition} summarizes the energy flow over the course of the event. Stored magnetic energy is initially converted into the kinetic energy of the CME, direct heating in the \edit1{supra-arcade} region, and accelerated particles due to reconnection.  These quantities then cascade into other forms of energy as the event progresses, such as the energy deposited in the chromosphere that eventually drives heated plasma into flare loops, and accelerated particles across the CME shock. We do not discuss the accelerated particles escaping anti-sunward due to reconnection in this paper. These particles represent a small fraction of the total event energy; a detailed discussion of these particles is given by \citet{Klein2022}.

\begin{table}
\begin{center}
\caption{Energy partition into flare- and CME-associated strucures. \label{energy_partition.tab}}
\begin{tabular*}{\linewidth}{@{\extracolsep{\fill}} l c r}
\tableline\tableline
 Type & Energy\tablenotemark{a} & \multicolumn{1}{c}{Method}  \\
\tableline
{\it Primary} & &\\
\hspace{0.2cm}Magnetic free & 8.2$\pm$2.6  & NLFFF\\
{\it Flare} & & \\
\hspace{1em}Nonthermal Energy & & \\
\hspace{2em}electrons (footpoints) & 1.6$\pm$ 0.4 & STIX \\
\hspace{2em}electrons (loops) & 0.05$^{+0.01}_{-0.02}$ & EOVSA \\
\hspace{2em}\edit1{electrons (decay)} & \edit1{0.01--0.02} & \edit1{Fermi/GBM} \\
\hspace{2em}ions & 0.6$\pm 0.8$  & Scaling Law \\
\hspace{1em}Bolometric Energy  &  & \\
\hspace{2em}total & 1.6--3.0 & Scaling Law \\
\hspace{1em}Deposited in Chrom.  &  & \\
  \hspace{2em}prompt & 1.2 $\pm$ 0.4 & UFC \\
  \hspace{2em}tail & 0.5 $\pm$ 0.15 &  UFC \\
\hspace{1em}Thermal Loop Energy  &  & \\
\hspace{2em}peak (\S\ref{subsec:thermal_DWL}) & 0.6 & GOES Flux \\
\hspace{2em}peak (\S\ref{sec:AIA_erg}) & $0.49^{+0.07}_{-0.06}$ & DEMs\\
\hspace{2em}tail & 0.3 & GOES Flux\\
{\it CME} & & \\
\hspace{1em}\edit2{Kinetic @ 14.87 R$_{\odot}$} & \edit2{4.12$\pm$2.77} & \edit2{COR2 data}\\
\hspace{1em}\edit2{Potential @ 14.87 R$_{\odot}$} & \edit2{4.73$\pm$2.33} & \edit2{COR2 data}\\
\hspace{1em}EUV Wave (total) & 0.9$\pm$0.4 & Patsourakos \& \\
 & & Vourlidas (2012) \\
\tableline\tableline
\tableline\tableline
Total & Energy\tablenotemark{a} & \multicolumn{1}{c}{Fraction}  \\
\tableline
Magnetic free & 8.2$\pm$2.6  & 100\% \\
Flare & 
{2.25$\pm$0.9} & 
{13--56\%}\\
CME & \edit2{9.75$\pm$3.6} & \edit2{$>$56\%} \\
\tableline\tableline
\end{tabular*}
\end{center}
\vspace{-3ex}\tablenotetext{a}{Energies are in units of 10$^{31}$ ergs.}
\end{table}

Using the available observations coupled with scaling laws derived from from previous studies of ensemble events \citep[i.e.][]{Kretzschmar,emslie12}, we make estimates of the following quantities, organized by section:

\begin{itemize}[label={}, noitemsep, leftmargin=1em]
    \item \S\ref{section:mag}: Magnetic energy
    \item \S\ref{section:nonthermal}: Energy in non-thermal electrons 
    \item \S\ref{subsec:ions}: Energy in the non-thermal ions
    \item \S\ref{subsec:bolo}: Bolometric energy
    \item \S\ref{section:chromosphere}: Energy deposited in the chromosphere
    \item \S\ref{section:thermal}: Thermal energy radiated in the flare loops
    \item \S\ref{subsec:wave_energetics}: Energy in the EUV wave
    \item \S\ref{section:kinetic}: CME kinetic \& gravitational potential energy
    \item \S\ref{section:cme}: CME energy fluxes in the heliosphere
    \item \S\ref{section:shocks}: Energy partition at CME shock
\end{itemize}
These calculations give us a very detailed diagnosis of the energy partition in this event, tracking energy flow all the way from storage, prior to the onset of the eruptive flare, through many key phenomena during the event, and beyond, to the energy flux in the CME as it passes through the heliosphere. In several cases, we use multiple methods or estimates of related parameters to validate our energy calculations and find consistent values, which provides confidence that the estimates we report here are reasonable. 

\edit1{Table~\ref{energy_partition.tab} summarizes the \edit2{full accounting of the} energy partition into the primary forms of energy present in flare- and CME-associated structures for this event.} \edit2{It also presents total energies for the flare and CME components.}   \edit1{The \edit2{total} flare energy is derived from the energy in the nonthermal electrons (\S\ref{section:nonthermal}) and ions (\S\ref{subsec:ions}).  The \edit2{total} energy partitioned into the CME is given by the kinetic and potential energy (\S\ref{section:kinetic}) and the energy dissipated from the EUV wave (\S\ref{subsec:wave_energetics}). The range in the Fraction column is derived based on the extrema of the values, given the errors.  Direct heating of plasma in the supra-arcade region is also a possible primary result of the eruption reconnection (see Figure~\ref{fig:energy_partition}), but we cannot measure this quantity directly due to the superposition of supra-arcade structures and flare loops, which are a secondary structure formed due to chromospheric evaporation.  However, studies examining limb flares have shown that the thermal energy release due to the direct heating component is small compared to the thermal energy found in the flare loops \citep{roy_thermal}. }

\edit2{We note that the total energy expended during the event is not simply the sum of the items in Table~\ref{energy_partition.tab}, as many lines essentially count the same energy in different ways. For example, much of the flare-associated heating is driven by the energetic particles, and thus adding these two lines would double count much of this energy. Details of the calculations of each of the energy terms are given in the following sections. }

\subsection{Magnetic Energy}
\label{section:mag}

\begin{figure}[h!]
\includegraphics[scale=1]{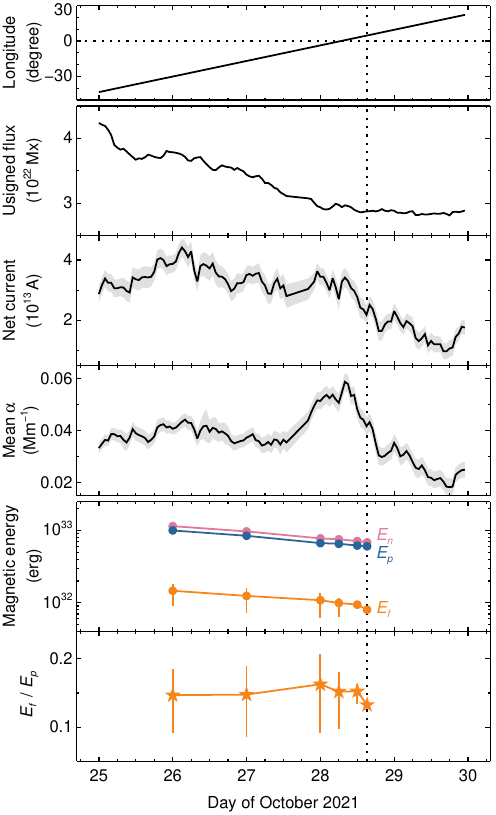}
\caption{\label{mag_quantites.fig}
Magnetic properties of AR\,12887. From top to bottom: Stonyhurst longitude of the AR centroid, unsigned magnetic flux, net electric current, mean torsional parameter $\alpha$, magnetic energy from extrapolations, and ratio between the magnetic free energy ($E_f$) and the potential energy ($E_p$). Blue, pink, and orange symbols show the results for NLFFF, potential field, and magnetic free energy, respectively. The error bands for the second to fourth panels are $3\sigma$ values from the SHARP database. The six magnetic field model times are $1$, $2$, $3$, $3.25$, $3.5$, and $3.63$ days after Oct 25 00~UT. The error bars for the last two panels show the minimum and maximum of all the extrapolation models that pass the solenoidality test at each time. No error bar is provided for the last time step because only one model passes the test. The vertical dotted line indicates the GOES flare start time.
}
\end{figure}

\begin{figure}[h!]
\includegraphics[scale=1]{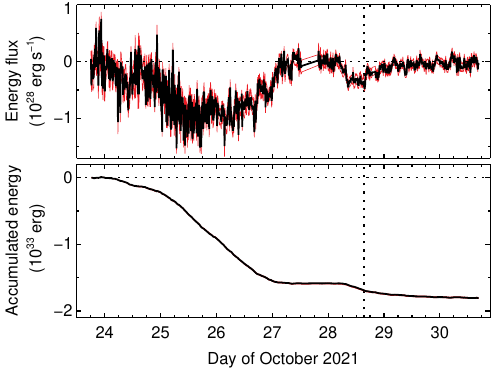}
\caption{\label{energy_flux.fig}
Magnetic energy flux of AR\,12887 based on the PDFI method. Top: instantaneous energy flux. Bottom: accumulated energy since 2021\,October\,24\,00:00\,UT. The values are from the PDFI database \citep{hoeksema2020}. Red curves show $2\sigma$ formal uncertainty estimate based on a Monte-Carlo method; the uncertainty is small for the accumulated energy. Flux is taken to be zero during HMI data gaps. The vertical dotted line indicates the GOES flare start time.
}
\end{figure}

We evaluated the photospheric magnetic properties of AR\,12887 using the Space-weather HMI Active Region Patches \citep[SHARP;][]{bobra2014} over five days, shown in Figure\,\ref{mag_quantites.fig}. As a consequence of magnetic flux cancellation, the extensive properties (e.g., totals) generally decreased over time. For example, the unsigned magnetic flux and the net electric current decreased by about $34\%$ and $66\%$, respectively. On the other hand, the intensive properties (e.g., means), such as the mean torsional parameter (measure of magnetic twist), exhibited an increase starting late on 2021\,October\,27, when the central small bipole started to form. As apparent flux cancellation commenced in the small bipole on 2021\,October\,28, mean $\alpha$ started to decrease as well, dropping by about $69\%$ over the next $1.4$\,days.

To assess the coronal magnetic field energy, we performed nonlinear force-free field (NLFFF) and potential field extrapolations \citep{wiegelmann2012} at selected times leading to the flare. For each time step, four different models were calculated using different sets of free model parameters following \citet{Thalmann2020}. The differences between these four models allow for an estimate of the systematic error. Further, we calculated the energy for the non-solenoidal component of the modeled coronal field \citep{Valori2013}, and rejected those with large errors. The spread is generally small for the NLFFF ($E_n$) and PF ($E_p$) energy: the maximal difference between models is $2\%$--$8\%$ of the mean. However, the spread is much larger for the magnetic free energy $E_f$, defined as the difference between the NLFFF and potential field energy ($E_n-E_p$). The spread can be as large as $70\%$ of the mean: significant systematic uncertainty is expected. Additionally, uncertainty due to spectropolarimetric noise can be estimated from pseudo Monte Carlo method. For HMI magnetograms, different noise realizations will lead to a typical relative error of $10\%$-$20\%$ for $E_f$ \citep[e.g.,][]{sun2012}.

\edit2{We note that the systematic uncertainty here is likely a lower bound. The widely used potential field extrapolation method adopts a Neumann boundary condition, that is, the vertical field $B_z$ on the lower boundary is required to match the observation exactly. This approach overfits the magnetogram in the presence of noise, and the consequence is not entirely clear. In an illuminating example, \citet{welsch2016} combine the Neumann and the Dirchlet (in the form of $\nabla_h \cdot \bm{B_h}$) boundary condition weighted by their relative noise level. The method seeks a solution that best fits the observed magnetic field \emph{vector} in a statistical sense, and the extrapolated $B_z$ is statistically similar to the observation in terms of noise. Surprisingly, the “hybrid” potential field solution has a much lower magnetic energy than the fiducial method (by $15\%$—$38\%$), which suggests that our free energy may be significantly underestimated.}

Over the $3.6$\,days leading to the flare, we found that $E_n$ decreases from $1.1 \times 10^{33}$ to $6.8 \times 10^{32}$ erg, by about $4.6\times10^{32}$~erg, or $40\%$ (Figure\,\ref{mag_quantites.fig}). This decrease is closely related to the apparent flux cancellation. The magnetic free energy at 2021\,October\,28\,15:12\,UT, right before the X-class flare, is about $8.0 \times 10^{31}$\,erg, about 13\% of the potential field energy. 

We additionally assessed the change of coronal magnetic energy by integrating the estimated magnetic energy flux through the photosphere. To this end, photospheric electric field $\bm{E}$ was to be inferred using the ``PTD-Doppler-FLCT Ideal'' \citep[PDFI;][]{kazachenko2014} technique. The vertical component of the Poynting flux vector $\bm{E} \times \bm{B}$ was then integrated over the AR and in time. We found significant negative energy flux from 2021\,October\,25--27, and on 2021\,October\,28 (Figure\,\ref{energy_flux.fig}), consistent with the drop in photospheric proxies (Figure\,\ref{mag_quantites.fig}) and is likely caused by the magnetic flux cancellation. The integrated energy flux from 2021\,October\,25\,00:06\,UT to 2021\,October\,28\,15:06\,UT is estimated to be $1.5\times10^{33}$\,erg, about $3.2$\,times the change of energy from the coronal extrapolation model $E_n$. We have performed an additional Monte Carlo test to estimate the effect of random noise on the estimate \citep{avallone2020}. The effect is rather small on the accumulated energy; we expect the system uncertainty that is not well understood to dominate.

\subsection{Energy in Non-Thermal Electrons}
\label{section:nonthermal}
\subsubsection{Non-thermal Energy \edit1{from X-rays}} \label{section:STIX}

The energy in the non-thermal electrons is estimated from the STIX observations. Figure~\ref{fig.stix_lc} shows that the majority of the emission from hard X-rays comes from the two footpoints in this event. Following \cite{emslie12}, we perform a spectral fit to the STIX observations, to constrain the injected electron spectra $F_{0}(E_{0})$. \edit1{We have used the thermal model `\texttt{f\_vth}' and cold thick target model `\texttt{thick2}' publicly available in the OSPEX package to fit the STIX spectrum as a function of time. Throughout the fitting process, only the high-energy cutoff was kept fixed at $\mathrm{3.2~\times~10^{4}~keV}$. The typical low energy cutoff is within the 10{--}15 keV range.} \citep[For further details regarding the fitting procedure, please refer to][]{roy_thermal}. The constrained electron energy distribution spectra is used to estimate the instantaneous energy deposited by the electrons in the foot points as: 
\begin{equation}
U_{e}(t=t')~=~A~\Delta t\int_{E_{min}}^{E_{max}}~E_{0}F_{0}(E_{0})(t=t')dE_{0},  
\label{eq:non_therm}
\end{equation} where the electron energy distribution in constrained by fitting the count spectra for the time interval 
\begin{equation}
t'-\Delta t/2 \le t \le t'+\Delta t/2.
\end{equation}
The cumulative non-thermal energy deposited at the foot points as a function of time is estimated by adding the instantaneous non-thermal energy from every time bin: 
\begin{equation}
U^{cumulative}_{e}(t)~=~\sum_{t'=t_{0}}^{t}U_{e}(t').
\end{equation}

\begin{figure}
    \includegraphics[width=0.5\textwidth,trim={1cm 0cm 1cm 1.8cm},clip]{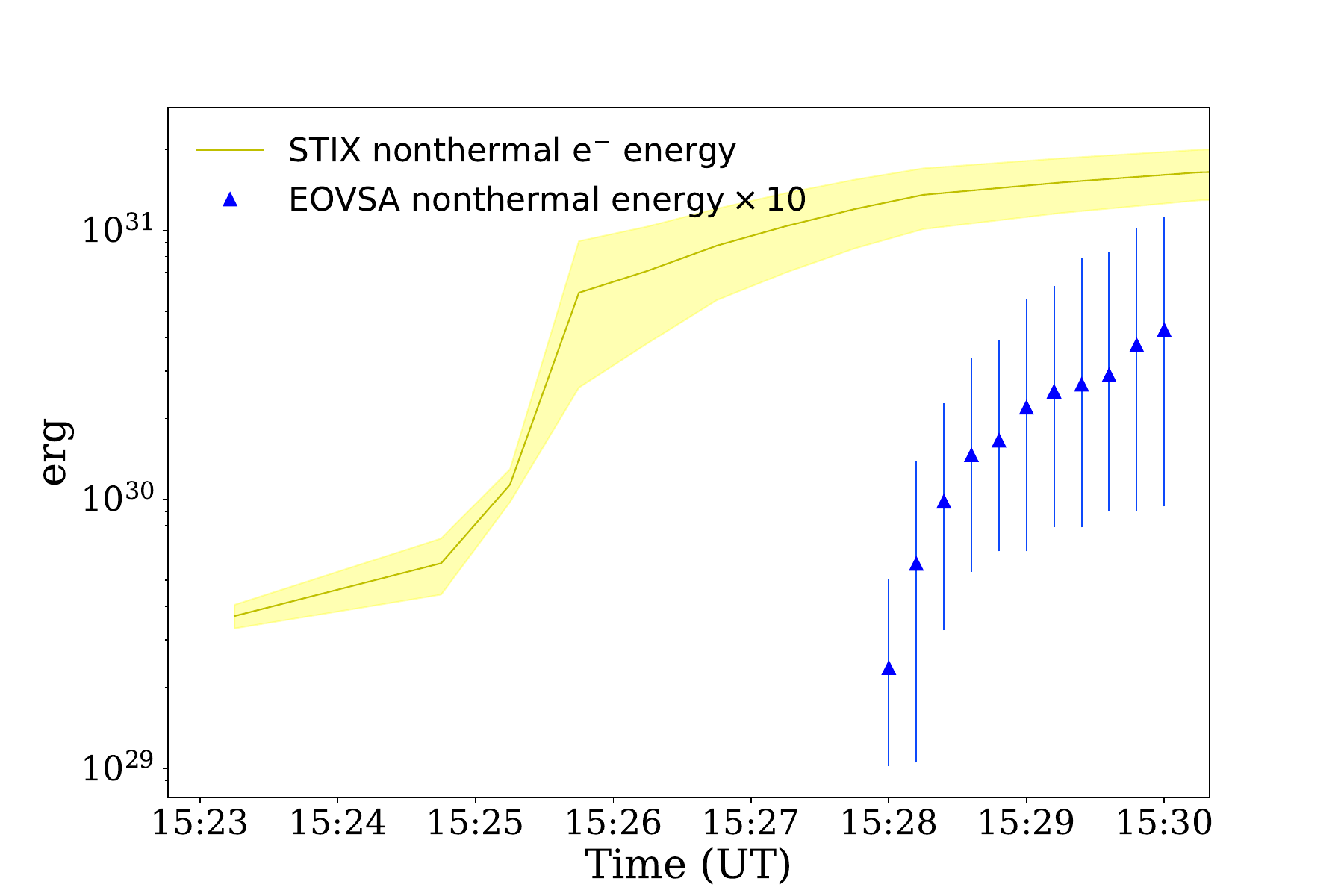}
    \caption{Comparison between nonthermal energies calculated from STIX data and EOVSA. Yellow shading on the STIX data indicates the error bounds. Note that the EOVSA energies are multiplied by a factor of 10. \label{stix_nonthermal.fig}}
\end{figure}

The cumulative non-thermal energy as a function of time, as inferred from STIX observations is shown in Figure \ref{stix_nonthermal.fig} (solid line with yellow shading) in comparison to cumulative non-thermal energy constrained from EOVSA microwave observations (discussed in \S\ref{sec:microwave}) shown with blue triangles. The energy deposited via the non-thermal \edit1{electrons} at the foot-points levels off at $\sim~(1.6~\pm~0.4)\times10^{31}$~ergs. The uncertainty in the energy deposited is calculated by propagating the fit parameter uncertainties in Equation~\ref{eq:non_therm}.

\edit1{During the decay phase starting at 15:50 UT, we use \textit{Fermi} GBM to calculate the cumulative energy for non-thermal energy in electrons above 15 keV.  We use two different cutoff energies fixed in the spectral fitting, one at 15 keV and one at 20 keV. These values represent the spread obtained when the cutoff energy is a free parameter in the fitting process. Both values lead to equally satisfying $\chi^2$ values. We find that the maximum value of the cumulative energy from non-thermal electrons in this phase is about $0.01-0.02 \times 10^{31}$~ergs. Unfortunately energetic particle hits on the STIX detector after 15:50~UT make that data unreliable, so it is not possible to directly compare the two measurements. }
\subsubsection{Microwave spectral diagnostics} 
\label{sec:microwave}
The microwave spectra obtained from EOVSA enable us to constrain the properties of the
nonthermal electron distribution during the flare based on the nonthermal gyrosynchrotron emission model.  The gyrosynchrotron emission spectrum depends on several parameters: the vector magnetic field, the nonthermal electron distribution, the density and temperature of the thermal plasma, the area of the emission source, and  the depth along the line of sight (LOS). Following \citet{chen2020a}, we perform a Markov chain Monte Carlo (MCMC) analysis to explore this parameter space and fit the observed spectra. For the spectral modeling, we use the fast gyrosynchrotron modeling codes developed in \citet{fleishman2010} and \citet{kuznetsov2021}. We assume that the nonthermal electron distribution is isotropic and and has a power-law form $f(E) = dn_{\rm nth}/dE \propto E^{-\delta'}$, where $n_{\rm nth}=\int_{E_{\rm min}}^{E_{\rm max}}f(E)dE$ is the total nonthermal electron density, with ${E_{\rm min}}$ and ${E_{\rm max}}$ being the low- and high-energy cutoff, respectively, and $\delta'$ is the power-law index. $E_{\rm max}$ are fixed at 10 MeV respectively. We denote the magnetic field as $B$, density of thermal electrons as $n_{th}$, angle between the magnetic field and LOS by $\theta$. We assume that the temperature of the gyrosynchrotron emitting plasma is 7 MK and its depth along the LOS is 7.2 Mm. Due to poor spatial resolution of the data, we have modeled the integrated flux density of the flaring source.  

While the flare source is inhomogeneous in general, we assume a homogeneous distribution of plasma parameters within the source to limit the degrees of freedom during the spectral modeling procedure to a manageable number.  This approximation holds reasonably well in case of well-resolved sources, but  the poor spatial resolution of the EOVSA data in this case was an issue. In particular, assuming the entire flare source has the same source area over 3--15 GHz, over the emission height is an over-simplification. We have devised a simple method to take into account the possible variation in source area during spectral modeling. We assume that the fractional error incurred due to our assumption of equal area over the entire frequency range will be same as the fractional change of the total area of the source over the same frequency range. The total source area was obtained from the images produced using standard techniques\footnote{\url{http://www.ovsa.njit.edu/wiki/index.php/Calibration_Overview},\url{https://github.com/suncasa/suncasa-src/blob/master/examples/eovsa_flare_slfcal_example.py}}.

\begin{figure}
    \centering
    \includegraphics[scale=0.5]{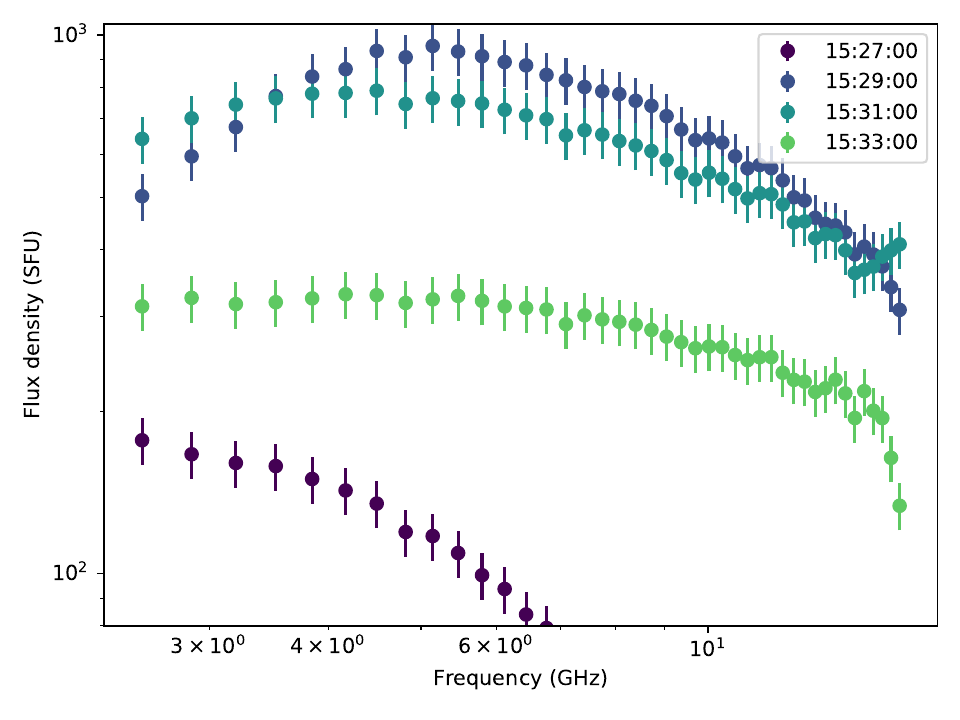}
    \caption{The temporal evolution of the microwave spectrum from EOVSA.  The flattening of the spectra as time goes on indicates a decreasing number of nonthermal electrons in the flare loop system.}
    \label{fig:spectral_evolution}
\end{figure}

In Figure \ref{fig:spectral_evolution}, we show the variation of the microwave spectra with time\footnote{The flux scale of EOVSA is determined regularly by matching the observed EOVSA flux with that reported by the Radio Solar Telescope Network (RSTN). The RSTN flux products do not include formal statistical uncertainties. Inter-comparisons between RSTN sites and with independent instruments indicate that the reported flux densities are typically accurate to within $\sim$5--10\% \citep{Giersch2022}. Therefore, we have taken a fiducial 10\% systematic uncertainty on the absolute flux, shown as the error bars in the figure.}. We find that around 15:27 UT, the spectral peak is at frequency $\lesssim 3$GHz, indicating small quantities of nonthermal electrons prior to 15:27 UT, which is also during the start of the flare. At 15:29 UT, the peak of the spectrum is about 5 GHz, indicating enhancement in the number density of nonthermal electrons. Around 15:31 UT, the spectrum becomes flat, a feature that persists at 15:33 UT. This flattening indicates that by this time, not many nonthermal electrons are present in the system, which is also consistent with the STIX results (see Section \ref{section:STIX}). 

We have used the MCMC chains to constrain the evolution of the energy flux of nonthermal electrons, with energy above 20 keV. The results are shown in Figure \ref{stix_nonthermal.fig}, compared with the energy derived from STIX. The error bars correspond to the $67.5$\% confidence interval. \edit1{We note that substantial uncertainties of the total energy flux of the microwave-emitting electrons may also arise from the assumed homogeneous source and single-power law distribution of the source electrons.} We find that while the total energy estimated from the microwave data is much smaller than that estimated from STIX data, likely
due to the fact that X-rays and microwaves probe different regions of the flare. While the microwave observations primarily probe nonthermal electrons trapped inside the flaring loop, at the loop top region, X-ray observations probe nonthermal electrons which are deposited at the chromosphere. 
Nevertheless, interestingly, we find that the temporal evolution of the nonthermal energy estimated from both X-rays and microwave data is qualitatively similar in that they both start to flatten around 15:30 UT.

\subsection{Energy in Non-Thermal Ions}
The best way to measure flare-accelerated ions is using the 2.223 MeV neutron-capture gamma-ray line, for example as observed by {\it RHESSI} \citep[e.g.][]{Shih2009}.  Unfortunately this data does not exist for the 28 October 2021 flare, so we use a scaling law found by \citep{emslie12} for an ensemble of fourteen events with suitable {\it RHESSI} data:
\begin{equation}
    E_{nti}/E_{nte} = 0.34 \pm 0.5
    \label{eq:ions}
\end{equation}
where $E_{nte}$ is the energy in the nonthermal electrons and $E_{nti}$ is the energy in the nonthermal ions.  This relation was also used by \citet{Aschwanden2017} in their estimates of nonthermal ion energy.

Using Equation \ref{eq:ions}, we find that the energy in the nonthermal ions is on the order of $0.6 \pm 0.8 \times 10^{31}$ ergs.  Together with the accelerated electron energies found from EOVSA and STIX, this estimate gives a total of $ \sim 2.25 \times 10^{31}$ ergs of energy in the accelerated particles in this event.

\edit1{It is worth noting that this event was detected with \textit{Fermi} Large Area Telescope \citep[LAT,][]{Atwood2009} showing the presence of ions\footnote{\edit1{See https://hesperia.gsfc.nasa.gov/fermi/lat/qlook/\\max\_likelihood/2021/10/lat\_maxlike\_20211027\_20211030.png}} with energies of $\ge 100$~MeV. \citet{Klein2022} speculate that the long duration gamma-ray emission in this event could be due to the partial trapping and mirroring of relativistic protons in large-scale coronal magnetic fields.}

\label{subsec:ions}

\subsection{Bolometric Energy}
\begin{figure}
\centering
\includegraphics[width=0.48\textwidth]{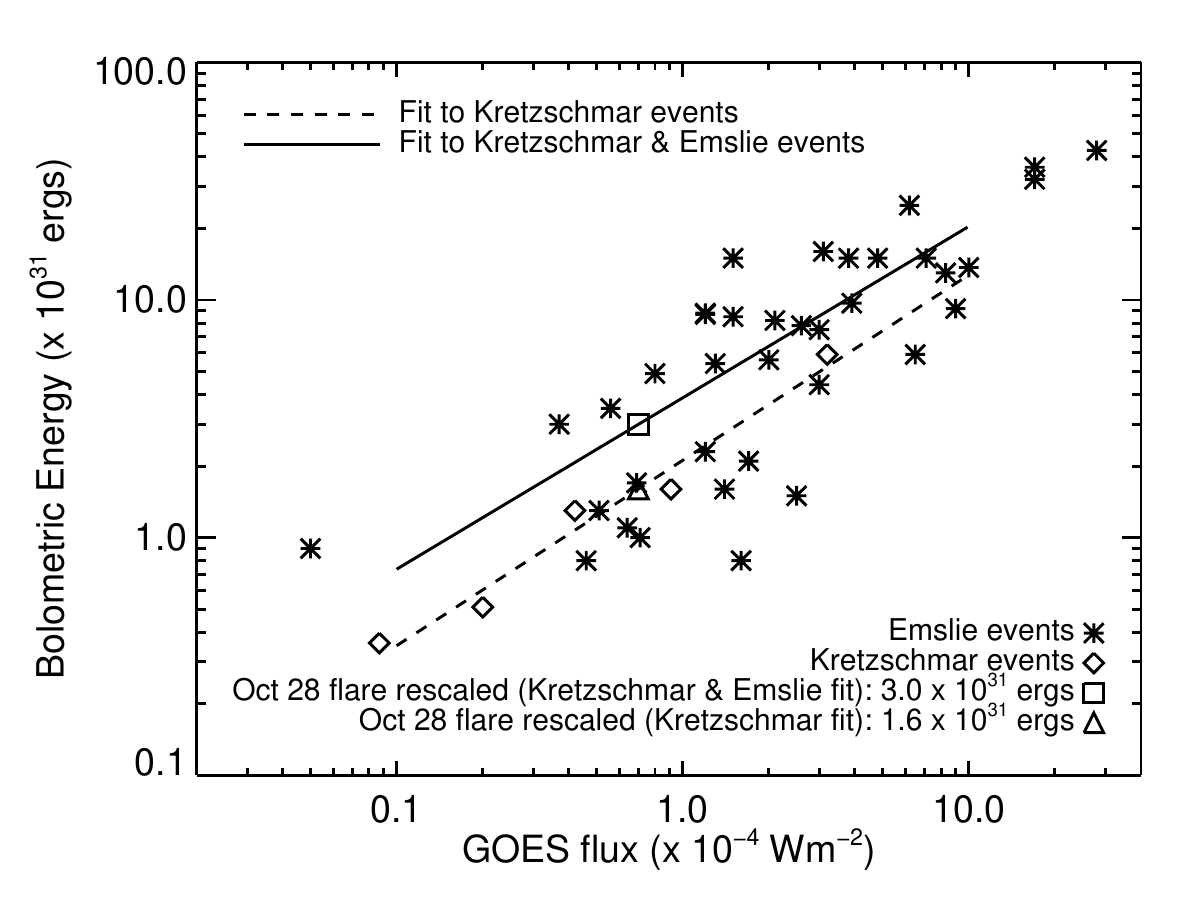}
\caption{\label{bolo.fig} Bolometric energy vs. GOES Flux.  Events used in \citet{emslie12} are shown as asterisks, and events used in \citet{Kretzschmar} are shows as diamonds.  The fits to all events and to the Kretzschmar events alone are shown as solid and dashed lines, respectively.  The estimate of the bolometric energy for the 28 October 2021 event from the Kretzschmar fit ($2.1 \times 10^{31}$ ergs) is shown as a square and that from the Emslie+Kretzschmar fit ($3.9 \times 10^{31}$ ergs) is shown as a triangle. The GOES flux of the 28 October flare has been rescaled to be consistent with the older observations.}
\end{figure}

Most of the energy radiated from the flare should come from the bolometric energy, which is the integrated irradiance over the entire electromagnetic spectrum.  Thus the bolometric energy provides a useful check on the energies of the non-thermal particles, since these particles hitting the lower atmosphere produces the radiation \citep[e.g][]{Warmuth2020}.  Previous accounts of energy partitioning in solar eruptions have found relationships between the bolometric energy and the peak GOES flux.  \citet{emslie12} uses measurements for a few flares from the TIM instrument on SOURCE combined with estimates from an empirical model for the rest of their events.  \citet{Aschwanden2017} uses the relationship found by \citet{Kretzschmar}, who calculated the total solar irradiance of set of 2100 superimposed flares.

We use these previous observations to determine the bolometric energy for the 28 October 2021 event.  Figure~\ref{bolo.fig} shows the 28 October flare compared to previous estimates of the relationship between X-ray irradiance and bolometric energy by \citep{Kretzschmar} and \citet{emslie12}. Following \citet[][see equation 13]{Aschwanden2017} we derive scaling-law relationships between X-ray flux and bolometric energy for both the Kretzschmar flare distribution and total population of flares reported by both Emslie and Kretzschmar. These relationships show that the bolometric energy for an X1.0 flare is between 2.1-3.9 $\times 10^{31}$ ergs.

Note that reanalysis of the relationship between operationally reported flare class and calibrated irradiance indicates a correction factor of 1.43 should be applied to X-ray flux measurements obtained by GOES-15 and earlier \citep[see the discussion in the introduction of][]{woods24}, while more recent data, including the irradiance reported for the October 28 flare, do not require this correction. To maintain the consistency of the previously data in the figure and its the source papers, we did not apply this factor to Figure~\ref{bolo.fig}. Instead, we reduce the flare class for the October 28 flare by a factor of 1.43 (to $7\times10^{-5}\,\mathrm{W/m^2}$) to account for the underestimate of irradiance in historical data; thus the bolometric intensities for the October 28 flare are $3.0\times10^{31}$~ergs for the Emslie+Kretzschmar fit and $1.6\times10^{31}$~ergs for the Kretzschmar-only fit. Both of these values fall well within the uncertainty of the overall distribution.

\label{subsec:bolo}

\subsection{Energy Deposition in the Chromosphere}
\label{section:chromosphere}

While there is not appropriate irradiance data to gauge the total bolometric energy in this event, we can infer the energy deposition in the chromosphere from the AIA\,1600\,\AA\ images of flare ribbons using the {\em Ultraviolet Footpoint Calorimeter} method \citep[UFC,][]{Qiu2012,Liu2013}.  Recent work by \citet{Qiu2021} has shown that this method gives energies that are comparable to the bolometric energies shown in Figure \ref{bolo.fig}.  Thus the UFC method serves as a useful check on the estimate of the bolometric energy provided in \ref{subsec:bolo}.

\begin{figure*}
(a)\includegraphics[width=3.1in,viewport=20 10 560 460,clip]{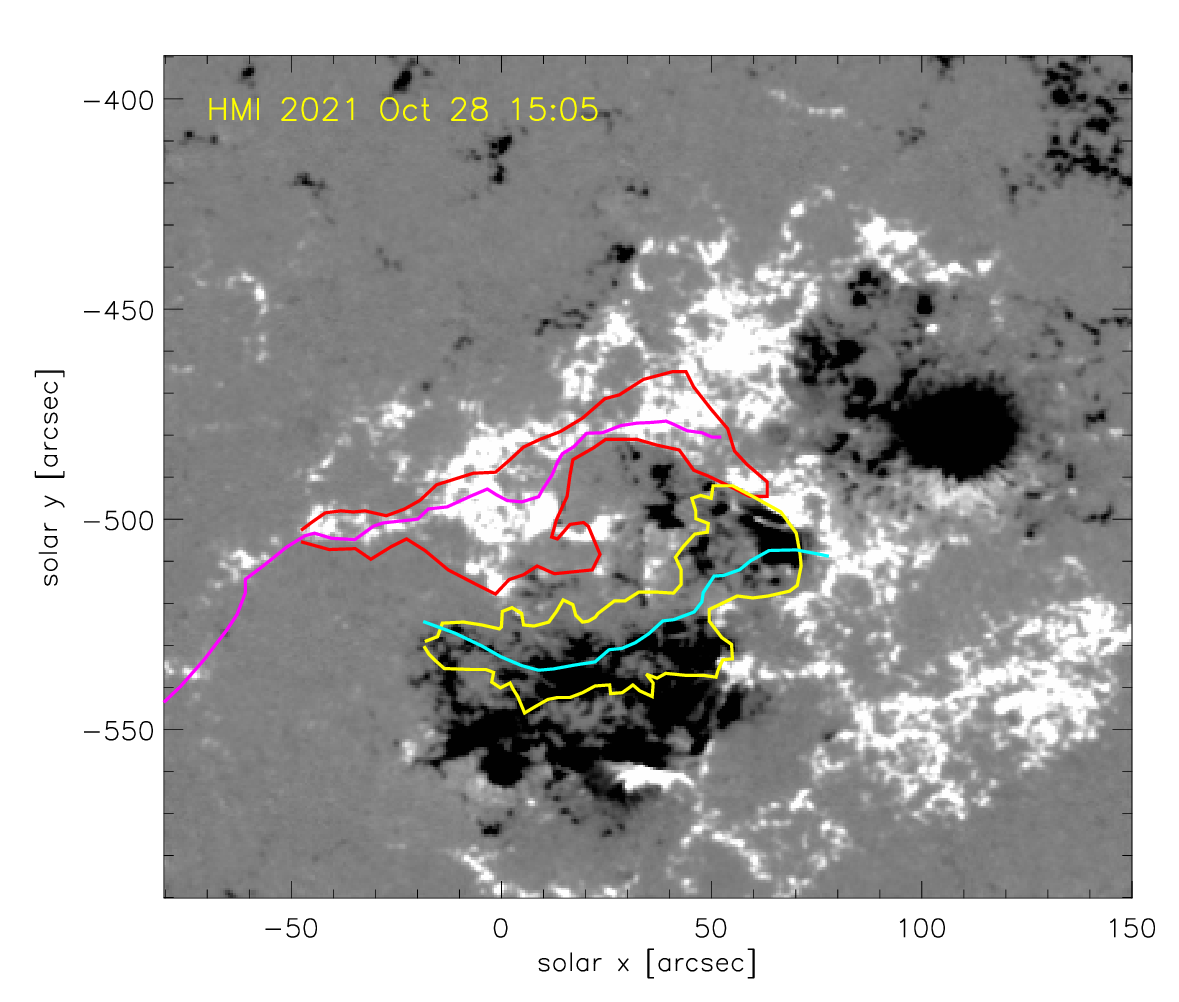}%
~~(b)\includegraphics[width=3.5in]{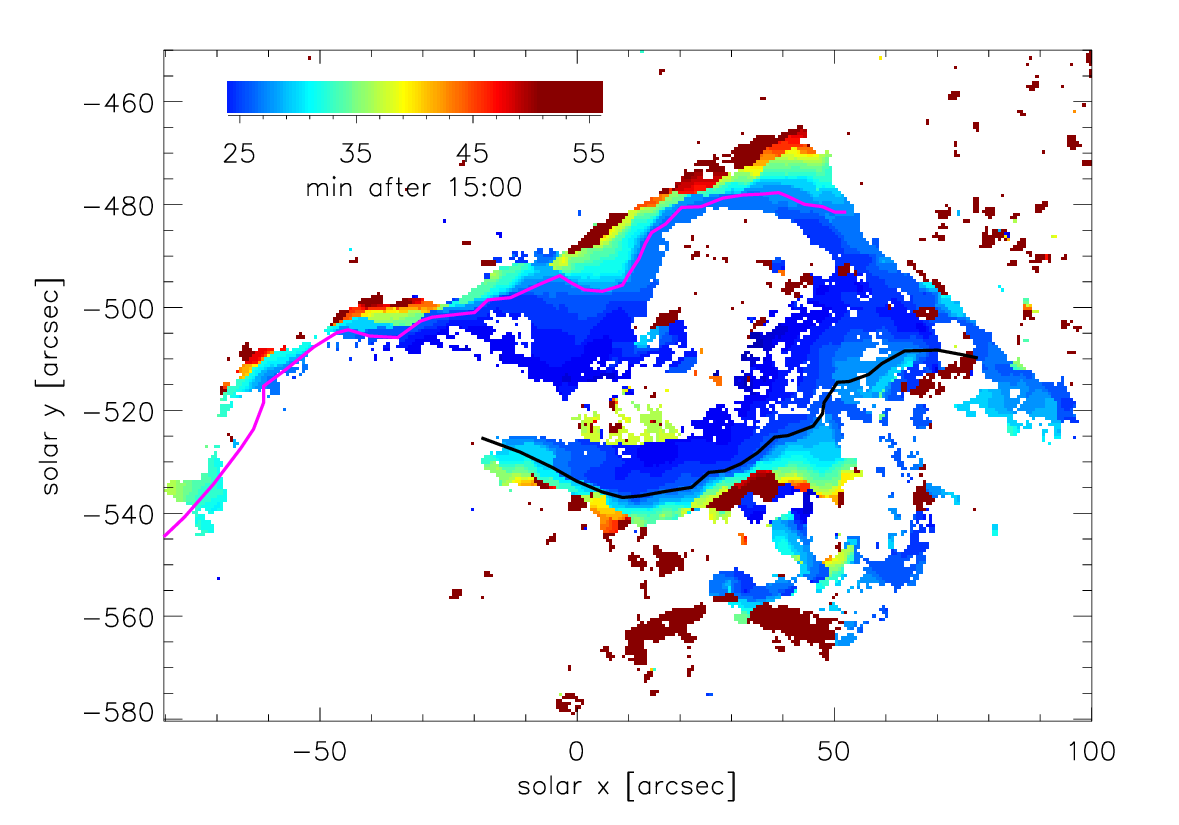}
\caption{The chromospheric flare ribbons. (a) is the HMI line-of-sight magnetogram between $\pm300$\,G in greyscale.  Ribbons from 15:31\,UT traced shown as magenta and cyan curves.  Regions enclosing most of the ribbons are outlined in red and yellow.  (b) A map of all ribbon pixels, showing the time of the peak emission as color, according to a scale shown in a color bar.  The ribbon locations from 15:31\,UT traced shown as magenta and black curves, matching those in panel (a).}
	\label{fig:HMI}
\end{figure*}

\begin{figure*}
\parbox[b]{3.7in}{(a)\includegraphics[width=3.5in,viewport=0 0 525 360,clip]{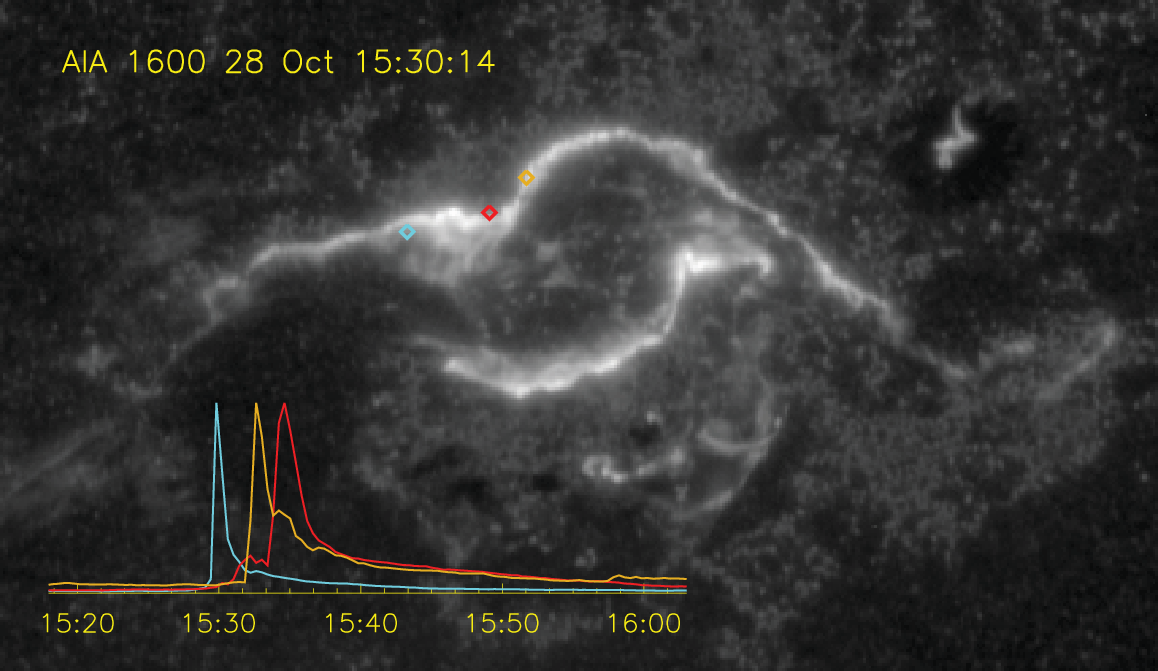}\\
(b)\includegraphics[width=3.3in,viewport=0 0 525 400,clip]{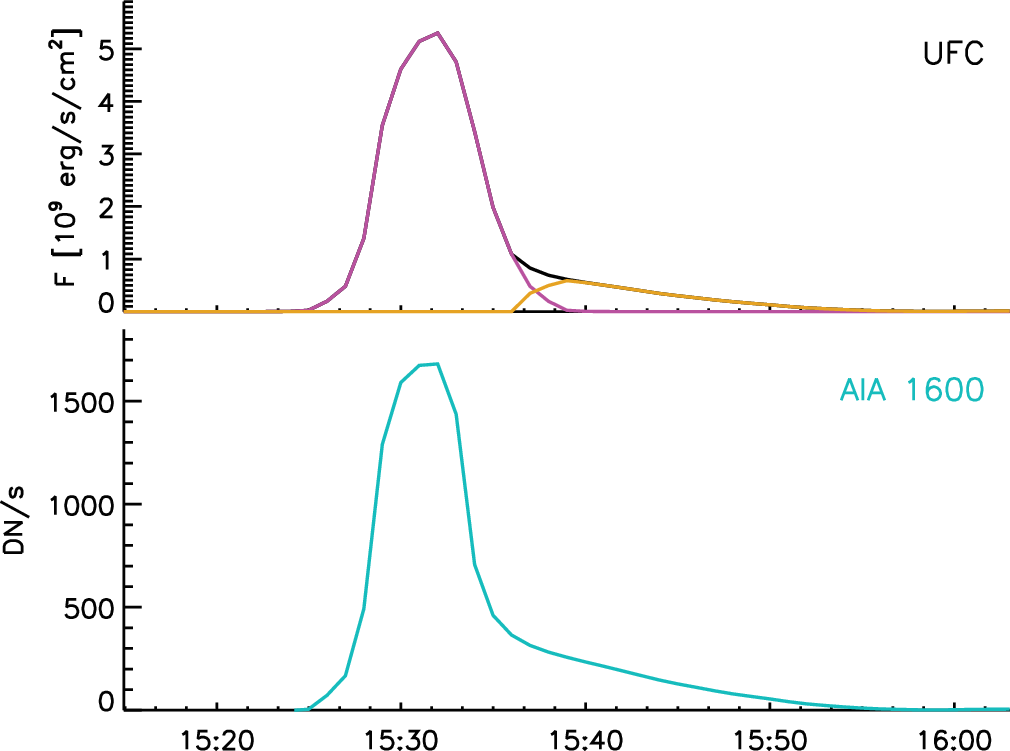}}%
~(c)\includegraphics[height=3.1in,angle=90,viewport=0 0 650 250,clip]{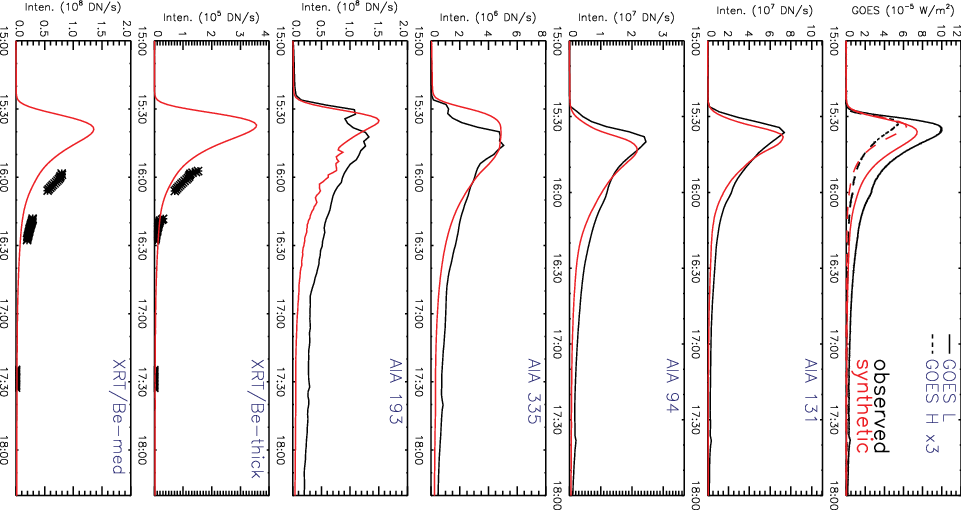}
\caption{The UFC procedure for inferring power from light curves.  (a) shows the ribbon in AIA\,1600\,\AA\ on a logarithmic gray scale.  Three pixels are identified by colored diamonds.  The light curves from these pixels are plotted in an inset, although without accounting for solar rotation.  (b) shows how the light curve of a given ribbon pixel, in this case the cyan pixel from (a), is translated into an energy flux.  The light curve is plotted in cyan in the bottom panel, after accounting for co-rotation.  Its conversion to energy flux is shown as a black curve in the upper panel.  The prompt and tail components are over-plotted in magenta and \edit1{orange} respectively.  (c) shows the full set of GOES, AIA, and XRT light curves (black) along with the version synthesized using UFC.}
	\label{fig:1600}
\end{figure*}

For the 28 October X1.0 flare, there are two roughly parallel flare ribbons traced in magenta and cyan over the HMI magnetogram in Figure\,\ref{fig:HMI}a.  The north (south) ribbon lies in positive (negative) flux.  Ribbon pixels are identified by their brief, impulsive brightening in AIA\,1600\,\AA, as illustrated in Figure\,\ref{fig:1600}a.  We find 14,760 different pixels fitting this criteria. Following the classical pattern of a two-ribbon flare, successive brightening sweeps outward, as seen in Figure\,\ref{fig:HMI}b.  The ribbons move very rapidly from 15:25 -- 15:30\,UT to cover the innermost regions (i.e.\ the blue pixels) and have mostly ended by 15:40\,UT (yellow).

Each of the 14,760 ribbon pixels is interpreted as the footpoint of a flare loop which has been impulsively energized.  The light curve of each pixel (DN/s), computed after accounting for solar rotation, is converted to energy flux (${\rm erg\, cm^{-2}\, s^{-1}}$) as shown in Figure\,\ref{fig:1600}b.  This calculation is done using two empirical parameters \citep{Qiu2012}.  The resulting heating profile is fed as input to the highly efficient, zero-dimensional Enthalpy-Based Thermal Evolution of a Loop model \citep[EBTEL,][]{Klimchuk2008} to derive the time evolution of the half-loop.  The time profiles of temperature and density are used to synthesize that half-loop's contribution to each of the AIA passbands, XRT filters, and both GOES X-ray channels.  This procedure is repeated for each of the 14,760 ribbon pixels, to build up the full-flare light-curve of each passband, as shown in Figure\,\ref{fig:1600}c.  The empirical parameters are then adjusted to optimize the agreement between observed and synthesized light curves.  After a number of iterations we obtain the reasonable agreement shown in Figure\,\ref{fig:1600}c, and take the energy profile (i.e.\ Figure\,\ref{fig:1600}b) as the actual flare energy flux incident on the chromosphere.  Owing to the nature of EBTEL, this flux may take the form of either thermal conduction or non-thermal particles; it is simply the flux of flare energy into that pixel.

\begin{figure}
\centerline{\includegraphics[width=3.5in]{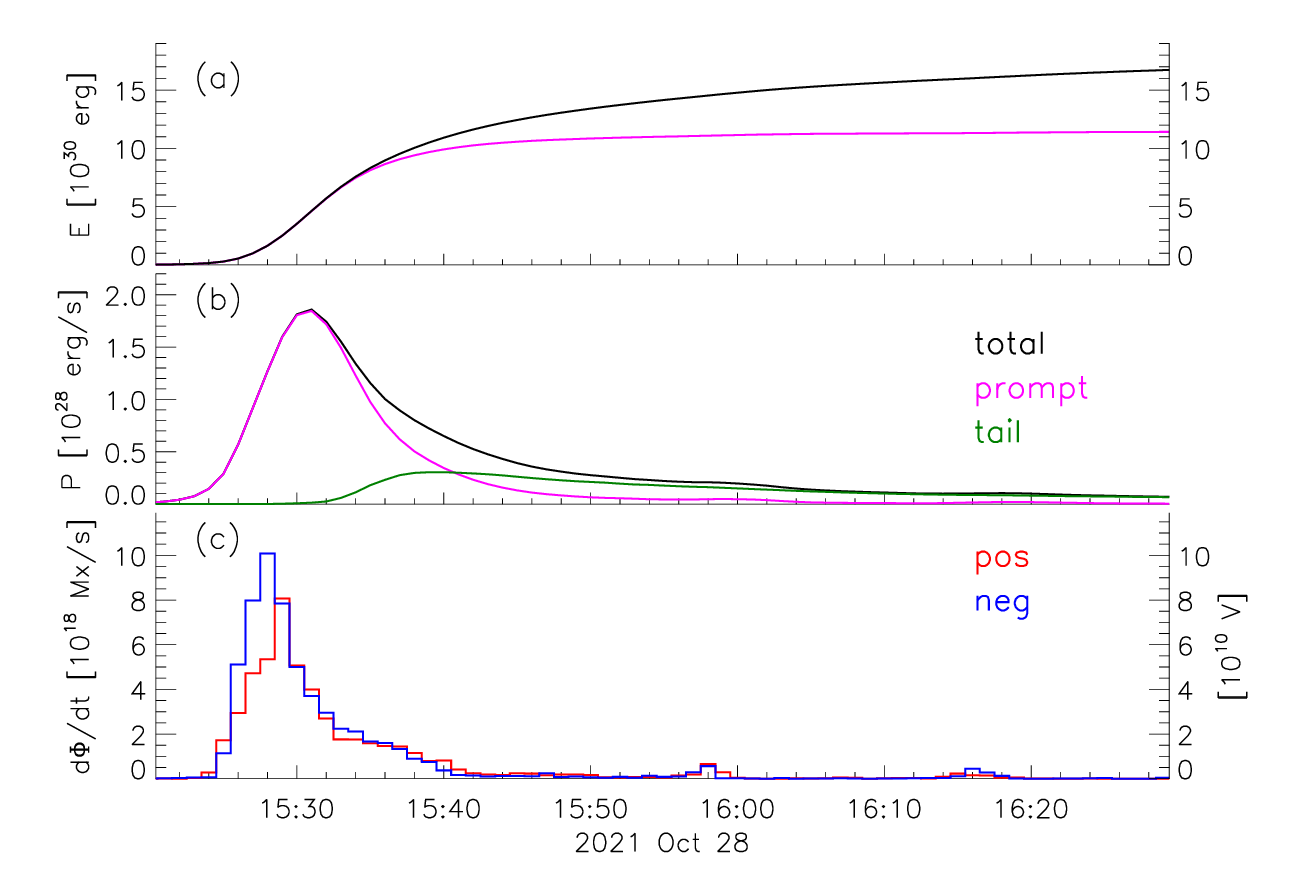}}
\caption{Time history of the energy inferred from UFC.  (a) shows the accumulated energy of the total (black) and the prompt components (magenta).  (b) is the power of the total input (black) prompt component (magenta) and tail component (green).  (c) shows the rate of magnetic reconnection in the positive (red) and negative (blue) ribbon footpoints.}
	\label{fig:UFC_erg}
\end{figure}

The light curve of a ribbon pixel typically consists of a short, intense peak, followed by a slower tail (see the bottom panel of Figure\,\ref{fig:1600}b).  These two components have been interpreted as the results of {\em prompt} energy flux from magnetic reconnection, followed by a long {\em tail} of lower energy input \citep{Qiu2016,Zhu2018}.  The energetic contributions of these two components are distinguished as magenta and green curves in the upper panel of Figure\,\ref{fig:1600}b.  Combing the UFC energy fluxes from all 14,760 ribbon pixels gives the power plotted in Figure\,\ref{fig:UFC_erg}b, with the separate contributions of the prompt and tail components plotted in magenta and green respectively.

The ribbons reveal the progress of magnetic reconnection which is assumed to power the flare.  The reconnection rate is computed by mapping the HMI magnetogram to each ribbon pixel, and attributing that pixel's flux to the time the AIA\,1600\,\AA\ light curve first crosses half maximum \citep{Forbes1984,Poletto1986}.  The resulting reconnection rate is shown separately for the positive (red) and negative (blue) ribbon in Figure \ref{fig:UFC_erg}c.  The two versions agree to a reasonable extent, although the stronger, more compact nature of the negative flux makes that slightly more reliable.  In the end, $\Delta\Phi\simeq3.3\times10^{21}$ Mx of flux is reconnected --- about 20\% of all flux in the active region.  The peak rate, $\dot{\Phi}\simeq10^{19}$\,Mx\,s$^{-1}$, is at the higher end of rates observed in other large flares \citep{Jing2005,Zhu2020}, confirming our assumption of an unusually impulsive flare.  The reconnection occurs primarily during the flare's impulsive phase (15:25 -- 15:35\,UT), and has largely subsided by 15:40\,UT, when the radiative cooling phase begins.

The energy flux inferred from UFC, plotted in Figure\,\ref{fig:UFC_erg}b, reflects the reconnection history shown in Figure\,\ref{fig:UFC_erg}c.  It is not surprising that the prompt component, plotted in magenta, resembles the reconnection curves, and has largely subsided by 15:40\,UT.  Its integral, plotted in the same color on Figure\,\ref{fig:UFC_erg}a, levels off at $\Delta E\simeq 1.2\times10^{31}$ ergs. The tail contribution, on the other hand, builds up over the evaporative phase (15:30 -- 15:40\,UT), peaking at $P_{\rm tail}=0.3\times10^{28}\, {\rm erg/s}$ at 15:40\,UT.  Its integrated contribution accounts for the difference between the black and magenta curves in Fig.\ \ref{fig:UFC_erg}a.  This difference reaches $\Delta E_{\rm tail} = 0.5\times10^{31}$ ergs by 16:40.  The combination of the peak and tail energies calculated in this manner is 1.7 $\times10^{31}$ ergs, very similar to the lower value \edit1{of 1.6 $\times10^{31}$ ergs} estimated for the bolometric energy in \S \ref{subsec:bolo}.

\subsection{Thermal Energy}
\label{section:thermal}
We use two different methods to approximate the thermal energy \edit1{of the plasma heated} during the eruption.  In the first method (section \ref{subsec:thermal_DWL}), we use the flux from the GOES XRS and a single-loop approximation to estimate the physical quantities necessary to derive the thermal energy.  In section \ref{sec:AIA_erg}, we employ differential emission measure calculations using AIA and SUVI imaging data to estimate these quantities.  

\subsubsection{Single loop approximation}
\label{subsec:thermal_DWL}

\begin{figure}
\centerline{\includegraphics[width=3.5in]{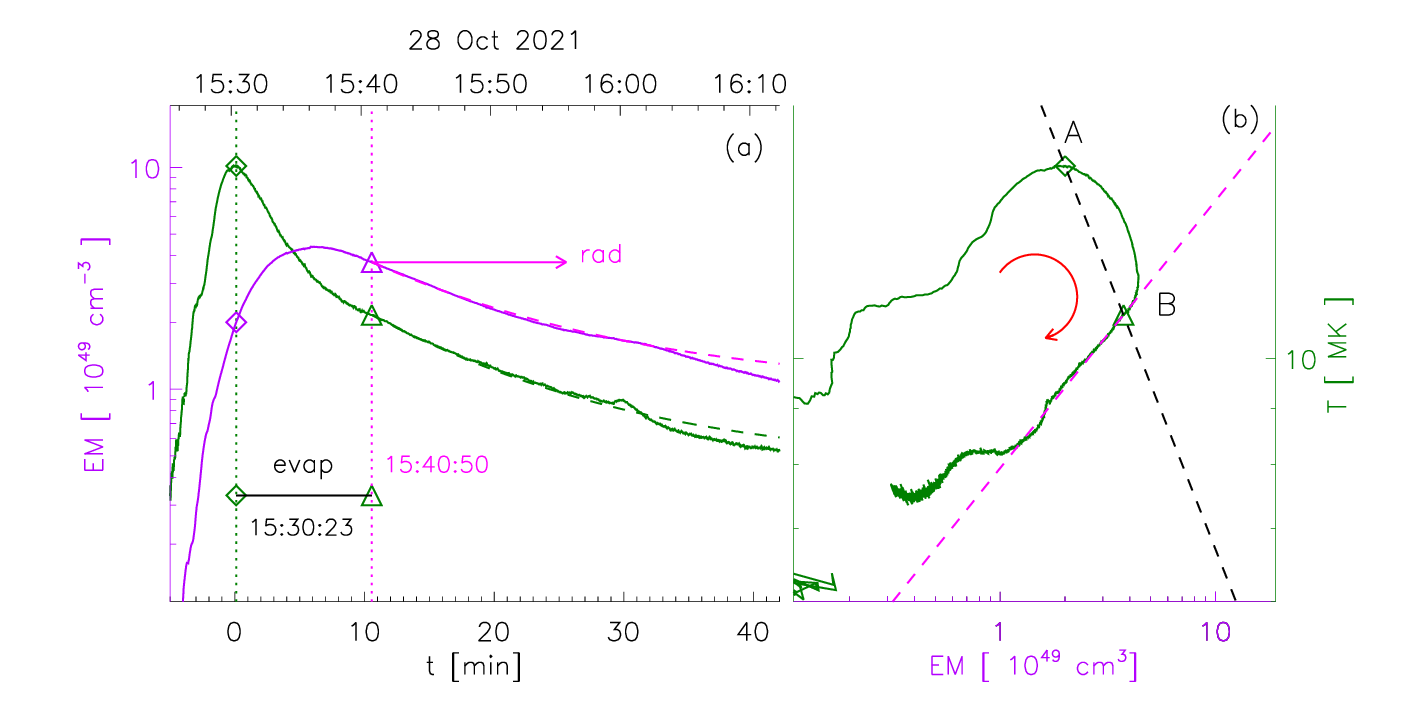}}
\caption{Global characteristics of the flare from \textit{GOES}.  (a) shows the evolution of the temperature (green) and emission measure (violet) derived from the two X-ray bands.  The bottom axis is in minutes from temperature peak, while the top is UT on 2021\,October\,28. Evolution under mechanical equilibrium is shown by dashed curves of corresponding colors following the triangle.  (b) is a phase portrait formed by plotting $T$ {\em vs.} $EM$ in green.  The red arc indicates the clockwise sense of evolution.  Dashed lines show the ideal characteristics of evaporation (black) and radiative cooling (magenta).  The beginning of each phase is marked by a symbol, which appears in (a) to show the times of these phases.}
	\label{fig:phase}
\end{figure}

The thermal energy of the flare loops can be obtained by using the two \textit{GOES} X-ray channels to derive temperature, $T$, and \edit1{volume} emission measure, $EM$, \citep{Thomas1985,Garcia1994} shown in Figure\,\ref{fig:phase}a, after assuming coronal abundances.   A phase portrait of $T$ {\em vs.}\ $EM$, shown in Figure\,\ref{fig:phase}b, reveals this flare's evolution (clockwise as indicated by the red arc) to be unusually similar to that of a single impulsively-heated loop \citep{Jakimiec1992}.  Adopting this interpretation provides an estimate of the flare's energetics without recourse to imaging.

The phase portrait shows that evaporation ends at approximately 15:41\,UT, with $T_B=10.9$ MK as indicated by the triangles in Figure\,\ref{fig:phase}.  After this the loop remains in mechanical equilibrium, \edit1{with no gradient in the plasma pressure $p$  ($\partial p/\partial\ell\simeq0$)}, cooling radiatively to reduce \edit1{pressure} $p$ and peak temperature $T_{\rm pk}$.
While traditional equilibrium treatments assume a uniform volumetric heating, $h$, time evolution under mechanical equilibrium differs through the addition of $-\partial p/\partial t$, which must be uniform,  \edit1{like the pressure itself.}  
Thus, even as it remains in mechanical equilibrium, the loop cools according to
\begin{equation}
  {d\,T_{\rm pk}\over dt} = {2\,T_{\infty}^{3/2}\over \tau_{c,B}\,T_B^{1/2}}\left[\left({T_{\infty}\over T_{\rm pk}}\right)^{2} - 
  \left({T_{\rm pk}\over T_{\infty}}\right)^{3/2}\right]~~,
    \label{eq:RTV_cooling}
\end{equation}
where 
\begin{equation}T_{\infty}~=~0.69\left({hL^2\over\kappa_0}\right)^{2/7} ~~,
  	\label{eq:Tinfty}
\end{equation}
is the asymptotic temperature (i.e.\ when $\partial p/\partial t=0$) and $\tau_{c,B}$ is the conductive time at the beginning of the radiative cooling phase.  The green dashed curve in Figure\,\ref{fig:phase}a, is the solution with $\tau_{c,B}=92$ min and an asymptotic temperature $T_{\infty}=8.0$\,MK, found by fitting the curve.     Demanding that the cooling leg remains in equilibrium (details are relegated to an appendix) yields an estimate of the full loop length $L=102$\,Mm, and requires $V/L^2\simeq3.0$\,Mm, from which we obtain the flare volume, $V=3.1\times10^{28}\, {\rm cm}^3$, without using an image.  

\begin{figure}
\centerline{\includegraphics[width=3.5in]{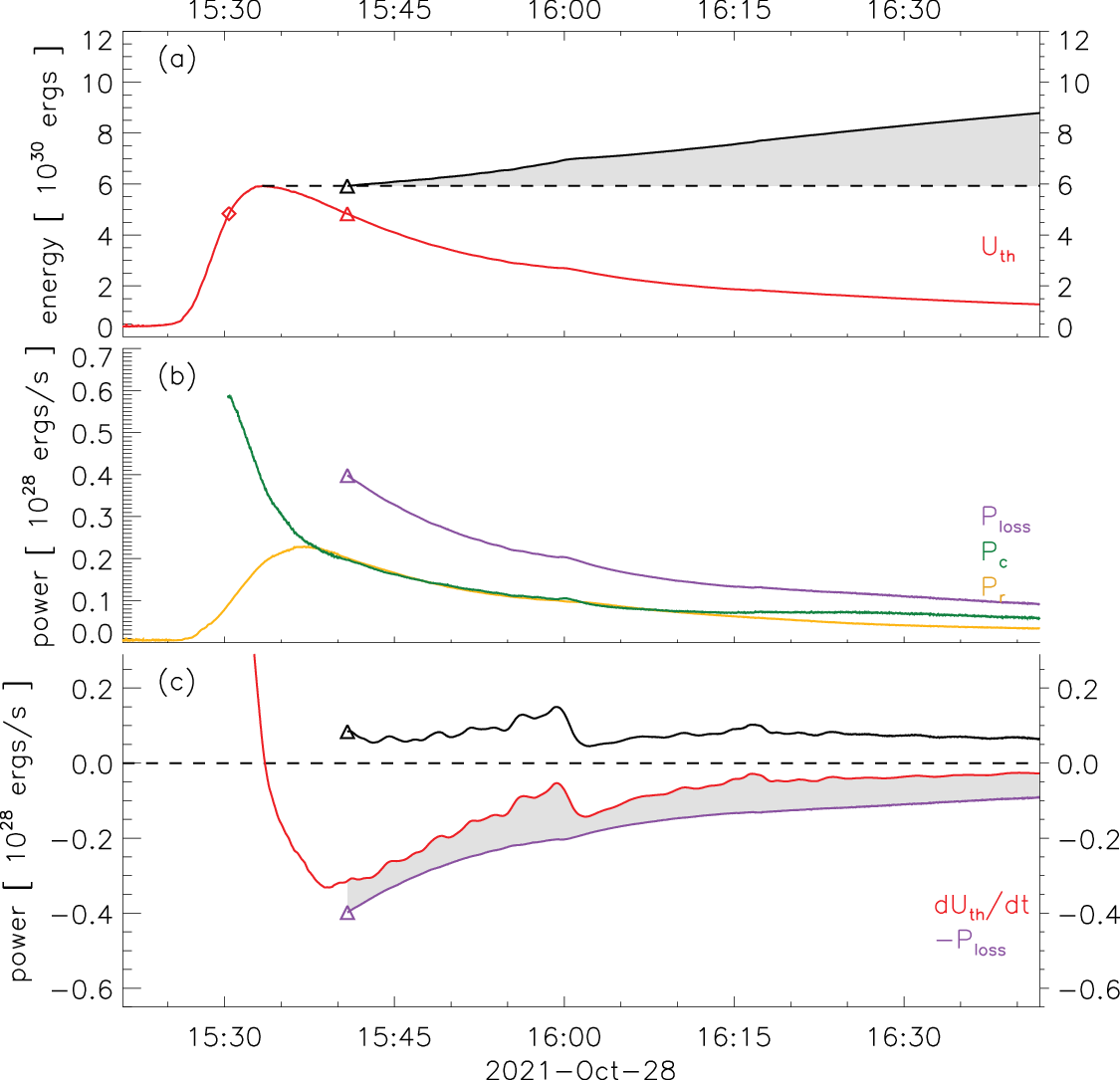}}
\caption{Evolution of energy (a) and power (b) and (c), as computed from \textit{GOES} light curves.  (a) shows the evolution of thermal energy, $U_{\rm th}$, in red.  The diamond and triangle mark the beginning and end of the evaporative phase, as in Figure\,\ref{fig:phase}.  The shaded region integrates the power plotted as shaded in (c), and is added to the peak of $U_{\rm th}$.  (b) shows the power lost to radiation (blue) and thermal conduction (green).  During the evaporative phase these combine to a net loss, shown in violet.  (c) plots the change in thermal energy (red) and the losses as a negative contribution.  The grey region between the curves is the power that must be supplied to keep the thermal energy from falling too rapidly during radiative cooling.  This difference is plotted as a black curve.}
	\label{fig:GOES_energy}
\end{figure}

The analysis above and in the appendix can be used to compute the energy evolution of the flare.  The net thermal energy, 
\begin{equation}
  U_{\rm th} ~={3\over 2}\, p\, V ~=~ 3\,k_{\rm b}\, T\, {\sqrt{EM\,V}} ~~,
        \label{eq:DWL_Uth}
\end{equation}
plotted as a red curve on Figure\,\ref{fig:GOES_energy}a, peaks at $U_{\rm th}\simeq 0.6\times10^{31}\, {\rm ergs}$ at 15:33:20\,, which is about half of the peak energy calculated via the UFC method in \S \ref{section:chromosphere} and shown in Figure \ref{fig:UFC_erg}b.   The power lost through conduction or radiation, \edit1{$P_c$ or $P_r$ respectively}, is computed by dividing $U_{\rm th}$ by the corresponding time \citep{Longcope2010}, as shown in Figure\,\ref{fig:GOES_energy}b.  The evaporative phase is dominated by conduction ($P_c\gg P_r$) driving the evaporation which carries the energy back into the loop --- no energy is lost.  The radiative cooling phase is characterized by $P_c\simeq P_r$ (see Figure\,\ref{fig:GOES_energy}b) both of which represent losses from the system: $P_r$ is lost to radiation from the corona while $P_c$ is conducted out of the corona to the feet where it is also radiated.  

During the radiative cooling phase, the combined losses exceed the observed decrease in thermal energy, as indicated by the gray region in Figure\,\ref{fig:GOES_energy}c.  This difference must be somehow supplied to the flare plasma even during the so-called cooling phase to account for the slow rate at which $U_{\rm th}$ is observed to decrease.  This extra heating accounts for the fact that the loop is asymptotically approaching a very high temperature, $T_{\infty}=8.0$\,MK, as shown in  Figure\,\ref{fig:phase}a.  The net power required to maintain this temperature is
\begin{eqnarray}
  P_{\rm tail} &=& h\, V ~=~ 3.6\, \kappa_0\, T_{\infty}^{7/2}\, {V\over L^2} \nonumber \\
  &\simeq & ~0.16\times10^{28}\, {\rm erg \,s^{-1}} ~~,
\end{eqnarray}
after inverting Equation\,\ref{eq:Tinfty} and using $V/L^2=3$\,Mm.  This power is consistent with the observed difference, plotted as a black solid line on Figure\,\ref{fig:GOES_energy}c.  When that power is integrated we obtain a persistent contribution to the energy shown as a shaded region in Figure\,\ref{fig:GOES_energy}a.  This contribution continues to grow, even as $U_{\rm th}$ decreases.  As of 16:40\,UT about $0.3\times10^{31}$\,ergs has been supplied, raising the total flare energy by 50\% to $0.9\times10^{31}$\,ergs.  Even at this late time the energy appears to be continually growing. 

The cooling-phase contribution inferred from the \textit{GOES} light curves ($0.3\times 10^{31}$\,ergs, Figure\,\ref{fig:GOES_energy}a), is comparable to that derived from the UFC method in \S \ref{section:chromosphere} ($0.5\times 10^{31}$\,ergs, Figure \ref{fig:UFC_erg}a).  Indeed, the qualitative and quantitative similarity of Figures\,\ref{fig:UFC_erg}a and \ref{fig:GOES_energy}a, derived using independent analyses, gives support to their energy profiles, \edit1{including the tail heating component, which contributes to the energetics of each loop.  Interestingly, these values are about one order of magnitude greater than the energy in the non-thermal electrons during the decay phase as estimated from the Fermi GBM data.}
\subsubsection{differential emission measure approximation}
\label{sec:AIA_erg}

\begin{figure*}
    \centering
    \includegraphics[width=1.\textwidth,trim={0 5cm 0.5cm 0},clip]{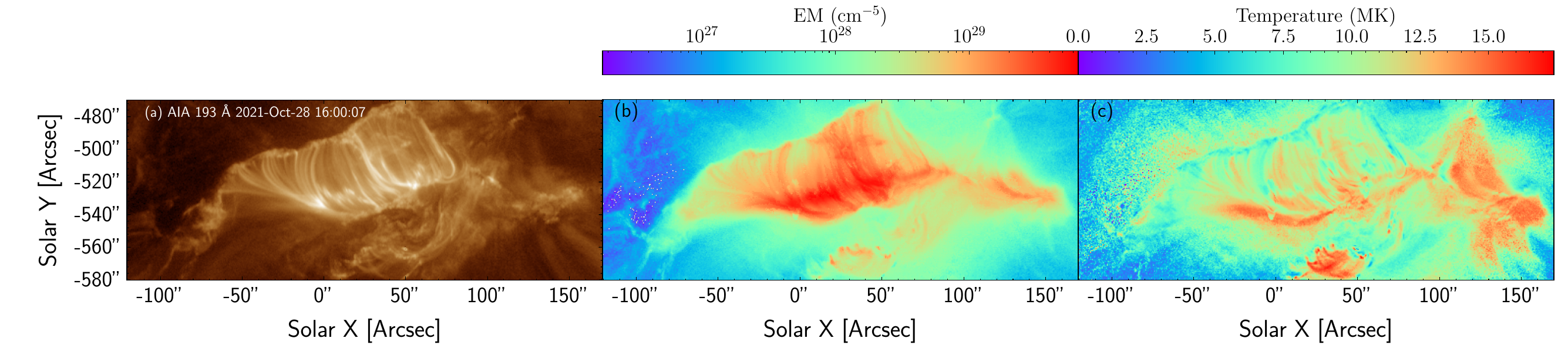}
    \caption{Panel (a): The flare arcade in AIA 193~\AA. Panel (b): the emission measure of the region for $5~<log(T)~<7.4$. Panel (c): the DEM weighted temperature of the region.}
    \label{fig:flare}
\end{figure*}

As a second method of calculating the thermal energy, we estimate the temperatures and emission measures of the event by calculating the differential emission measure (DEM) of the flaring region using imaging data. To compute DEMs we use both AIA EUV passbands (94, 131, 171, 193, 211, 335\,\AA) and the high-dynamic-range SUVI passbands (94, 131, 171, 195\,\AA) in frames where the AIA observations showed signs of major saturation. We used the regularized inversion method \citep{hannah&kontar12} to calculate the DEMs. In Figure~\ref{fig:flare} we show the flare observed in AIA 193~\AA\ during the decay phase (panel a) and the corresponding inverted emission measure (EM) (panel b) and DEM-weighted mean temperature (panel c). 

The main goal of calculating the DEM maps was to estimate the thermal energy arising form various parts of the flare arcade. The thermal energy of plasma of temperature $T$ and volume $V$ is given by

\begin{equation}
    U_{Th} = 3n_{e}k_{B}TfV
\end{equation}
where \textit{$n_{e}$} is the electron number density of the plasma and \textit{f} is the volume filling factor. Throughout this paper we have assumed \textit{f}=1. For our analysis the volume of an individual pixel is given by $V_{j}~=~A_{AIA}\times~LOS_{j}$, where $LOS_{j}$ is the line of sight (LOS) for the $j$-th pixel in the FOV and $A_{AIA}$ is the physical area of an AIA pixel and similarly for SUVI. To calculate the thermal energy in the FOV we sum over the thermal energy arising from all the pixels in the FOV which gives us,

\begin{equation}
    U_{Th} = \sum_{j} \frac{3k_{B}A_{AIA}\sqrt{LOS_{j}}}{\sqrt{EM_{j}}}\int_{T_{min}}^{T_{max}}\,DEM(T)_{j}\,T\,dT
    \label{eq:therm_eneg}
\end{equation}
where we have used the DEM-weighted temperature $\bar{T}$ and $n_{e}^{2}\times LOS = EM$. Simultaneous observations from a different perspective, e.g., observations from STEREO-A, can be used to infer the line of sight across the pixels in the field of view. For further details regarding the line of sight estimation, and DEM inversion refer to \cite{roy_thermal}.

\begin{figure}
    \includegraphics[width=0.5\textwidth,trim={1cm 0cm 1cm 1.8cm},clip]{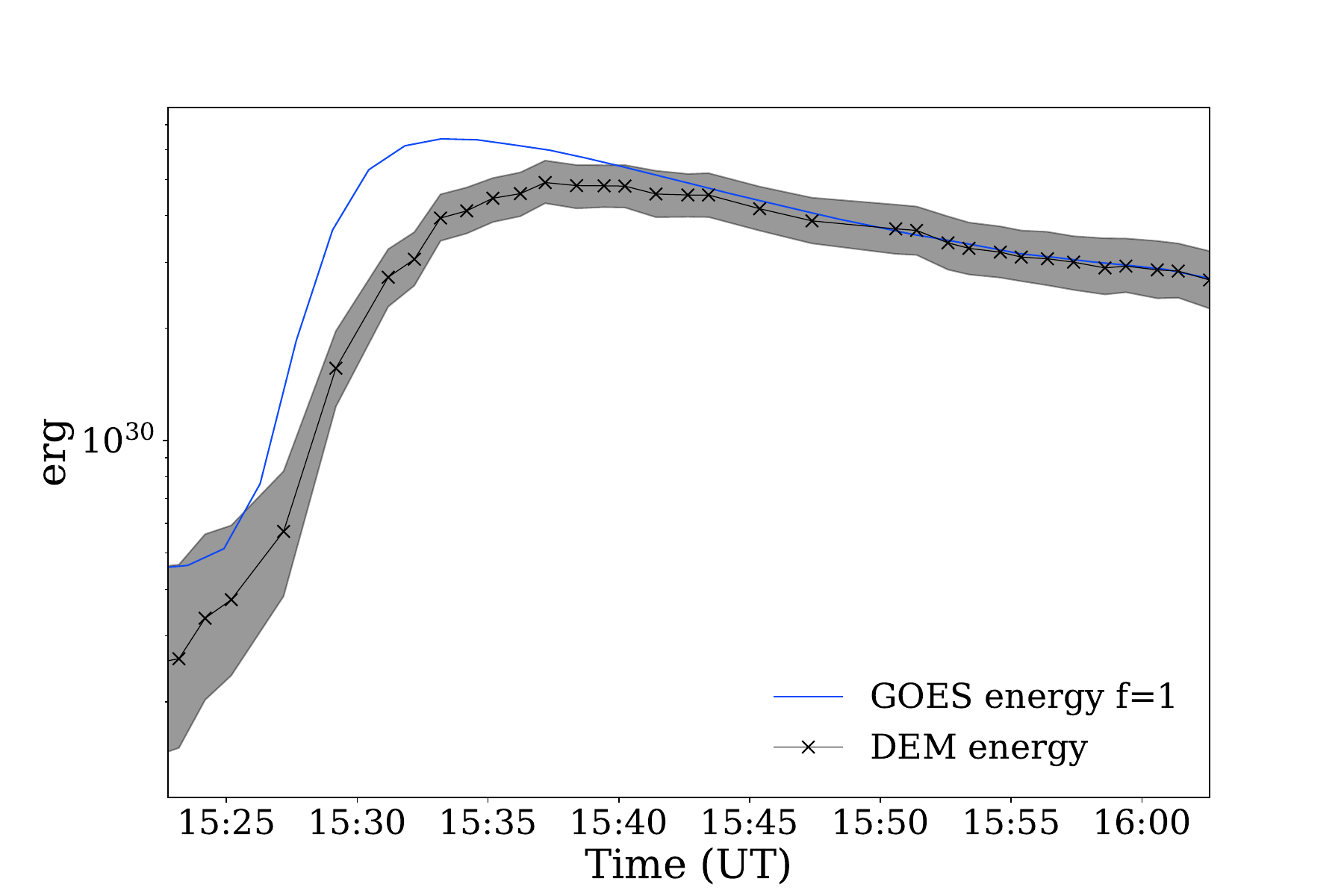}
     \caption{Comparison of thermal energies calculated from \textit{GOES} flux and from AIA/SUVI DEMs. The shaded black region shows the uncertainties from DEM inversion propagated through the Eq.~\ref{eq:therm_eneg}.}
     \label{fig:therm_eneg}
\end{figure}

In Figure~\ref{fig:therm_eneg}, we compare the thermal energy derived from AIA/SUVI DEMs (solid black line with crosses) with that estimated from \textit{GOES} observations under the single-loop assumption from \S \ref{subsec:thermal_DWL} (solid blue line, also shown as the solid red line in Figure~\ref{fig:GOES_energy}a). The uncertainty in the thermal energy (shaded black region) is calculated by propagating the uncertainty in the inverted DEM through Eq.~\ref{eq:therm_eneg}. While both estimates exhibit similar trends during the decay phase, there is a discrepancy during the impulsive phase and near the flare peak. This difference arises primarily from the assumption of a constant flaring volume in the single-loop treatment, which does not account for the rapid change in volume of the emitting plasma volume during the early stages of the flare. Despite the differences in temporal evolution, the peak thermal energy estimated using the single-loop approximation ($\sim0.6 \times 10^{31}$~erg) is very similar to that derived from the AIA/SUVI DEMs ($\sim 0.49^{+0.07}_{-0.06} \times 10^{31}$~erg). 

The thermal energy derived in this section is primarily due to the plasma that is driven into the flare loops via chromospheric evaporation.  However, previous studies have shown that the plasma sheet above the flare loops is likely directly heated during the reconnection process \citep{Reeves2017,Reeves2019,Warren2018}.  Since the viewpoint of this event is top down and the plasma sheet is superimposed on the flare loops in this viewing angle, our thermal energy calculations using the DEMs also includes the contribution from direct heating in the plasma sheet. A detailed analysis of the thermal energies of the flare loops and the plasma sheet region for a different event observed at the limb indicated that the directly heated plasma contributes a small fraction to the total thermal energy \citep{roy_thermal}.

\subsection{Energy Dissipated by the EUV Wave}
This event generated a powerful EUV wave that traversed the entirety of the visible solar disk.  Multiple authors have examined the energy flux of similar large-scale waves in the corona \citep[e.g.,][]{Aschwanden2004, Patsourakos2012, Long2015}. Many of these treatments assume that such waves can be treated as blast waves. Such an approach appears to be appropriate here, given the kinematic evolution of the wave associated with this event. The surface EUV wave appears to be the footprint of an essentially spherical 3D wave with many characteristics of a blast wave. This section addresses the wave energy analysis; the kinematic evolution of the wave that supports this analysis appears in Appendix~\ref{sec:wave_kinematics}.

Here we follow the method discussed in \citet{Patsourakos2012}. The wave energy flux has three main contributions: kinetic energy flux, radiative loss flux, and thermal conduction flux.

The kinetic energy flux of the wave is given by \citep{Patsourakos2012} as:
\begin{equation}
    F_{\mathrm{kin}} = \rho \left(\delta v\right)^2 v_{\mathrm{gr}}/2,
        \label{eq:F_kin1}
\end{equation}
where $\rho$ is the mass density, $v_{\mathrm{gr}}$ is the group speed, and $\delta v$ the velocity perturbation. \citet{Liu2011} show how it is possible to rewrite Equation~\ref{eq:F_kin1} for small magnetosonic perturbations as:
\begin{equation}
        F_{\mathrm{kin}} = \rho \left(\delta I/I\right)^2 v_{\mathrm{gr}}^3/8,
        \label{eq:F_kin2}
\end{equation}
where $\delta I / I$ is the change in intensity resulting from the perturbation.

Following \citet{Aschwanden2004}, $\rho = \mu m_\mathrm{p} n_\mathrm{e}$, where $m_\mathrm{p}$ is the mass of the proton, $n_\mathrm{e}$ is the electron number density, $\mu = 1.27$ is the mean molecular weight assuming a 10:1 ratio of hydrogen to helium. We assume a typical quiet sun density of $n_\mathrm{e} = 5\times10^8\;\mathrm{cm^{-3}}$, while, for this event, we estimate the increase in intensity resulting from the wave is $\delta I/I \sim30\%$ and the peak $v_\mathrm{gr}\sim9.1\times10^7\;\mathrm{cm\,s^{-1}}$. From Equation~\ref{eq:F_kin2} the kinetic energy flux is then $F_\mathrm{kin} = 9.0\pm6.7\times10^6\;\mathrm{erg\,cm^{-2}\,s^{-1}}$.

The energy flux due to radiative losses for a structure with length $L$ in the corona is given by:
\begin{equation}
    F_\mathrm{rad} = n_\mathrm{e}^2 \Lambda(T) L,
    \label{eq:F_rad}
\end{equation}
where $\Lambda(T)$ is temperature-dependent radiative loss function. Meanwhile, the energy flux due to thermal conduction is given by:
\begin{equation}
    F_\mathrm{cond} = -\frac{2}{7}\kappa T^{7/2}/L,
    \label{eq:F_cond}
\end{equation}
where $\kappa=9.2\times10^{-7}\;\mathrm{erg s^{-1}\,cm^{-1}\,K^{-7/2}}$ is the Spitzer conductivity.

We must estimate the temperature and emission measure, $T$ -- including the change from the ambient background -- to compute the two values in Equations \ref{eq:F_rad} and \ref{eq:F_cond}. To estimate these values, we use observations from SDO/AIA and the \textsf{simple\_reg\_dem} package \citep[in \textsf{SolarSoft IDL};][]{Plowman2020} to determine the differential emission measure of the plasma both before (at about 15:00~UT) and during (at 15:28~UT) the passage of the wave. Figure~\ref{fig:aia_overview} shows the observations used in this calculation.

\begin{figure}
\centering
\includegraphics[width=0.48\textwidth]{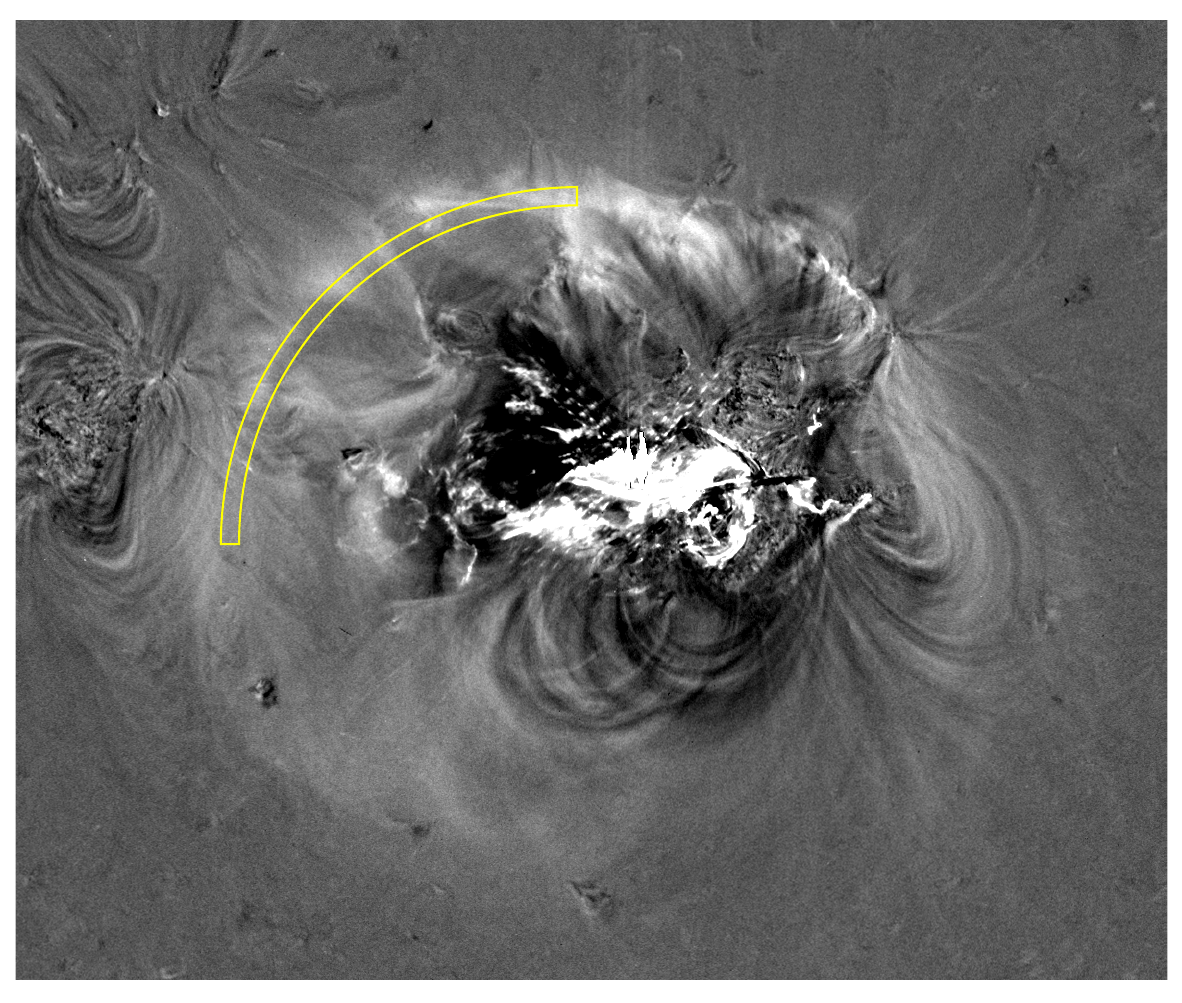}
\caption{\label{fig:dem_roi} AIA 193\,\AA\ difference image showing the change in brightness resulting from the wave and the region we used to compute average DEMs (yellow contour).
}
\end{figure}

\begin{figure}
\centering
\includegraphics[width=0.48\textwidth]{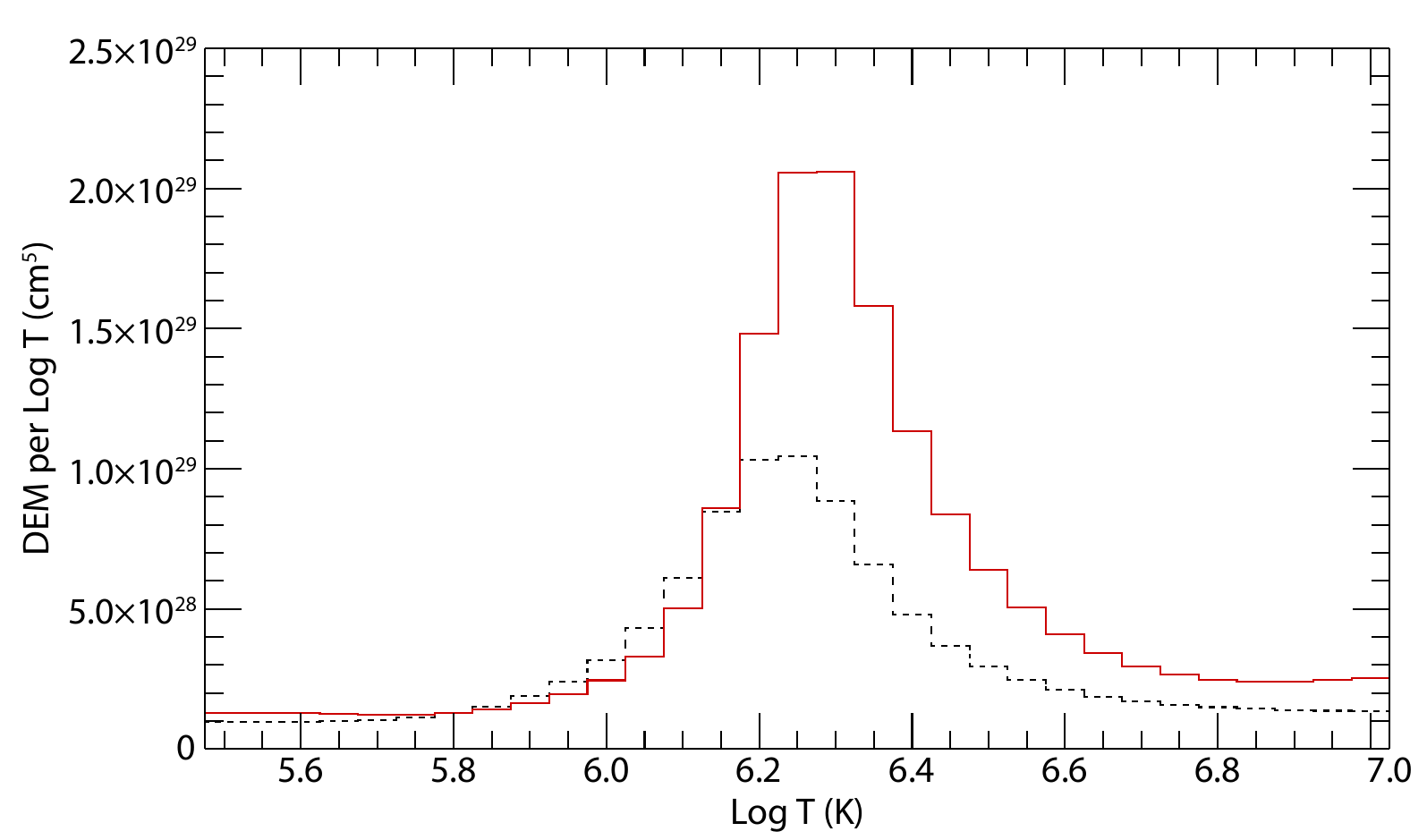}
\caption{\label{fig:dem_comparison} Spatially averaged DEMs before (dashed black line) and during (solid red line) the wave's passage through the region highlighted in Figure~\ref{fig:dem_roi}.
}
\end{figure}

We computed a spatially-averaged DEM over the portion of the wave that was most clearly differentiated from the background. Figure~\ref{fig:dem_roi} shows the region for which we computed the DEMs, while Figure~\ref{fig:dem_comparison} shows the resulting DEMs, both before and during the wave.

Both DEMs are strongly peaked at single temperatures, so it is reasonable to compute a DEM-weighted average temperature for the region before and during the wave. Prior to the event, we find that $T\sim1.8\pm0.9$ MK, while during the wave's passage, the temperature increases to $T\sim2.1\pm0.7$ MK, an increase of about 14\%.

\citet{Reeves2002} provide the radiative loss functions as a piecewise-defined function of temperature. For temperatures in the range of $10^{6.2}\leq T \le 10^{6.6}$ K, the loss function $\Lambda(T) = 3.7\times10^{-12}T^{-1.6}$. Given our estimated wave temperature, $\Lambda(T\sim10^{6.323}) \sim 2.8\times10^{-22} \mathrm{erg\,cm^{-3}s^{-1}}.$

We use this information and Equations~\ref{eq:F_rad} and \ref{eq:F_cond} to compute $F_\mathrm{rad}$ and $F_\mathrm{cond}$. The increase in total emission measure due to the wave shown in Figure~\ref{fig:dem_comparison} is about $60\pm10$\%; because emission measure is proportional to density squared, we can estimate the density increase due to the wave is $n_\mathrm{wave}\sim\sqrt{1.6} n_\mathrm{qs}$, where $n_\mathrm{qs}$ is our previous estimate of the quiet sun density. Here we take $L$ to be the approximate width of the wave, which we measure to be about $L\sim5.4\pm0.9\times10^9$ cm.
Thus, $n_\mathrm{wave}\sim1.26\pm0.25\times10^8 \mathrm{cm^{-3}}$ and $F_\mathrm{rad} \sim 6.1\pm3.5\times10^5\;\mathrm{erg cm^{-2}\,s^{-1}}$.

Likewise, using $L$ and $T$ from above, we obtain $F_\mathrm{cond}\sim6.5\pm7.7\times10^5\;\mathrm{erg\,cm^{-2} s^{-1}}$. Both $F_\mathrm{rad}$ and $F_\mathrm{cond}$ are only a small fraction of the total wave energy flux, which is dominated by the kinetic energy. In fact, some of these radiative and conductive losses are the result of the background corona itself, and would occur regardless of the wave. Repeating the calculation above, assuming that $n_\mathrm{e}$ prior to the wave is the background quiet sun electron density, and using the pre-wave DEM-weighted average temperature of $T\sim1.8$ MK allows us to estimate that the wave represents $\sim30\%$ of the total radiative energy flux and $\sim50\%$ conductive energy fluxes we computed above. This calculation yields final values of $F_\mathrm{rad} \sim 1.2\pm0.7\times10^5 \mathrm{erg\,cm^{-2}\,s^{-1}}$ and $F_\mathrm{cond}\sim3.2\pm3.8\times10^5 \mathrm{erg\,cm^{-2}\,s^{-1}}$. We note that, although obtained in a different way, these values are roughly consistent with those obtained by \citet{Patsourakos2012}.

To estimate the total wave energy, we must integrate the flux over the total volume and time of the event. We find:
\begin{equation}
    E_\mathrm{wave} = \left( F_\mathrm{kin} + F_\mathrm{rad} + F_\mathrm{cond} \right) 2\pi \, R \, dR \Delta t.
    \label{eq:wave_total_energy}
\end{equation}
Here $R=1.6\times10^{10}$ cm is the radius of the spherical wave shell at the time of its peak energy flux, $dR=L=5.4\pm0.9\times10^9$ cm is the thickness of the wave at this time, and $\Delta t=1800$ s is the time elapsed from the wave onset at this point, all of which are determined directly from the SUVI images of the wave. Inserting the appropriate values into Equation~\ref{eq:wave_total_energy} yields $E_\mathrm{wave} = 9.4\pm4.0\times10^{30}$ erg.

\citet{Long2015} provide a different approach to computing the energy of a wave, applying the Sedov-Taylor approximation, which is used to compute the energy of a spherical blast wave. These authors derive a relation between blast wave radius, $R$, time, $t$, energy, $E$, and a parameter, $\alpha$, which describes the rate of fall-off of atmospheric density with height in the corona,
\begin{equation}
    \log R = \frac{2}{5-\alpha}\log t+\log \left(\frac{E}{n} \right)^{1/(5-\alpha)},
    \label{eq:long_sedov_taylor}
\end{equation}
where $n$ is the density at the base of the corona.

We found that this approach is extremely sensitive to the choice of exact start time of the wave's acceleration (that is, the choice of $t=0$), and a wide range of $\alpha$ and total energies can be obtained depending on small changes in this choice. However, if we set $E = 9.3\times10^{30}$ erg and $n=1.26\times10^8 \mathrm{cm^{-3}}$, as above, and assume the wave's expansion began at 15:20~UT, roughly the start time of the flare's initial rise, we can then solve Equation~\ref{eq:long_sedov_taylor} to find $\alpha\sim3.5$. This yields an acceleration profile for the wave that matches the data, but implies a much steeper falloff in density than is typically observed in the corona \citep[see, e.g.,][]{Pascoe2019}. It could be that the wave, expanding into the region evacuated by the CME, indeed encounters a steep density depletion, or could suggest other problems with this approach: that it remains too sensitive to $t=0$ for us to accurately fit the wave or that the Sedov-Taylor approximation is not appropriate for this event. In either case, it is prudent to exercise caution when attempting to characterize wave energetics using the \citet{Long2015} approach.

\label{subsec:wave_energetics}

\subsection{CME Kinetic \edit2{and Gravitational Potential Energies} }
\label{section:kinetic}

Determining the kinetic energy of the CME after the eruption is possible -- as long as reliable estimates for the mass and velocity can be determined -- using the formula

\begin{equation}
    KE= \frac{1}{2} mv^2.
    \label{eq:kinetic}
\end{equation}

The mass is calculated using remote sensing imaging of the CME in white light, which measures Thomson scattering by electrons, to infer the electron density along the line of sight. Making some assumptions about the location and width of the structure that is causing the scattering along the line of sight, we convert these electron densities into mass in each given pixel of a coronagraph image \citep{Vourlidas2010}. If we select a particular region of interest inside an image and perform this procedure on each pixel corresponding to the CME, the sum of the masses provides an estimate of the total CME mass. 

Given the assumptions that go into each individual measurement, performing multiple measurements of the CME mass as the CME traverses the field of view of an instrument helps to establish a more accurate estimate of the mass. However, it is important to note that in a coronagraph the mass will not be constant, due both to mass continuously emerging from behind the occulter and from mass pile-up at the front of the CME. \citet{Colaninno2009} and \citet{Bein2013} developed a height-mass formula to account for this mass evolution, given by
\begin{equation}
    m(h)= m_0+\Delta m(h-h_{occ},)
    \label{eq:mass_evol}
\end{equation}
where $m$ is the mass at a given height, $h$, and  $h_{occ}$ represents the height of the occulter edge, where mass from the CME first becomes measurable.  The mass of the CME when the leading edge is first visible beyond this occulter is given by $m_0$, and $\Delta m$ is the mass accumulated by pile-up as the CME propagates.

\begin{figure}
\centering
\includegraphics[scale=0.25]{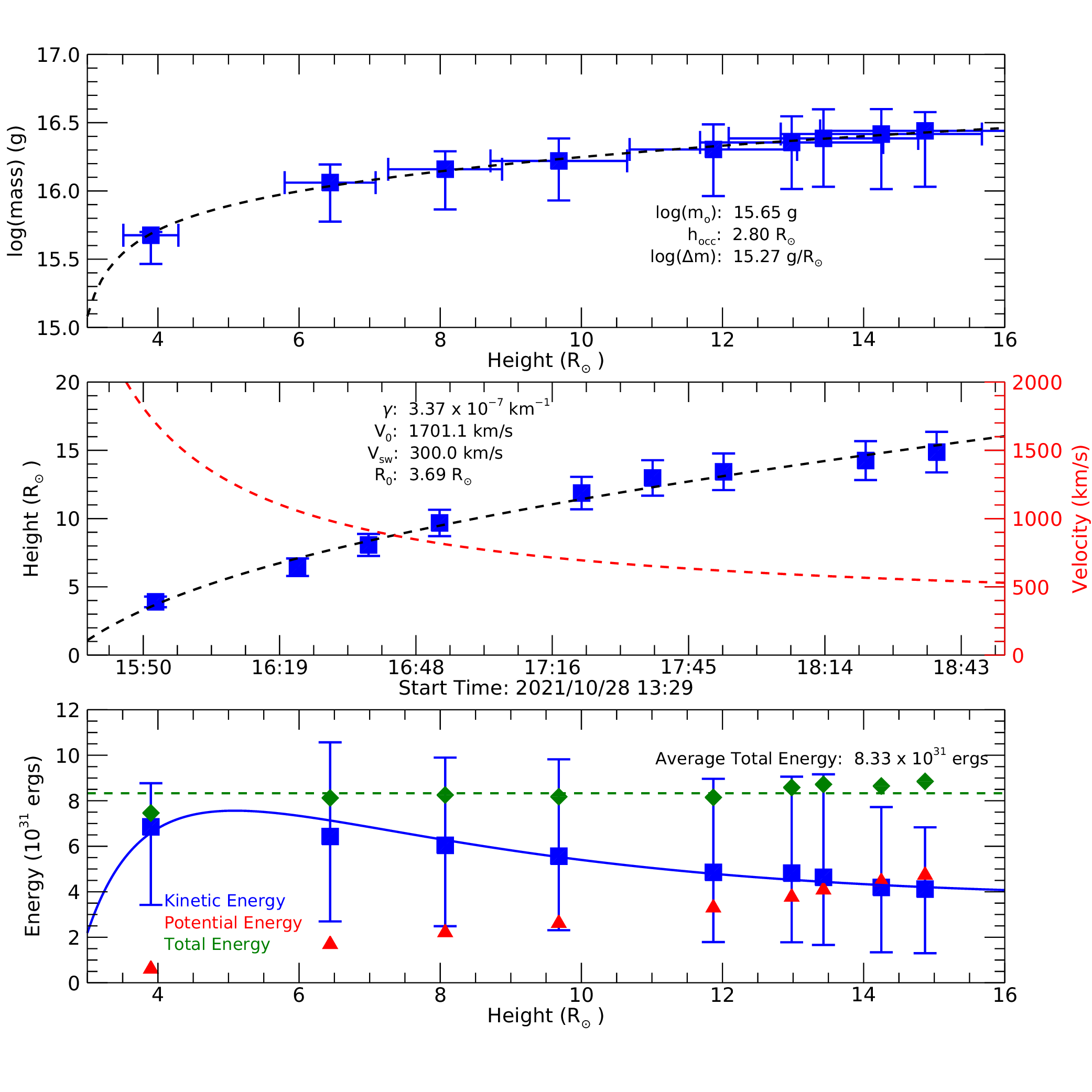}
\caption{\label{masske.fig} Top: Measured CME Mass for the individual images in which the front is visible in COR2 (blue squares). The heights are the outermost edge of the manually determined region of interest, representing the CME. The dashed line is the fit to the mass evolution equation. The extracted coefficients are provided. The error bars on the height are assumed to be $\pm10~\%$. The mass error bars were generated by calculating the mass of the CME in 1000 instances, assuming the same $10~\%$ error along the CME\,ROI boundaries. Middle: CME height (blue squares; same as top \edit2{panel}) as a function of image time. The black dashed line is the fit to the DBM (coefficients provided). The red dashed line shows the velocity profile, determined from the analytical derivative of the height profile. The error bars are the same as above. Bottom: Kinetic energy in each image (blue squares), based on the measured mass from the top plot and the velocity profile in the middle plot. The blue line is determined by from the mass evolution profile and the velocity at all heights. The red triangles are the gravitation potential energy, based on the distribution of masses in each pixel within the manually selected region of interest. The green diamonds are the sum of the the kinetic and potential energies. The dashed line is the mean value of these points. The kinetic energy error bars are determined from the mass error in the top plot, and a $10~\%$ error on the velocity. To avoid over-crowding the plot, errors for the potential energy are not shown, but are of a comparable magnitude to the kinetic energy.}
\end{figure}

The top plot in Figure\,\ref{masske.fig} shows the masses we determined for this event using observations from \textit{STEREO}'s COR2A coronagraph, as well as the fitting result and derived terms from Equation\,\ref{eq:mass_evol}. The derived values for both $m_0$ and $\Delta m$ are towards the high end of the distribution reported in \citet{Bein2013}. On one hand, given the size and speed of the event, a mass that is initially above average and that continues to accumulate at a high rate as the CME propagates is not unexpected. However, we also cannot rule out the possibility that, despite our best efforts to isolate the CME in question, signal from the other event present in COR2 observations contributes additional mass to the selected regions of interest. Furthermore, because the nose of this CME is estimated to be just $\sim40^{\circ}$ from \textit{STEREO-A}, the bulk of the electrons are far from the plane of the sky, where Thomson scattering efficiency would be maximized. These two factors contribute significant uncertainties to the already volatile mass calculations. Especially given the confounding effects of the other event, we consider this mass estimate to be an upper bound on the actual mass. 

To provide some estimate of uncertainty, we focus on the error introduced by the manually selected boundaries of the CME region of interest (ROI) in each image. We consider a $10~\%$ error on both the height (Figure\,\ref{masske.fig} top panel) and the position angles that we determine to constrain the CME width. For this event, we constrain the CME to be between $180^{\circ}$ and $280^{\circ}$ from solar north. Randomly altering these boundaries 1000 times and finding the total mass within the resulting ROI, we get a distribution of masses around the manually selected region. We use the minimum and maximum values for the error. The error bars Figure\,\ref{masske.fig} are clearly asymmetrical because of this method of determining error. If we assume the originally determined height is correct, moving the boundary inwards will remove pixels containing mass from the measurement. However, expanding the boundary beyond the CME adds only background pixels, which contain significantly lower mass values. 

The same COR2 images are also used to determine the velocity of the CME. The middle panel of Figure\,\ref{masske.fig} shows the height of the CME at each time in the COR2 image used to generate the mass measurements. In order to obtain a height-time profile from which an analytic derivative could be used to determine the velocity, these points were fit with the equation for the Drag Based Model \citep[DBM;][]{Vrsnak2013}:
\begin{equation}
    R(t)=\frac{1}{\gamma}ln[1+\gamma(v_{0}-w)t]+wt+R_0,
\end{equation}
which provides a radial distance for the leading edge, $R$ as a function of time $t$, assuming that the deceleration of a fast CME is caused by drag from the solar wind. The drag parameter, $\gamma$, governs the rate at which the solar wind will slow the CME, and $v_0$ and $w$ are the speed of the CME at height $R_0$ and the upstream speed of the ambient solar wind, respectively. 

While the DBM can be used for forecasting purposes to project the full CME kinematic evolution to 1\,au \citep{Calogovic2021}, we are less concerned with obtaining accurate predictions of the CME behavior in the heliosphere than in obtaining a reasonable velocity profile for the CME in the coronagraph observations. This profile is determined by taking the analytic solutions for velocity in \citet{Vrsnak2013}, given by
\begin{equation}
    v(t)=\frac{(v_{0}-w)}{1+\gamma(v_{0}-w)t}+w,
    \label{eq:dbmv}
\end{equation}
which produces a velocity profile that will asymptotically approach that of the solar wind. The DBM model terms, height time profile, and resulting velocity are included with the measured heights in the middle panel of Figure\,\ref{masske.fig}. By combining this velocity profile with the mass measurements and mass function from Equation\,\ref{eq:mass_evol}, we determine the evolution of the CME kinetic energy throughout the COR2\,FOV.

The kinetic energy is shown in the bottom panel of Figure\,\ref{masske.fig} (blue squares).  This quantity is calculated using Equation\,\ref{eq:kinetic} based on the mass measurements of the blue squares in the top panel of Figure\,\ref{masske.fig} and velocities at each height from Equation\,\ref{eq:dbmv}. The blue curve in the bottom panel of Figure\,\ref{masske.fig} is a continuous representation of the kinetic energy, based on Equation\,\ref{eq:mass_evol} and the velocity profile from the middle panel. The sharp initial increase in kinetic energy is unrealistic, as it is caused by the mass that is initially unaccounted for behind the occulter when the CME leading edge is at low heights. The gradual decline in the kinetic energy as the CME speed declines, despite the slight increase in mass, could be explained by the momentum transfer between the CME and the solar wind as the CME decelerates from its initial impulsive phase.

We also use the mass measurements to study the evolution of gravitational potential energy as the CME propagates. The determination of the mass allows us to estimate the potential energy, by taking computed mass in each pixel and calculating the radial height of that pixel from the center of the Sun. This procedure essentially treats every pixel as an individual particle with gravitation potential $U$, given by
\begin{equation}
    U=\frac{GMm}{R},
\end{equation}
where $G$ is the gravitational constant ($6.67\times 10^{-8}$\ erg\ cm\ g$^{-2}$), $M$ is the mass of the Sun ($2\times 10^{33}$ g) and $R$ is distance from the center of the Sun. With the mass and height at each pixel, we determine the potential energy in that pixel and sum these values together to determine a total potential energy for the CME. This method of calculating the gravitational potential energy while ignoring the distribution of mass throughout the CME volume is notably simplistic, but it still provides a useful estimate for the potential energy and allows us to see how it evolves with CME height. 

The red triangles in the bottom plot of Figure\,\ref{masske.fig} show the potential energy determined from each image. It is very clear that as the CME propagates outwards and the kinetic energy declines, the gravitational potential increases because of the increasing CME mass and height. The green diamonds show the total of the kinetic and potential energies, which becomes roughly constant through all the measurements. 

Conceptually, this outcome is easy to understand. The CME initially begins with a significant amount of kinetic energy as it rapidly accelerates and leaves the corona. Eventually, the acceleration phase of the CME \citep{Zhang2004} ends, and the CME starts to decelerate. This deceleration corresponds to a decrease in the kinetic energy, which is caused by the resistance of the ambient solar wind on the CME itself. In the process of this interaction between the solar wind and the CME, solar wind plasma gets piled up in front the CME, causing the additional mass and the conversion of the kinetic energy into additional potential energy. 

The mean value of the total (kinetic+potential) energy at each point is $8.33\times10^{31}$~ergs, just under the value of $9.1\times10^{31}$~ergs of free magnetic energy calculated in Section\,\ref{section:mag}, indicating that nearly all of the initial available energy goes into the CME. Because we consider the mass values to be an upper bound, resulting energies are also an upper bound, and thus it is likely that the actual total CME energy in the corona is somewhat lower.

The error bars on the kinetic energy are determined by taking the same mass errors shown in the top panel of Figure\,\ref{masske.fig}, as well as extending the $10~\%$ error on the height to the velocity and inserting the resulting mass and velocity values into Equation\,\ref{eq:kinetic} to get minimum and maximum values. We do not include potential and total energy errors, but the potential errors scale directly with the mass, and the total error would be the combination of both. The resulting errors are all on the order of $\sim50~\%$ of the magnitude.

\subsection{CME Energy Fluxes in the Heliosphere}
\label{sec:CME}

\begin{figure*}
\centering
\includegraphics[width=0.47\textwidth]{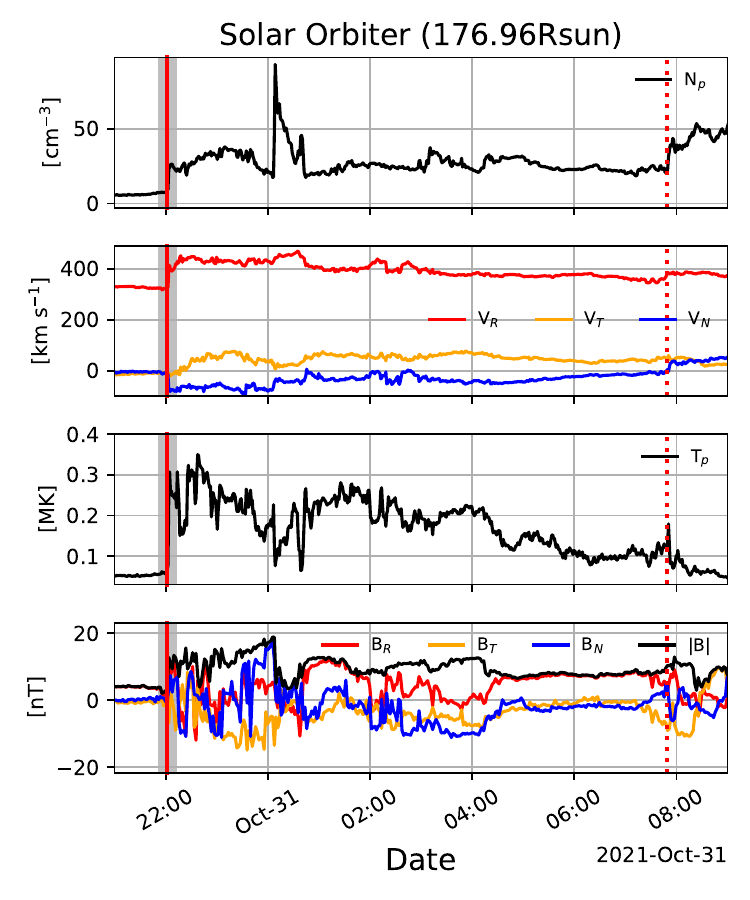}
\includegraphics[width=0.47\textwidth]{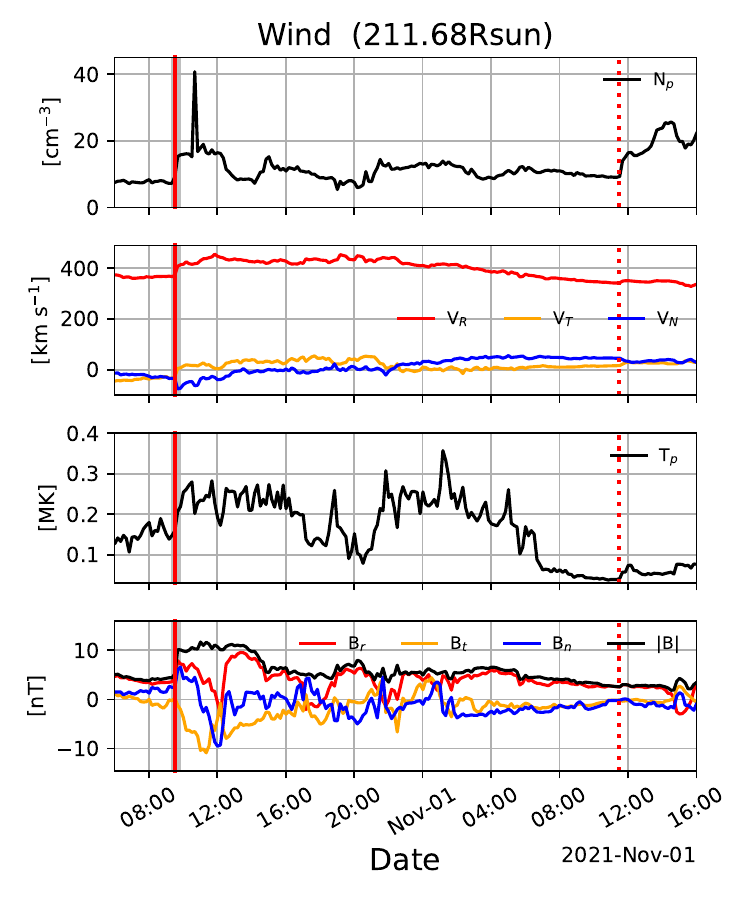}
\caption{Properties of the CME at \textit{Solar Orbiter}, measured at 176.96\,R$_{\Sun}$ (left), and \textit{Wind}, measured at 211.68\,R$_{\Sun}$ (right), showing the proton density, speed components in the RTN frame, proton temperature, and magnetic field components and magnitude. The vertical solid and dotted red lines indicate the shock and trailing edge of the CME, respectively. The shaded region prior to and after the shock front are the time periods considered for the shock analysis in Section \ref{sec:shocks}.}
\label{fig:CMEprop}
\end{figure*}

The CME was observed in the heliosphere by several spacecraft near the Sun-Earth line. In order to calculate the energy fluxes of this event in the heliosphere, we include observations from \textit{Solar Orbiter} and \textit{Wind} that are closely aligned in longitude, as noted in Figure~\ref{fig:orbit}. \textit{Solar\,Orbiter} was at a heliocentric distance of $\sim 0.81$~au (176.96\,R$_{\sun}$) at a heliographic longitude of $-2.052^\circ$ at time of CME arrival at \textit{Wind} was at $\sim0.98$~au (211.68\,R$_{\sun}$) at a heliographic longitude of 0$^\circ$ at CME arrival. Figure~\ref{fig:CMEprop} shows observations of the CME at each spacecraft in their typical units, top to bottom, the proton number density, RTN velocity components, proton temperature, and RTN magnetic field components and magnitude. The plots contain the location of the shock (solid red line) and end of the ejecta (dashed red line).  

The gray shaded region around the shock in Figure~\ref{fig:CMEprop} is the time frame used to determine the shock properties discussed in Section\,\ref{sec:shocks}. At each spacecraft, the shock was identified by the concurrent enhancement of the density, bulk speed, temperature, and field strength \citep{Zurbuchen2006,Richardson2010}. The trailing end of the CME was determined by identifying the location where the plasma parameters returned to upstream, ambient solar wind conditions, or at a discontinuity indicating a distinct stream. At \textit{Solar\,Orbiter}, the shock arrived around 22:02\,UT on 2021\,October\,30 and the trailing boundary is set to 07:49\,UT on 2021\,October\,31. At \textit{Wind}, the shock arrived at 9:33\,UT on 2021\,October\,31 and the trailing boundary is set to 11:29\,UT on 2021\,November\,1. 

Table\,\ref{cme_prop.tab} shows the mean and peak values observed between the red vertical boundaries at each spacecraft location in Figure~\ref{fig:CMEprop}.  The CME exhibits properties typical of transients near 1~au, showing a pronounced enhancement in the density, speed, and magnetic field forming the shock that is immediately followed by variability in the magnetic field component that form the sheath. The speed and temperature monotonically decrease across the CME body, indicating rapid expansion. Some points of interest include an enhancement in the proton density at \textit{Solar Orbiter} covering $\sim40$\,minutes that is also observed near the front during its passage at \textit{Wind}. The magnetic field components exhibit similar profiles, showing multiple changes in polarity exhibited by the B$_R$ component which becomes \edit1{almost} radial towards the trailing half of the CME at both heliocentric locations. Their similarities are likely due to the small longitudinal separation of the two spacecraft.

In line with the findings of \citet{Liu2005}, we find that the bulk speed remains \edit1{steady} between the two spacecraft and while we expect 
the temperature to decrease \edit1{due to expansion of the CME}, there is a \edit1{significant} standard deviation, suggesting \edit1{correspondingly} significant \edit1{temperature}
variations across \edit1{the CME's} 
substructure.  We also find that the temperature profile is high compared to the upstream solar wind, deviating from its expected cooler temperature profile \citep{Lopez1986}, as the 
temperature panel in Figure\,\edit1{\ref{fig:CMEprop}} shows. This result suggests significant heating at the CME source, similar to other X-flare-associated CMEs whose ionic compositions reflect extreme temperatures near the eruption site and a proton temperature that remains comparable to $\sim10^{5-6}$\,K near 1\,au \citep{Wang2001, Liu2008, Rivera2019, Rivera2023}.

\begin{figure*}
\centering
\includegraphics[width=0.98\textwidth]{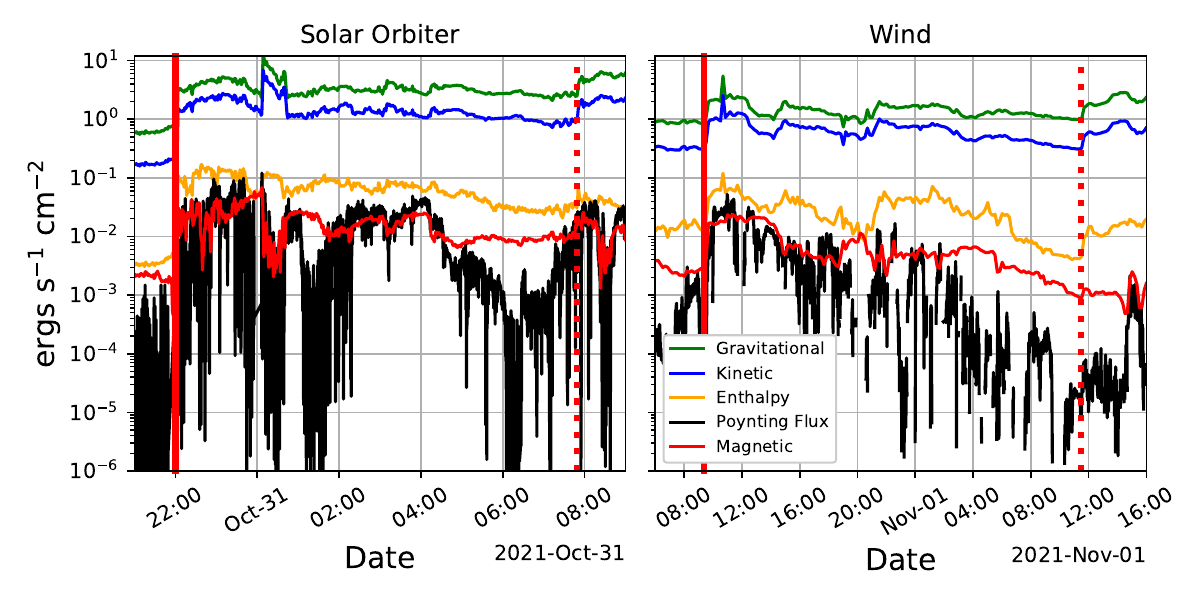}
\caption{Energy and Poynting fluxes computed in time for \textit{Solar Orbiter} (left) and \textit{Wind} (right). The red vertical lines are the boundary of the CME at both spacecraft discussed in Section \ref{sec:CME}.}
\label{fig:CMEenergyflux}
\end{figure*}

Figure~\ref{fig:CMEenergyflux} shows the fractional energy flux components at the two spacecraft, \edit2{and Table\,\ref{cme_prop.tab} presents the individual in situ  energy fluxes of the CME at the different spacecraft locations. The energies are calculated in the same manner as \citet{Liu2021, Rivera2024, Rivera2025}, but only considering the proton population. The values are an average within the respective boundaries at each spacecraft and the variability shown in the standard deviation. The enthalpy and magnetic terms decrease, as expected, from the cooling and expansion experienced as the CME propagated from \textit{Solar Orbiter} to \textit{Wind}. The kinetic and gravitational terms dominate the energy fluxes, but they also decrease between the two spacecraft. This decrease is likely driven by the density at both spacecraft, which indicates an expansion profile faster than the $1/r^2$ that is expected for self-similar expansion, where a mean density of 26.8\,cm$^{-3}$ would be expected to decrease to $\sim18.73$\,cm$^{-3}$ at \textit{Wind} but is instead much lower, measured to be a mean value of 11.2\,cm$^{-3}$. This rapid expansion is consistent with the observed statistical trend of CMEs measured throughout the inner heliosphere \citep{Liu2005}.}
Across the CME body, we see that all the energy fluxes decrease towards the trailing end of the CME but jump back up after the red dashed line that marks the end of the CME. The enhancement after the end of the CME is driven by a compression region trailing the CME. 

\begin{table}
\begin{center}
\caption{CME properties \edit2{and mean energy fluxes measured at \textit{Solar Orbiter} and \textit{Wind}}.   \label{cme_prop.tab}}
\begin{tabular*}{\linewidth}{@{\extracolsep{\fill}} l r r}
\tableline\tableline

{\it Solar\,Orbiter} 176.96$R_{\Sun}$ & &\\
 Parameter & Mean & Max \\
\hspace{0.2cm} Density (cm$^{-3}$) & 26.8 & 93.2 \\
\hspace{0.2cm} Speed (km s$^{-1}$) & 402 & 471 \\
\hspace{0.2cm} Temperature (MK) & 0.18 & 0.35 \\
\hspace{0.2cm} Magnetic field (nT) & 9.8 & 19.9 \\ 
Energy Fluxes\tablenotemark{a} & Mean & Std. Dev. \\
\hspace{0.2cm} Kinetic & 1.53 & $\pm0.73$ \\
\hspace{0.2cm} Enthalpy & 0.07 & $\pm0.03$ \\
\hspace{0.2cm} Gravitational & 3.46 &$\pm1.17$ \\
\hspace{0.2cm} Magnetic & 0.017 &$\pm0.011$ \\
\hspace{0.2cm} Poynting & 0.014 &$\pm0.015 $ \\ 
\\
{\it Wind} 211.68$R_{\Sun}$ & & \\
 Parameter & Mean & Max \\
\hspace{0.2cm} Density (cm$^{-3}$) & 11.2 & 40.7 \\
\hspace{0.2cm} Speed (km s$^{-1}$) & 408 & 456 \\
\hspace{0.2cm} Temperature (MK) & 0.18 & 0.36 \\
\hspace{0.2cm} Magnetic field (nT) & 6.2 & 11.7 \\
Energy Fluxes\tablenotemark{a} & Mean & Std. Dev. \\
\hspace{0.2cm} Kinetic & 0.66 &$\pm0.27$ \\
\hspace{0.2cm} Enthalpy & 0.03 &$\pm0.019$ \\
\hspace{0.2cm} Gravitational & 1.46 &$\pm0.49$ \\
\hspace{0.2cm} Magnetic & 0.007 & $\pm0.006$ \\
\hspace{0.2cm} Poynting & 0.006 & $\pm0.008$ \\
\tableline\tableline
\end{tabular*}
\tablenotetext{a}{Energy fluxes are in units of ergs s$^{-1}$ cm$^{-2}$.}

\end{center}
\end{table}

\label{section:cme}

\subsection{Energy Partition at the CME Shock}
\label{sec:shocks}

To calculate the energy partition at the CME shock, we use the shock analysis method described by \citet{David2022}, where not only are the upstream and downstream energy fluxes partitioned between kinetic, thermal, and magnetic components, but energetic particles are also extracted and compared. The present method was applied for the first time to include the contribution of energized particles (both protons and electrons) partial pressure to the total enthalpy flux to a sample of $1$\,au IPs that crossed \textit{Advanced Composition Explorer} \citep[\textit{ACE}, ][]{1998SSRv...86....1S} and \textit{Wind} spacecraft and an IPs that crossed Parker\,Solar\,Probe at $\sim 0.8$\,au. The magnetohydrodynamic (MHD) Rankine-Hugoniot (RH) jump conditions for energy and momentum normal to the shock surface are evaluated in the shock frame. 
We also include the enthalpy flux of suprathermal (only for \textit{Wind}) and energetic electrons in our analysis since \citet{David2022} unexpectedly found that the energetic electron energy fluxes is comparable to that of energetic protons for a significant fraction of the subsample of \textit{Wind} shocks at $1$\,au. 

We disentangle the solar wind energy fraction from the energized protons (both supra-thermal and energetic) via the cutoff $v/V_{sh} \gtrsim 3$ \citep{David2022}, where $v$ ($V_{sh}$) is the proton (shock) speed\footnote{Here $V_{sh}$ is calculated along the normal to the shock surface.}, both evaluated in the spacecraft frame, or equivalently, for a kinetic energy $10$ times the ram energy of the bulk plasma. Such a criterion implies a proton energy cutoff of $8.91\,\text{keV}$ for \textit{SolO} and $10.0\,\text{keV}$ for \textit{Wind}. These cutoffs manifest as removing the lowest STEP-p energy bin while the minimum 3DP/SOSP-p bin is above this threshold (see Figure~\ref{solo energy ranges.fig}). The 3DP/PLSP ions instrument cannot be used because its highest energy (3~keV) is smaller than the 10\,keV energy cutoff. The contribution of energized heavy ions is neglected here. 

As for the electrons, we take an additional step to avoid double-counting the electron energy contribution at the \textit{Wind} shock. Since ELSP-e extends down to 5.18\,eV and overlaps with the thermal core of the electron distribution, within the lowest nine energy bins, extending up to 65.2\,eV, we verified that the ELSP-e flux matches the flux of a Maxwellian with temperature 17\,eV and density equal to the measured thermal proton density. 
Thus, these nine bins are removed from the present analysis. Furthermore, we subtract the expected thermal flux from the remaining six energy bins to count only the energized (supra-thermal) component. These steps are not necessary for the \textit{SolO}/STEP-e data since the minimum energy of 4.1\,keV is well above the thermal core. This analysis does not include \textit{SolO} electron data below 4.1\,keV.
 
The \textit{SolO} upstream ($\sim 30$\,minutes) flow velocity, magnetic field, and thermal pressure and plasma density time series show no \edit1{meaningful} amplitude fluctuations, except within a few minutes of the shock. Thus, the jump conditions use the 10\,minute average of these quantities before the time of the shock passage, ending 1\,minute before the shock. Although excluding time intervals with \edit1{significant} fluctuations narrows error-bars
, we find that jump conditions are satisfied to a within 1-$\sigma$, \edit2{and there is} good agreement in the conservation of the flux of energy density and the normal component of momentum density across the jump. Unlike the upstream plasma, the downstream flow speed, magnetic field, plasma temperature, and density show larger and longer time scale ($\sim$ hours) fluctuations (Figure\, \ref{SolO_Oct30_time_series.fig}). 

The radial component of the flow velocity, $V_R$, grows about 15\%, from seconds to 30 minutes, downstream of the shock. Within an hour, the density grows from about $25\, {\rm particle}\, \text{cm}^{-3}$ to $35\, {\rm particle}\, \text{cm}^{-3}$ and the temperature strongly fluctuates between 12\,eV and 32\,eV. The magnetic field varies by a factor of five within the same timeframe and the $B_R$ and $B_T$ components change sign multiple times within 30~minutes after the shock, possibly due to the crossing of a large-scale structure such as a flux tube. We average these quantities for 10~minutes beginning one minute after the shock, where they are approximately constant. This process yields a good agreement between the upstream and downstream energy and momentum fluxes. We applied the same averaging interval to all energetic particle data. We possess errors for all quantities except the flow velocities, temperatures, and thermal densities, for which we assumed 10\% measurement uncertainty.

The \textit{Wind} flow velocity, thermal proton density, and proton temperature have much lower cadences -- about 100 seconds -- those of the \textit{SolO} data. The time series remain approximately constant within 20~minutes both upstream and downstream of the shock. These two facts lead us to choose a 20-minute averaging interval for \textit{Wind}, which we also used for the high-cadence (11 per second) \textit{Wind} magnetic field data and all \textit{Wind} energetic particle data.

As discussed by \citet{David2022}, to avoid double-counting energy gained at the shock surface, the enthalpy flux of energized particles is not added to the upstream energy flux to verify the RH jump conditions. A more detailed analysis has to disentangle the upstream flux of energized incoming particles from the those reflected by the shock; here we do not analyze the 3D\,velocity distribution function necessary to such separation.

\begin{figure*}
\includegraphics[width=0.96\textwidth, ,viewport=0 200 800 430,clip]{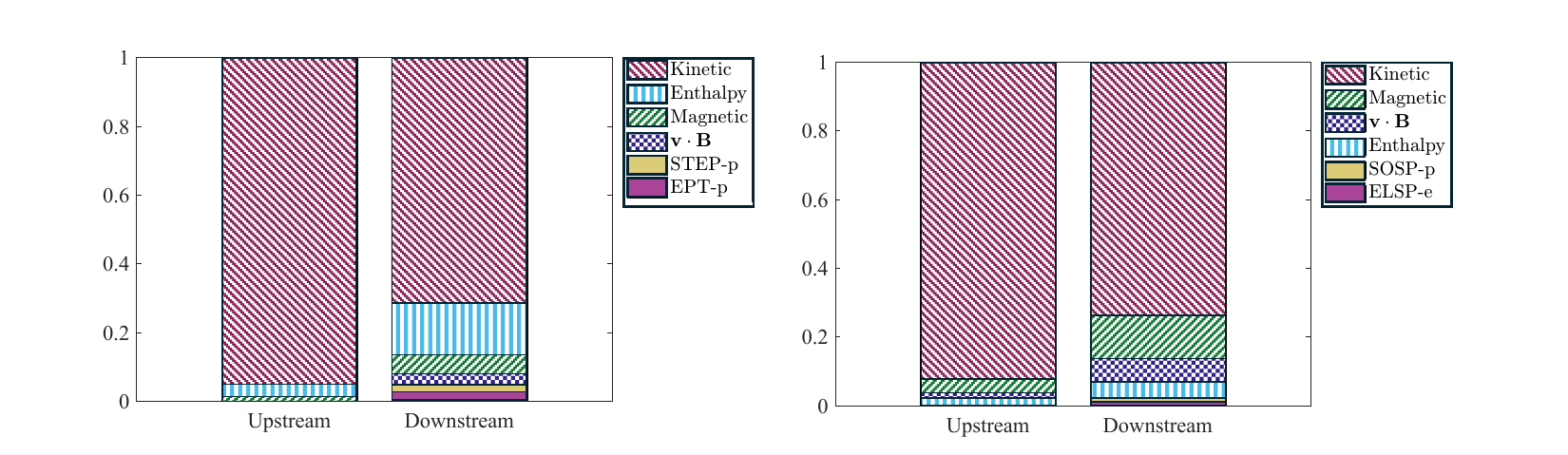}
\caption{\label{energy partition oct30.fig} Energy partition for the 2021\,October\,30 shock at \textit{SolO} (left) and \textit{Wind} (right), normalized to unity and ordered based on downstream contribution. The conversion, from upstream to downstream, of the dominant kinetic component into other forms (i.e., thermal and magnetic) with a small fraction available to energize charged particles is apparent.}
\end{figure*}

Figure~\ref{energy partition oct30.fig} shows the energy partition across the shock at \textit{Solar Orbiter} and \textit{Wind}. Quite surprisingly the partial energy flux into proton and electron energization compared to the total downstream energy flux is extremely small ($<5$\%), in comparison with the typical 15--20\% measured by \cite{David2022} over a broad IPs sample, and as a consequence even smaller in comparison with the upstream ram pressure.

\label{section:shocks}

\section{Discussion and Conclusions}

 We find for the 28 October 2021 eruption that the total energy associated with the CME  is $9.75\pm3.6\times10^{31}$~erg, \edit2{amounting to a large fraction of the released magnetic energy}, while the nonthermal energy \edit2{in the flare} is 2.25 $\pm$ 0.9$\times10^{31}$~erg. A previous survey of 38 eruptive flares by \citet{emslie12} found that CME energy represented about four times the energy contained in accelerated ions and electrons for the same events. On the other hand, \citet{Aschwanden2017} argued that the CME represented only about 10\% of the energy of accelerated ions and electrons. Thus, for this event at least, our findings are much more consistent with the \citeauthor{emslie12} result.  \edit2{Aschwanden et al. used an unconventional dimming-based method to estimate CME properties from EUV images, while we estimate CME properties using a more traditional coronagraph-based approach. One possible explanation for the disparity is that the CME estimation method used by \citet{Aschwanden2017} is producing an underestimate for typical CME kinetic energy.}


The fact that the total of CME and nonthermal energies for this event exceeds the \edit2{estimated} free magnetic energy suggests some of the values in Table~\edit2{\ref{energy_partition.tab}} may be overestimates. The CME-associated energies, which are difficult to constrain and have error bars \edit1{$>40$\%}, are good candidates -- though even a reduction by a factor of two would still support the conclusion of \citeauthor{emslie12} that CMEs represent the majority of the energy released by an eruptive flare. There is evidence that EUV waves are CME-driven, and that more energetic CMEs are associated with faster, more impulsive EUV waves \citep{Murh2014}. Thus the strong, fast EUV wave associated with the event is also consistent with the energy distribution for an event in which the CME carries the majority of the total energy. \edit2{Additionally, it is possible that our magnetic free energy is an underestimate (see discussion in Section \ref{section:mag}.)}



\citet{Warmuth2020} point out that the bolometric energy is a useful check on the total energy contained in the flare part of the eruption. We did not measure bolometric energy directly, but it can be estimated using the previous result of \citet{emslie12} and \citet{Kretzschmar}.  Using this method we found a bolometric energy in the range 1.6--3.0~$\times 10^{31}$\,ergs.  \edit2{Additionally}, the UFC method has been found to be a reasonable proxy for the bolometric energy \citep{Qiu2021}, and \edit2{using this method,} we find an energy of 1.2~$\times 10^{31}$\,ergs at the peak of the flare, with another 0.5~$\times 10^{31}$\,ergs in the tail, consistent with the \edit2{estimate of the bolometric energy} within errors.

\edit2{For our event, the total energy in the nonthermal particles is estimated to be $\sim$2.25~$\times 10^{31}$\,ergs, which is within the calculated range of  bolometric energies, thus verifying that our estimates of the nonthermal energy are reasonable. Previous estimates of the ratio of nonthermal energy to the bolometric energy have been in the range of 10-60\% \citep{Warmuth2020}.  An outlier is  the study of \citet{Aschwanden2017}, which finds that the nonthermal energy is over 5 times the bolometric energy. However, the nonthermal energies for the events in this sudy were revised significantly downward in a subsequent paper \citep{Aschwanden2019}, leading to more typical ratios of $\sim$35\%.}

\edit2{Our estimates put the ratio of bolometric energy to thermal energy for this event somewhere between 1.5 and 6, depending on which end of the bolometric energy range is included and whether only the peak or the peak + tail is considered. } \citet{Aschwanden2017} find that the bolometric energy is approximately equal to the thermal energy \edit2{and} \citet{emslie12} report bolometric energies that are between 1.6--11 times greater than the thermal energy, with the mean falling somwehere around 5.  \edit2{Our ratio of bolometric to thermal energy} is \edit2{thus more} consistent with \edit2{ratios} reported by \citet{emslie12}. 

\citet{Warmuth2020} argued that DEM-based methods tend to overestimate the amount of high temperature plasma, and thus yield unrealistic levels of X-ray emission for many events. This overestimation may explain why \citet{Aschwanden2017} concluded bolometric and thermal energies should be approximately equal, while \citet{emslie12} concluded that the energy radiated by the SXR plasma is often only 20\% of the bolometric energy or less. Here we have estimated the thermal energy of the Oct.~28 eruption in two different, independent ways, including one that directly leverages the SXR emission. That we arrive at consistent results suggests our thermal energy estimates may be sufficiently reliable to conclude that -- for at least some events -- bolometric energy is likely to exceed thermal energy by a factor of at least a few.


\edit2{From the in situ analysis, we find that the leading energy fluxes are the gravitational and kinetic terms at both \textit{SolO} and \textit{Wind}. The magnetic energy flux and the Poynting flux are of the same order of magnitude at both heliocentric distances. The enthalpy flux is an order of magnitude larger than both the magnetic and Poynting flux, or the electromagnetic wave energy, in both cases.  Similar to the solar wind at large heliocentric distances, under energy conservation, the thermal, magnetic, Poynting flux energies have translated {\edit1{the majority of}} their energy to accelerating the CME plasma against gravity via thermal pressure gradients and magnetic forces. }

\edit2{Despite the CME carrying a large amount of kinetic energy, the CME shock does not produce a large number of accelerated particles.}  The 2021~October~30 shock at \textit{SolO} exhibits a conversion of upstream kinetic energy into mostly thermal energy, with the fraction of energetic particle components much higher than later observed at \textit{Wind}, as shown in Figure~\ref{energy partition oct30.fig}. 
The partial \edit2{conversion of} energy flux into energization of protons and electrons is surprisingly small compared to other events.
Previous cases of IPs inefficiently accelerating protons have been reported at 1~au \citep{Lario.etal:03}. Although a specific geometric configuration could explain the lack of $>$100~keV protons as a local effect at a single spacecraft \citep{Dimmock2023,Trotta2023c}, the less efficient energization at \textit{Wind} (corresponding to a smaller $M_{fms}$, as shown in Table \ref{tab:IPs}), reported here for this event, requires that a shock region with comparable plasma properties is observed at 1~au by \textit{Wind} and the efficiently accelerating shock regions do not cross either spacecraft\footnote{STEREO-A measured a $< 200$~keV ion enhancement in the Solar Electron Proton Telescope \citep[SEPT,][]{MuellerMellin.etal:08},  sunward (anti-sunward) looking direction started at $\sim$00:00 UT ($\sim$06:00 UT) of 30 October 2021 (not shown here), unlikely related to the eruption discussed herein.}.
Full 3D simulations of the expansion of the CME-driven shock will be necessary to fully interpret the joint \textit{SolO}-\textit{Wind} measurements.

The sum total of the evidence presented here indicates that the majority of the energy released in the 28 October 2021 eruption is partitioned into the kinetic and gravitational potential energy of the CME. This energy division continues to hold even as the CME evolves in the heliosphere, where in situ measurements (Figure~\ref{energy partition oct30.fig}) show that the kinetic energy still represents a majority of the total energy flux in the shock, and is not redirected into energetic particles to a significant extent. The overall partition in this event is similar to the distribution reported by \citet{emslie12}, where the CME kinetic energy represents about 3/4 of the total event energy.

This project was made possible by the wide variety of measurements available in the Heliophysics System Observatory (HSO) suite of missions and instruments, and the NASA HSO Connect Program that brought together experts from many different fields to do the analysis. Projects of this scope can only be realized by dedicated investments across multi-disciplinary research fields.
\label{section:discussion}

\newpage
\begin{acknowledgments}
The authors would like to thank the anonymous referee for insightful comments that improved this paper.  This work was supported by the NASA Heliophysics System Connect Program, grant number 80NSSC20K1283. PH acknowledges additional support from the Office of Naval Research. DT acknowledges funding from ISRO/RESPOND for the project “Solar Flares: Physics and Forecasting for Better Understanding of Space Weather” ISRO/RES/2/438/22-23. DP acknowledges the support by the National Natural Science Foundation of China (Grant Nos. 42188101, 42521007, 42474221). SY acknowledges additional support from NASA grant 80NSSC24K1242 to NJIT.
We acknowledge the use of AIA and HMI data. AIA and HMI are instruments onboard SDO, a mission for NASA’s Living With a Star (LWS) program.  Hinode is a Japanese mission developed and launched by ISAS/JAXA, with NAOJ as domestic partner and NASA and STFC (UK) as international partners. It is operated by these agencies in co-operation with ESA and NSC (Norway).   The STEREO/SECCHI data are produced by an international consortium of the NRL, LMSAL, NASA GSFC (USA), RAL and the University of Birmingham (UK), MPS (Germany), CSL (Belgium), and IOTA and IAS (France). The SOHO/LASCO data used here are produced by a consortium of the Naval Research Laboratory (USA), Max-Planck-Institut fuer Aeronomie (Germany), Laboratoire d'Astronomie (France), and the University of Birmingham (UK). SOHO is a project of international cooperation between ESA and NASA.
Solar Orbiter is a mission of international cooperation between ESA and NASA, operated by ESA. The Solar Orbiter magnetometer was funded by the UK Space Agency (grant ST/T001062/1). The Solar Orbiter STIX instrument is an international collaboration between Switzerland, Poland, France, Czech Republic, Germany, Austria, Ireland, and Italy. This paper uses Solar Orbiter EPD data, generated and maintained by the EPD team. The Solar Orbiter Solar Wind Analyser (SWA) scientific sensors, SWA-EAS, SWA-PAS, SWA-HIS, and the SWA-DPU have been designed and created, and are operated under funding provided in numerous contracts from the UK Space Agency (UKSA), the UK Science and Technology Facilities Council (STFC), the Agenzia Spaziale Italiana (ASI), the Centre National d’Etudes Spatiales (CNES, France), the Centre National de la Recherche Scientifique (CNRS, France), the Czech contribution to the ESA PRODEX programme, and NASA.  The authors thank the National Space Science Data Center of the Goddard Space Flight Center for the use permission of Wind data and the NASA CDAWeb team for making these data available at http://cdaweb.gsfc.nasa.gov/istp\_public/. 

\end{acknowledgments}

\facilities{GOES~(EXIS, SUVI), OVRO:SA, SDO~(HMI, AIA), SOHO (LASCO), SolO, STEREO (EUVI, Cor), WIND, Hinode (XRT), Fermi~(GBM, LAT)}

\appendix
\section{Treatment of the Flare as a Single Loop }
Interpreting the phase portrait of GOES-derived $T$ {\em vs.} $EM$, shown in Figure\,\ref{fig:phase}b, as a single impulsively-heated loop allows us to derive properties of the flare without recourse to imaging.  Impulsive energy input drives the temperature to a peak ($T=15.0$\,MK marked by a diamond) too rapidly for density -- i.e.\ $EM$ -- to change.  

After impulsive input, the radiative time, $\tau_{\rm r}$, remains longer than the conductive cooling time $\tau_{\rm c}$, so thermal conduction drives evaporation which returns the energy to the loop, along with additional mass.  As a consequence the energy and volume of the loop remain roughly fixed, as does its pressure \citep{Antiochos1978,Cargill1995},
\begin{equation}
  p~\simeq~ 2\,k_{\rm b}\, T\, n_e ~\simeq~ 2\,k_{\rm b}\, T\, {\sqrt{EM\over V}} ~~,
  	\label{eq:ideal_gas}
\end{equation}
where $V$ is the flare volume. This evaporative phase, characterized by $T\sim EM^{-1/2}$ (black dashed line in Figure\,\ref{fig:phase}b), persists until the density reaches a level sufficient to shorten the radiative cooling time to match the conductive time.  It appears that some additional energy release occurs during this phase, and the actual evolution arcs away from the constant-$p$ curve over the interval from point $A$ to $B$.

Evaporation ends at approximately 15:41\,UT, with $T_B=10.9$ MK and $EM_B=3.7\times10^{49}\,{\rm cm}^{-3}$, shown by the triangles in Figure\,\ref{fig:phase}.  After this the loop remains in mechanical equilibrium, $\partial p/\partial\ell\simeq0$, cooling radiatively to reduce $p$.  We assume classical thermal conduction, $F_c=\kappa_0T^{5/2}(\partial T/\partial\ell)$, where $\kappa_0\simeq 10^{-6}$ is the Spitzer-H\"arm coefficient.  Balancing this with radiative losses while maintaining mechanical equilibrium yields a loop described by the scalings of Rosner, Tucker and Vaianna \citep[RTV,][]{Rosner1978}
\begin{equation}
  p~\sim~ {T_{\rm pk}^3\over L} ~~,
  	\label{eq:RTV_p}
\end{equation}
where $T_{\rm pk}$ is loop's the apex temperature, and $L$ is its full length.  Mechanical equilibrium is maintained during the cooling, which proceeds according to eq.~(\ref{eq:RTV_cooling}).  Fitting the time evolution of $T$, shown in Fig.\ \ref{fig:phase}a, provides an estimate of the effective loop length of the flare.  An RTV loop is characterized by \citep{Serio1991}
\begin{equation}
  \tau_c ~=~ \tau_r ~\simeq~ 2\times 10^{-3}\, L \, T^{-1/2} ~~,
\end{equation}
when all quantities are expressed in CGS\,units.  Setting this to $\tau_c=92$ min at $T=T_B =10.9$\,MK, yields a full loop length $L=102$\,Mm.

The fit to the evaporative leg of the phase portrait can be used to find the flare volume without recourse to imaging.  To remain in mechanical equilibrium, the cooling loop must drain thereby decreasing $EM$ along with $T$
\begin{equation}
  EM ~\simeq~ C_{\rm rtv}\, {V\over L^2}\, \left({T\over 10^7\, {\rm K}}\right)^{\nu} ~~.
  	\label{eq:RTV_EM}
\end{equation}
applying the RTV scaling of eq.\ (\ref{eq:RTV_p}) yields $\nu=4$ with $C_{\rm rtv}=0.86\times 10^{41}\, {\rm cm}^{-4}$.  This relation, adjusted to pass through point $B$, is shown as the dashed magenta curves in both Fig.\ \ref{fig:phase}a and \ref{fig:phase}b.  Both fit extremely well until about 16:00, when some change to the equilibrium loop seems to occur.  Applying Equation (\ref{eq:RTV_EM}) to the values $T_B=10.9$\,MK and $EM_B=3.7\times10^{49}\,{\rm cm}^{-3}$ yields $V/L^2\simeq3.0$\,Mm.  Using the length, $L=102$ Mm found above from time evolution, yields a flare volume $V=3.1\times10^{28}\, {\rm cm}^3$.

The net thermal energy is given in eq.\ (\ref{eq:DWL_Uth}) and
plotted as a red curve on Figure\,\ref{fig:GOES_energy}a.  The conductive and radiative cooling times are \citep{Aschwanden2001}
\begin{eqnarray}
  \tau_{c} &=& 2.7\times10^{-10}\, T^{-5/2}\, L^2\, \sqrt{EM\over V} \\[7pt]
  \tau_{r} &=& 2.3\times10^{3}\, T^{3/2}\, \sqrt{V\over EM} ~~,
\end{eqnarray}
respectively, when all quantities are expressed in CGS\,units.  The power lost through either process is computed by dividing $U_{\rm th}$ by the corresponding time \citep{Longcope2010}, as shown in Figure\,\ref{fig:GOES_energy}b.  In Figure\,\ref{fig:GOES_energy}b these curves fall on top of one another because we match an RTV equilibrium when we set $V/L^2=3$\,Mm.
\label{section:dana_appendix}

\section{Analysis of EUV Wave Kinematics}
\begin{figure*}
\centering
\includegraphics[width=1.0\textwidth]{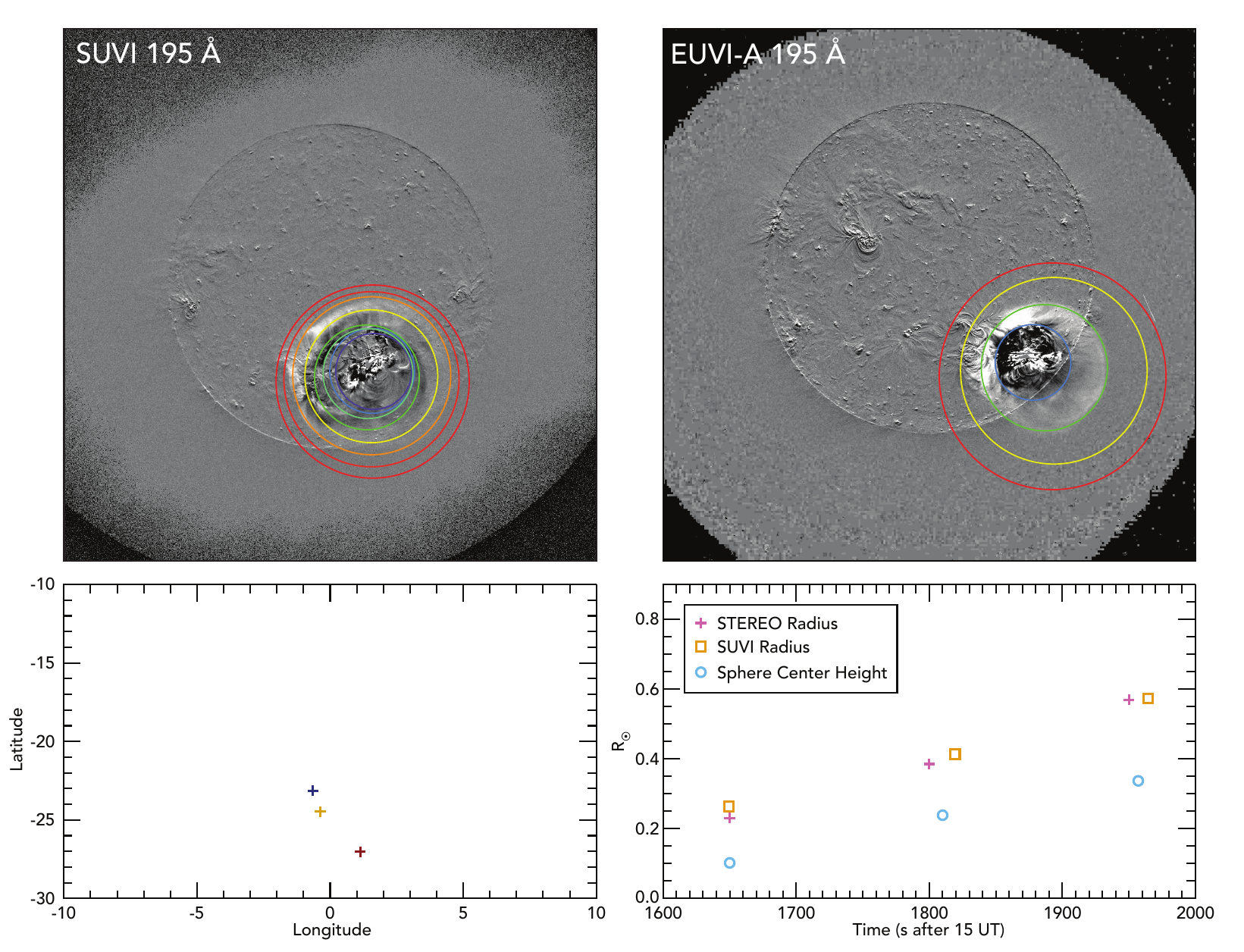}
\caption{\label{fig:wave_fits} Circular fits to the wave in SUVI (top left) and EUVI-A (top right) observations; color indicates time, with blue earliest and red latest. Bottom panels show the corresponding reconstructed position in latitude and longitude (bottom left; color indicates time as above) and radius and center height above the photosphere (bottom right).
}
\end{figure*}

The EUV wave associated with the 28 October 2021 eruption appears as an expanding circle both from the GOES and STEREO-A vantage points, which are separated by roughly $37.5\degree$ longitude at the time of the event (see Figure~\ref{fig:wave_fits}, top panels). The fact that the wave appears circular independent of viewpoint for much of its evolution strongly suggests it is essentially an expanding spherical shell, at least until parts of it begin to interact significantly with the ambient corona and magnetic field.

\begin{figure}
\centering
\includegraphics[width=0.7\textwidth]{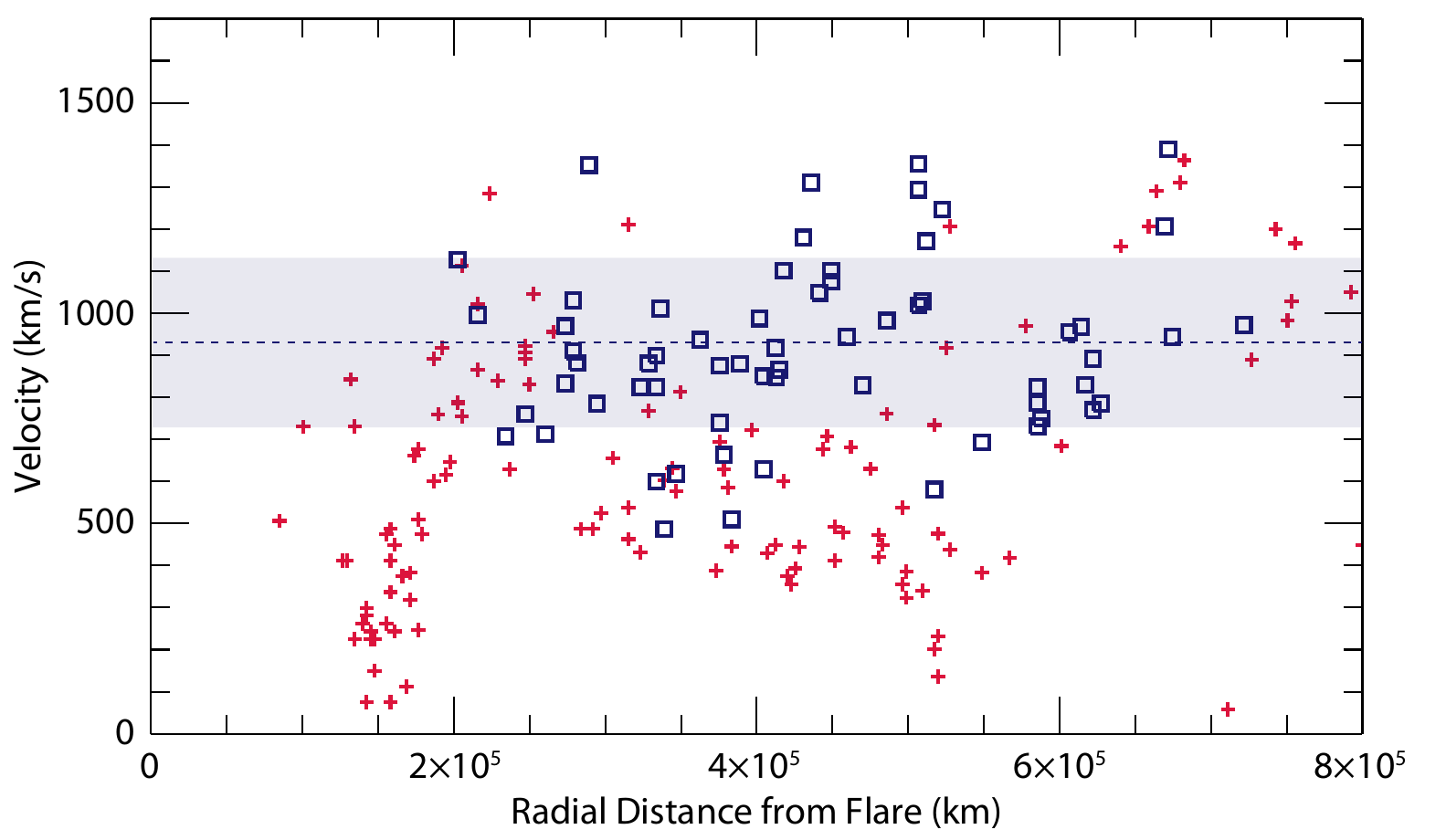}
\caption{\label{fig:wave_radial_vel} Wave radial velocity as a function of distance from the center of the flare site as observed in SUVI's 195~\AA\ passband. Red + symbols indicate points at which the wave is observed on-disk, while blue $\square$ symbols indicate points where the wave is observed above the limb. The dashed line indicates the wave's final average velocity of $\sim910\;\mathrm{km\,s^{-1}}$, while the shaded band indicates the standard deviation of speeds, which is about $\pm200\;\mathrm{km\,s^{-1}}$.
}
\end{figure}

We used SUVI 195~\AA\ images to estimate the evolution of the wave's radial propagation velocity away from its origin at the flare site in $20^\circ$ bins spanning all azimuthal angles around the center of the flare itself. Figure~\ref{fig:wave_radial_vel} shows a summary of the wave's propagation both on the disk (red + symbols) and above the limb (blue $\square$). The wave first becomes clearly visible at a radial distance of about $1.5\times10^5$\,km from the flare site, and it rapidly expands and accelerates until it reaches a speed of $910\pm200\;\mathrm{km\,s^{-1}}$ at a distance of $2-2.5\times10^5$\,km from the flare site. At this point, the part of the wave that travels off-the disk continues to propagate outwards at a \edit1{steady} velocity, while the part of the wave observed on-disk begins to interact with the ambient corona and slows and dissipates.

To understand how the wave propagates in 3D space, we fit circles to corresponding images of the wave's evolution in time from both STEREO-A and GOES perspectives and used epipolar geometric methods \citep{Inhestero2006} to reconstruct the size and location of a spherical shell resulting from these fits. Figure~\ref{fig:wave_fits} shows these fits and the reconstructed positions of wave in 3D. Our reconstruction indicates that the wave is simultaneously rising and traveling towards the south as it expands outwards. Fitting a line to the successive center point positions of the sphere, we find that the center is rising from the solar surface at about $530\pm40$ km s$^{-1}$, while the sphere's early expansion is consistent with the velocities shown in Figure~\ref{fig:wave_radial_vel}.

This result is not entirely straightforward to understand, but might be explained if the wave were expanding within an outflowing region, perhaps as the result of a large-scale flow related to the outgoing CME itself. The wave's early evolution appears to be insensitive to any variation in the magnetic field of the surrounding corona. Such behavior appears to support the interpretation that EUV waves are initiated as blast waves due to the rapid expansion of the erupting structures. This early behavior, plus the wave's later interactions with the low corona, appear to be consistent with the conclusion of \citet{Long2015} and \cite{Long2017}, that EUV Waves likely evolve from blast waves to fast-mode large-amplitude waves or shocks over their lifetimes.

\label{sec:wave_kinematics}

\newpage

\bibliographystyle{apj}

\begin{thebibliography}{}
\expandafter\ifx\csname natexlab\endcsname\relax\def\natexlab#1{#1}\fi

\bibitem[{{Antiochos} \& {Sturrock}(1978)}]{Antiochos1978}
{Antiochos}, S.~K., \& {Sturrock}, P.~A. 1978, ApJ, 220, 1137

\bibitem[{{Atwood} {et~al.}(2009){Atwood}, {Abdo}, {Ackermann}, {Althouse},
  {Anderson}, {Axelsson}, {Baldini}, {Ballet}, {Band}, {Barbiellini},
  {Bartelt}, {Bastieri}, {Baughman}, {Bechtol}, {B{\'e}d{\'e}r{\`e}de},
  {Bellardi}, {Bellazzini}, {Berenji}, {Bignami}, {Bisello}, {Bissaldi},
  {Blandford}, {Bloom}, {Bogart}, {Bonamente}, {Bonnell}, {Borgland},
  {Bouvier}, {Bregeon}, {Brez}, {Brigida}, {Bruel}, {Burnett}, {Busetto},
  {Caliandro}, {Cameron}, {Caraveo}, {Carius}, {Carlson}, {Casandjian},
  {Cavazzuti}, {Ceccanti}, {Cecchi}, {Charles}, {Chekhtman}, {Cheung},
  {Chiang}, {Chipaux}, {Cillis}, {Ciprini}, {Claus}, {Cohen-Tanugi},
  {Condamoor}, {Conrad}, {Corbet}, {Corucci}, {Costamante}, {Cutini}, {Davis},
  {Decotigny}, {DeKlotz}, {Dermer}, {de Angelis}, {Digel}, {do Couto e Silva},
  {Drell}, {Dubois}, {Dumora}, {Edmonds}, {Fabiani}, {Farnier}, {Favuzzi},
  {Flath}, {Fleury}, {Focke}, {Funk}, {Fusco}, {Gargano}, {Gasparrini},
  {Gehrels}, {Gentit}, {Germani}, {Giebels}, {Giglietto}, {Giommi}, {Giordano},
  {Glanzman}, {Godfrey}, {Grenier}, {Grondin}, {Grove}, {Guillemot}, {Guiriec},
  {Haller}, {Harding}, {Hart}, {Hays}, {Healey}, {Hirayama}, {Hjalmarsdotter},
  {Horn}, {Hughes}, {J{\'o}hannesson}, {Johansson}, {Johnson}, {Johnson},
  {Johnson}, {Johnson}, {Kamae}, {Katagiri}, {Kataoka}, {Kavelaars}, {Kawai},
  {Kelly}, {Kerr}, {Klamra}, {Kn{\"o}dlseder}, {Kocian}, {Komin}, {Kuehn},
  {Kuss}, {Landriu}, {Latronico}, {Lee}, {Lee}, {Lemoine-Goumard}, {Lionetto},
  {Longo}, {Loparco}, {Lott}, {Lovellette}, {Lubrano}, {Madejski}, {Makeev},
  {Marangelli}, {Massai}, {Mazziotta}, {McEnery}, {Menon}, {Meurer},
  {Michelson}, {Minuti}, {Mirizzi}, {Mitthumsiri}, {Mizuno}, {Moiseev},
  {Monte}, {Monzani}, {Moretti}, {Morselli}, {Moskalenko}, {Murgia},
  {Nakamori}, {Nishino}, {Nolan}, {Norris}, {Nuss}, {Ohno}, {Ohsugi}, {Omodei},
  {Orlando}, {Ormes}, {Paccagnella}, {Paneque}, {Panetta}, {Parent}, {Pearce},
  {Pepe}, {Perazzo}, {Pesce-Rollins}, {Picozza}, {Pieri}, {Pinchera}, {Piron},
  {Porter}, {Poupard}, {Rain{\`o}}, {Rando}, {Rapposelli}, {Razzano}, {Reimer},
  {Reimer}, {Reposeur}, {Reyes}, {Ritz}, {Rochester}, {Rodriguez}, {Romani},
  {Roth}, {Russell}, {Ryde}, {Sabatini}, {Sadrozinski}, {Sanchez}, {Sander},
  {Sapozhnikov}, {Parkinson}, {Scargle}, {Schalk}, \& {Scolieri}}]{Atwood2009}
{Atwood}, W.~B., {Abdo}, A.~A., {Ackermann}, M., {et~al.} 2009, \apj, 697, 1071

\bibitem[{{Aschwanden}(2004)}]{Aschwanden2004}
{Aschwanden}, M.~J. 2004, in ESA Special Publication, Vol. 575, SOHO 15 Coronal Heating, ed. R.~W. {Walsh}, J.~{Ireland}, D.~{Danesy}, \& B.~{Fleck}, 97

\bibitem[{{Aschwanden} \& {Alexander}(2001)}]{Aschwanden2001}
{Aschwanden}, M.~J., \& {Alexander}, D. 2001, Solar~Phys., 204, 91

\bibitem[{{Aschwanden} {et~al.}(2017){Aschwanden}, {Caspi}, {Cohen}, {Holman},
  {Jing}, {Kretzschmar}, {Kontar}, {McTiernan}, {Mewaldt}, {O'Flannagain},
  {Richardson}, {Ryan}, {Warren}, \& {Xu}}]{Aschwanden2017}
{Aschwanden}, M.~J., {Caspi}, A., {Cohen}, C. M.~S., {et~al.} 2017, \apj, 836,
  17

\bibitem[{{Aschwanden} {et~al.}(2019){Aschwanden}, {Kontar}, \& {Jeffrey}}]{Aschwanden2019}
{Aschwanden}, M.~J., {Kontar}, E.~P., \& {Jeffrey}, N. L.~S. 2019, \bibinfo{title}{{Global Energetics of Solar Flares. VIII. The Low-energy Cutoff},} \apj, 881, 1, DOI:10.3847/1538-4357/ab2cd4


\bibitem[{{Avallone} \& {Sun}(2020)}]{avallone2020}
{Avallone}, E.~A., \& {Sun}, X. 2020, \apj, 893, 123

\bibitem[{{Bein} {et~al.}(2013){Bein}, {Temmer}, {Vourlidas}, {Veronig}, \&
  {Utz}}]{Bein2013}
{Bein}, B.~M., {Temmer}, M., {Vourlidas}, A., {Veronig}, A.~M., \& {Utz}, D.
  2013, \apj, 768, 31

\bibitem[{{Benkhoff} {et~al.}(2021){Benkhoff}, {Murakami}, {Baumjohann},
  {Besse}, {Bunce}, {Casale}, {Cremonese}, {Glassmeier}, {Hayakawa}, {Heyner},
  {Hiesinger}, {Huovelin}, {Hussmann}, {Iafolla}, {Iess}, {Kasaba},
  {Kobayashi}, {Milillo}, {Mitrofanov}, {Montagnon}, {Novara}, {Orsini},
  {Quemerais}, {Reininghaus}, {Saito}, {Santoli}, {Stramaccioni}, {Sutherland},
  {Thomas}, {Yoshikawa}, \& {Zender}}]{2021SSRv..217...90B}
{Benkhoff}, J., {Murakami}, G., {Baumjohann}, W., {et~al.} 2021, \ssr, 217, 90

\bibitem[{{Benz}(2017)}]{Benz2017}
{Benz}, A.~O. 2017, Living Reviews in Solar Physics, 14, 2

\bibitem[{{Bobra} {et~al.}(2014){Bobra}, {Sun}, {Hoeksema}, {Turmon}, {Liu},
  {Hayashi}, {Barnes}, \& {Leka}}]{bobra2014}
{Bobra}, M.~G., {Sun}, X., {Hoeksema}, J.~T., {et~al.} 2014, \solphys, 289,
  3549

\bibitem[{{Brueckner} {et~al.}(1995){Brueckner}, {Howard}, {Koomen},
  {Korendyke}, {Michels}, {Moses}, {Socker}, {Dere}, {Lamy}, {Llebaria},
  {Bout}, {Schwenn}, {Simnett}, {Bedford}, \& {Eyles}}]{Brueckner1995}
{Brueckner}, G.~E., {Howard}, R.~A., {Koomen}, M.~J., {et~al.} 1995, \solphys,
  162, 357

\bibitem[{{Cargill} {et~al.}(1995){Cargill}, {Mariska}, \&
  {Antiochos}}]{Cargill1995}
{Cargill}, P.~J., {Mariska}, J.~T., \& {Antiochos}, S.~K. 1995, ApJ, 439, 1034

\bibitem[{{Chen} {et~al.}(2020){Chen}, {Shen}, {Gary}, {Reeves}, {Fleishman},
  {Yu}, {Guo}, {Krucker}, {Lin}, {Nita}, \& {Kong}}]{chen2020a}
{Chen}, B., {Shen}, C., {Gary}, D.~E., {et~al.} 2020, Nature Astronomy, 4, 1140

\bibitem[{{Colaninno} \& {Vourlidas}(2009)}]{Colaninno2009}
{Colaninno}, R.~C., \& {Vourlidas}, A. 2009, \apj, 698, 852

\bibitem[{{Darnel} {et~al.}(2022){Darnel}, {Seaton}, {Bethge}, {Rachmeler},
  {Jarvis}, {Hill}, {Peck}, {Hughes}, {Shapiro}, {Riley}, {Vasudevan}, {Shing},
  {Koener}, {Edwards}, {Mathur}, \& {Timothy}}]{Darnel2022}
{Darnel}, J.~M., {Seaton}, D.~B., {Bethge}, C., {et~al.} 2022, Space Weather,
  20, e2022SW003044

\bibitem[{David {et~al.}(2022)David, Fraschetti, Giacalone,
  Wimmer-Schweingruber, Berger, \& Lario}]{David2022}
David, L., Fraschetti, F., Giacalone, J., {et~al.} 2022, \apj, 928, 66

\bibitem[{{Dimmock} {et~al.}(2023){Dimmock}, {Gedalin, M.}, {Lalti, A.},
  {Trotta, D.}, {Khotyaintsev, Yu. V.}, {Graham, D. B.}, {Johlander, A.},
  {Vainio, R.}, {Blanco-Cano, X.}, {Kajdič, P.}, {Owen, C. J.}, \&
  {Wimmer-Schweingruber, R. F.}}]{Dimmock2023}
{Dimmock}, A.~P., {Gedalin, M.}, {Lalti, A.}, {et~al.} 2023, A\&A, 679, A106

\bibitem[{{Domingo} {et~al.}(1995){Domingo}, {Fleck}, \&
  {Poland}}]{1995SSRv...72...81D}
{Domingo}, V., {Fleck}, B., \& {Poland}, A.~I. 1995, SSR, 72, 81

\bibitem[{{Emslie} {et~al.}(2005){Emslie}, {Dennis}, {Holman}, \&
  {Hudson}}]{Emslie05}
{Emslie}, A.~G., {Dennis}, B.~R., {Holman}, G.~D., \& {Hudson}, H.~S. 2005,
  Journal of Geophysical Research (Space Physics), 110, A11103

\bibitem[{{Emslie} {et~al.}(2004){Emslie}, {Kucharek}, {Dennis}, {Gopalswamy},
  {Holman}, {Share}, {Vourlidas}, {Forbes}, {Gallagher}, {Mason}, {Metcalf},
  {Mewaldt}, {Murphy}, {Schwartz}, \& {Zurbuchen}}]{Emslie04}
{Emslie}, A.~G., {Kucharek}, H., {Dennis}, B.~R., {et~al.} 2004, Journal of
  Geophysical Research (Space Physics), 109, A10104

\bibitem[{{Emslie} {et~al.}(2012){Emslie}, {Dennis}, {Shih}, {Chamberlin},
  {Mewaldt}, {Moore}, {Share}, {Vourlidas}, \& {Welsch}}]{emslie12}
{Emslie}, A.~G., {Dennis}, B.~R., {Shih}, A.~Y., {et~al.} 2012, \apj, 759, 71

\bibitem[{{Fagundes} {et~al.}(2024){Fagundes}, {Pillat}, {Tardelli}, \&
  {Muella}}]{Fagundes2024}
{Fagundes}, P.~R., {Pillat}, V.~G., {Tardelli}, A., \& {Muella}, M.~T.~A.~H.
  2024, Journal of Geophysical Research (Space Physics), 129, e2024JA032597

\bibitem[{{Fleishman} \& {Kuznetsov}(2010)}]{fleishman2010}
{Fleishman}, G.~D., \& {Kuznetsov}, A.~A. 2010, \apj, 721, 1127

\bibitem[{Forbes \& Priest(1984)}]{Forbes1984}
Forbes, T.~G., \& Priest, E.~R. 1984, in Solar Terrestrial Physics: Present and
  Future, ed. D.~Butler \& K.~Papadopoulos (NASA), 35--39

\bibitem[{{Fox} {et~al.}(2016){Fox}, {Velli}, {Bale}, {Decker}, {Driesman},
  {Howard}, {Kasper}, {Kinnison}, {Kusterer}, {Lario}, {Lockwood}, {McComas},
  {Raouafi}, \& {Szabo}}]{2016SSRv..204....7F}
{Fox}, N.~J., {Velli}, M.~C., {Bale}, S.~D., {et~al.} 2016, SSR, 204, 7

\bibitem[{{Garcia}(1994)}]{Garcia1994}
{Garcia}, H.~A. 1994, Solar~Phys., 154, 275

\bibitem[{{Gary} {et~al.}(2018){Gary}, {Chen}, {Dennis}, {Fleishman},
  {Hurford}, {Krucker}, {McTiernan}, {Nita}, {Shih}, {White}, \&
  {Yu}}]{Gary2018eovsa}
{Gary}, D.~E., {Chen}, B., {Dennis}, B.~R., {et~al.} 2018, \apj, 863, 83

\bibitem[{{Giersch} \& {Kennewell}(2022)}]{Giersch2022}
{Giersch}, O., \& {Kennewell}, J. 2022, Radio Science, 57, e2022RS007456

\bibitem[{{Golub} {et~al.}(2007){Golub}, {DeLuca}, {Austin}, {Bookbinder},
  {Caldwell}, {Cheimets}, {Cirtain}, {Cosmo}, {Reid}, {Sette}, {Weber},
  {Sakao}, {Kano}, {Shibasaki}, {Hara}, {Tsuneta}, {Kumagai}, {Tamura},
  {Shimojo}, {McCracken}, {Carpenter}, {Haight}, {Siler}, {Wright}, {Tucker},
  {Rutledge}, {Barbera}, {Peres}, \& {Varisco}}]{Golub2007}
{Golub}, L., {DeLuca}, E., {Austin}, G., {et~al.} 2007, \solphys, 243, 63

\bibitem[{{Habarulema} {et~al.}(2022){Habarulema}, {Tshisaphungo},
  {Katamzi-Joseph}, {Matamba}, \& {Nndanganeni}}]{Habarulema2022}
{Habarulema}, J.~B., {Tshisaphungo}, M., {Katamzi-Joseph}, Z.~T., {Matamba},
  T.~M., \& {Nndanganeni}, R. 2022, Space Weather, 20, e2022SW003104

\bibitem[{{Hannah} \& {Kontar}(2012)}]{hannah&kontar12}
{Hannah}, I.~G., \& {Kontar}, E.~P. 2012, \aap, 539, A146

\bibitem[{{Hoeksema} {et~al.}(2020){Hoeksema}, {Abbett}, {Bercik}, {Cheung},
  {DeRosa}, {Fisher}, {Hayashi}, {Kazachenko}, {Liu}, {Lumme}, {Lynch}, {Sun},
  \& {Welsch}}]{hoeksema2020}
{Hoeksema}, J.~T., {Abbett}, W.~P., {Bercik}, D.~J., {et~al.} 2020, \apjs, 250,
  28

\bibitem[{{Horbury} {et~al.}(2020){Horbury}, {O'Brien}, {Carrasco Blazquez},
  {Bendyk}, {Brown}, {Hudson}, {Evans}, {Oddy}, {Carr}, {Beek}, {Cupido},
  {Bhattacharya}, {Dominguez}, {Matthews}, {Myklebust}, {Whiteside}, {Bale},
  {Baumjohann}, {Burgess}, {Carbone}, {Cargill}, {Eastwood}, {Erd{\"o}s},
  {Fletcher}, {Forsyth}, {Giacalone}, {Glassmeier}, {Goldstein}, {Hoeksema},
  {Lockwood}, {Magnes}, {Maksimovic}, {Marsch}, {Matthaeus}, {Murphy},
  {Nakariakov}, {Owen}, {Owens}, {Rodriguez-Pacheco}, {Richter}, {Riley},
  {Russell}, {Schwartz}, {Vainio}, {Velli}, {Vennerstrom}, {Walsh},
  {Wimmer-Schweingruber}, {Zank}, {M{\"u}ller}, {Zouganelis}, \&
  {Walsh}}]{Horbury.etal:20}
{Horbury}, T.~S., {O'Brien}, H., {Carrasco Blazquez}, I., {et~al.} 2020, \aap,
  642, A9

\bibitem[{{Howard} {et~al.}(2008){Howard}, {Moses}, {Vourlidas}, {Newmark},
  {Socker}, {Plunkett}, {Korendyke}, {Cook}, {Hurley}, {Davila}, {Thompson},
  {St Cyr}, {Mentzell}, {Mehalick}, {Lemen}, {Wuelser}, {Duncan}, {Tarbell},
  {Wolfson}, {Moore}, {Harrison}, {Waltham}, {Lang}, {Davis}, {Eyles},
  {Mapson-Menard}, {Simnett}, {Halain}, {Defise}, {Mazy}, {Rochus}, {Mercier},
  {Ravet}, {Delmotte}, {Auchere}, {Delaboudiniere}, {Bothmer}, {Deutsch},
  {Wang}, {Rich}, {Cooper}, {Stephens}, {Maahs}, {Baugh}, {McMullin}, \&
  {Carter}}]{2008SSRv..136...67H}
{Howard}, R.~A., {Moses}, J.~D., {Vourlidas}, A., {et~al.} 2008, SSR, 136, 67

\bibitem[{{Inhester}(2006)}]{Inhestero2006}
{Inhester}, B. 2006, arXiv e-prints, astro

\bibitem[{{Jakimiec} {et~al.}(1992){Jakimiec}, {Sylwester}, {Sylwester},
  {Serio}, {Peres}, \& {Reale}}]{Jakimiec1992}
{Jakimiec}, J., {Sylwester}, B., {Sylwester}, J., {et~al.} 1992, A\&A, 253, 269

\bibitem[{Jing {et~al.}(2005)Jing, Qiu, Lin, Qu, Xu, \& Wang}]{Jing2005}
Jing, J., Qiu, J., Lin, J., {et~al.} 2005, ApJ, 620, 1085

\bibitem[{{Kaiser}(2005)}]{2005AdSpR..36.1483K}
{Kaiser}, M.~L. 2005, Advances in Space Research, 36, 1483

\bibitem[{{Kazachenko} {et~al.}(2014){Kazachenko}, {Fisher}, \&
  {Welsch}}]{kazachenko2014}
{Kazachenko}, M.~D., {Fisher}, G.~H., \& {Welsch}, B.~T. 2014, \apj, 795, 17

\bibitem[{{Klein} {et~al.}(2022){Klein}, {Musset}, {Vilmer}, {Briand},
  {Krucker}, {Francesco Battaglia}, {Dresing}, {Palmroos}, \&
  {Gary}}]{Klein2022}
{Klein}, K.-L., {Musset}, S., {Vilmer}, N., {et~al.} 2022, \aap, 663, A173

\bibitem[{{Klimchuk} {et~al.}(2008){Klimchuk}, {Patsourakos}, \&
  {Cargill}}]{Klimchuk2008}
{Klimchuk}, J.~A., {Patsourakos}, S., \& {Cargill}, P.~J. 2008, ApJ, 682, 1351

\bibitem[{Koval \& Szabo(2023)Koval \& Szabo}]{Koval2023}
Koval, A., \& Szabo, A. 2023, Wind Magnetic Field Investigation (MFI) Full Resolution Data in RTN Coordinates, NASA Space Physics Data Facility, DOI:10.48322/S1KF-0B54

\bibitem[{{Kretzschmar}(2011)}]{Kretzschmar}
{Kretzschmar}, M. 2011, \aap, 530, A84

\bibitem[{Krimchansky {et~al.}(2004)Krimchansky, Machi, Cauffman, \&
  Davis}]{Krimchansky2004}
Krimchansky, A., Machi, D., Cauffman, S.~A., \& Davis, M.~A. 2004, in Sensors,
  Systems, and Next-Generation Satellites VIII, ed. R.~Meynart, S.~P. Neeck, \&
  H.~Shimoda, Vol. 5570, International Society for Optics and Photonics (SPIE),
  155 -- 164

\bibitem[{{Krucker} {et~al.}(2020){Krucker}, {Hurford}, {Grimm}, {K{\"o}gl},
  {Gr{\"o}belbauer}, {Etesi}, {Casadei}, {Csillaghy}, {Benz}, {Arnold},
  {Molendini}, {Orleanski}, {Schori}, {Xiao}, {Kuhar}, {Hochmuth}, {Felix},
  {Schramka}, {Marcin}, {Kobler}, {Iseli}, {Dreier}, {Wiehl}, {Kleint},
  {Battaglia}, {Lastufka}, {Sathiapal}, {Lapadula}, {Bednarzik}, {Birrer},
  {Stutz}, {Wild}, {Marone}, {Skup}, {Cichocki}, {Ber}, {Rutkowski}, {Bujwan},
  {Juchnikowski}, {Winkler}, {Darmetko}, {Michalska}, {Seweryn}, {Bia{\l}ek},
  {Osica}, {Sylwester}, {Kowalinski}, {{\'S}cis{\l}owski}, {Siarkowski},
  {St{\k{e}}{\'s}licki}, {Mrozek}, {Podg{\'o}rski}, {Meuris}, {Limousin},
  {Gevin}, {Le Mer}, {Brun}, {Strugarek}, {Vilmer}, {Musset}, {Maksimovi{\'c}},
  {F{\'a}rn{\'\i}k}, {Koz{\'a}{\v{c}}ek}, {Ka{\v{s}}parov{\'a}}, {Mann},
  {{\"O}nel}, {Warmuth}, {Rendtel}, {Anderson}, {Bauer}, {Dionies}, {Paschke},
  {Pl{\"u}schke}, {Woche}, {Schuller}, {Veronig}, {Dickson}, {Gallagher},
  {Maloney}, {Bloomfield}, {Piana}, {Massone}, {Benvenuto}, {Massa},
  {Schwartz}, {Dennis}, {van Beek}, {Rodr{\'\i}guez-Pacheco}, \&
  {Lin}}]{2020A&A...642A..15K}
{Krucker}, S., {Hurford}, G.~J., {Grimm}, O., {et~al.} 2020, \aap, 642, A15

\bibitem[{{Kuznetsov} \& {Fleishman}(2021)}]{kuznetsov2021}
{Kuznetsov}, A.~A., \& {Fleishman}, G.~D. 2021, \apj, 922, 103

\bibitem[{{Lario} {et~al.}(2003){Lario}, {Ho}, {Decker}, {Roelof}, {Desai}, \&
  {Smith}}]{Lario.etal:03}
{Lario}, D., {Ho}, G.~C., {Decker}, R.~B., {et~al.} 2003, in American Institute
  of Physics Conference Series, Vol. 679, Solar Wind Ten, ed. M.~{Velli},
  R.~{Bruno}, F.~{Malara}, \& B.~{Bucci}, 640--643

\bibitem[{{Lemen} {et~al.}(2012){Lemen}, {Title}, {Akin}, {Boerner}, {Chou},
  {Drake}, {Duncan}, {Edwards}, {Friedlaender}, {Heyman}, {Hurlburt}, {Katz},
  {Kushner}, {Levay}, {Lindgren}, {Mathur}, {McFeaters}, {Mitchell}, {Rehse},
  {Schrijver}, {Springer}, {Stern}, {Tarbell}, {Wuelser}, {Wolfson}, {Yanari},
  {Bookbinder}, {Cheimets}, {Caldwell}, {Deluca}, {Gates}, {Golub}, {Park},
  {Podgorski}, {Bush}, {Scherrer}, {Gummin}, {Smith}, {Auker}, {Jerram},
  {Pool}, {Soufli}, {Windt}, {Beardsley}, {Clapp}, {Lang}, \&
  {Waltham}}]{lemenet12}
{Lemen}, J.~R., {Title}, A.~M., {Akin}, D.~J., {et~al.} 2012, \solphys, 275, 17

\bibitem[{{Lepping} {et~al.}(1995){Lepping}, {Ac{\~{u}}na}, {Burlaga},
  {Farrell}, {Slavin}, {Schatten}, {Mariani}, {Ness}, {Neubauer}, {Whang},
  {Byrnes}, {Kennon}, {Panetta}, {Scheifele}, \& {Worley}}]{Lepping.etal:95}
{Lepping}, R.~P., {Ac{\~{u}}na}, M.~H., {Burlaga}, L.~F., {et~al.} 1995, \ssr,
  71, 207

\bibitem[{{Lin} {et~al.}(1995){Lin}, {Anderson}, {Ashford}, {Carlson},
  {Curtis}, {Ergun}, {Larson}, {McFadden}, {McCarthy}, {Parks}, {R{\`e}me},
  {Bosqued}, {Coutelier}, {Cotin}, {D'Uston}, {Wenzel}, {Sanderson}, {Henrion},
  {Ronnet}, \& {Paschmann}}]{Lin.etal:95}
{Lin}, R.~P., {Anderson}, K.~A., {Ashford}, S., {et~al.} 1995, \ssr, 71, 125

\bibitem[{{Liu} {et~al.}(2021){Liu}, {Issautier}, {Meyer-Vernet}, {Moncuquet},
  {Maksimovic}, {Halekas}, {Huang}, {Griton}, {Bale}, {Bonnell}, {Case},
  {Goetz}, {Harvey}, {Kasper}, {MacDowall}, {Malaspina}, {Pulupa}, \&
  {Stevens}}]{Liu2021}
{Liu}, M., {Issautier}, K., {Meyer-Vernet}, N., {et~al.} 2021, \aap, 650, A14

\bibitem[{{Liu} {et~al.}(2011){Liu}, {Title}, {Zhao}, {Ofman}, {Schrijver},
  {Aschwanden}, {De Pontieu}, \& {Tarbell}}]{Liu2011}
{Liu}, W., {Title}, A.~M., {Zhao}, J., {et~al.} 2011, \apjl, 736, L13

\bibitem[{Liu {et~al.}(2013)Liu, Qiu, Longcope, \& Caspi}]{Liu2013}
Liu, W.-J., Qiu, J., Longcope, D.~W., \& Caspi, A. 2013, ApJ, 770, 111

\bibitem[{{Liu} {et~al.}(2005){Liu}, {Richardson}, \& {Belcher}}]{Liu2005}
{Liu}, Y., {Richardson}, J.~D., \& {Belcher}, J.~W. 2005, \planss, 53, 3

\bibitem[{{Liu} {et~al.}(2008){Liu}, {Luhmann}, {M{\"u}ller-Mellin},
  {Schroeder}, {Wang}, {Lin}, {Bale}, {Li}, {Acu{\~n}a}, \&
  {Sauvaud}}]{Liu2008}
{Liu}, Y., {Luhmann}, J.~G., {M{\"u}ller-Mellin}, R., {et~al.} 2008, \apj, 689,
  563

\bibitem[{{Long} {et~al.}(2015){Long}, {Baker}, {Williams}, {Carley},
  {Gallagher}, \& {Zucca}}]{Long2015}
{Long}, D.~M., {Baker}, D., {Williams}, D.~R., {et~al.} 2015, \apj, 799, 224

\bibitem[{{Long} {et~al.}(2017){Long}, {Bloomfield}, {Chen}, {Downs},
  {Gallagher}, {Kwon}, {Vanninathan}, {Veronig}, {Vourlidas}, {Vr{\v{s}}nak},
  {Warmuth}, \& {{\v{Z}}ic}}]{Long2017}
{Long}, D.~M., {Bloomfield}, D.~S., {Chen}, P.~F., {et~al.} 2017, \solphys,
  292, 7

\bibitem[{Longcope {et~al.}(2010)Longcope, Des~{J}ardins, Carranza-Fulmer, \&
  Qiu}]{Longcope2010}
Longcope, D.~W., Des~{J}ardins, A.~C., Carranza-Fulmer, T., \& Qiu, J. 2010,
  Solar~Phys., 267, 107

\bibitem[{{Lopez} \& {Freeman}(1986)}]{Lopez1986}
{Lopez}, R.~E., \& {Freeman}, J.~W. 1986, \jgr, 91, 1701

\bibitem[{{Meegan} {et~al.}(2009){Meegan}, {Lichti}, {Bhat}, {Bissaldi},
  {Briggs}, {Connaughton}, {Diehl}, {Fishman}, {Greiner}, {Hoover}, {van der
  Horst}, {von Kienlin}, {Kippen}, {Kouveliotou}, {McBreen}, {Paciesas},
  {Preece}, {Steinle}, {Wallace}, {Wilson}, \& {Wilson-Hodge}}]{Meegan2009}
{Meegan}, C., {Lichti}, G., {Bhat}, P.~N., {et~al.} 2009, \apj, 702, 791


\bibitem[{{Muhr} {et~al.}(2014){Muhr}, {Veronig}, {Kienreich}, {Vr{\v{s}}nak},
  {Temmer}, \& {Bein}}]{Murh2014}
{Muhr}, N., {Veronig}, A.~M., {Kienreich}, I.~W., {et~al.} 2014, \solphys, 289,
  4563

\bibitem[{{M{\"u}ller} {et~al.}(2020){M{\"u}ller}, {St. Cyr}, {Zouganelis},
  {Gilbert}, {Marsden}, {Nieves-Chinchilla}, {Antonucci}, {Auch{\`e}re},
  {Berghmans}, {Horbury}, {Howard}, {Krucker}, {Maksimovic}, {Owen}, {Rochus},
  {Rodriguez-Pacheco}, {Romoli}, {Solanki}, {Bruno}, {Carlsson}, {Fludra},
  {Harra}, {Hassler}, {Livi}, {Louarn}, {Peter}, {Sch{\"u}hle}, {Teriaca}, {del
  Toro Iniesta}, {Wimmer-Schweingruber}, {Marsch}, {Velli}, {De Groof},
  {Walsh}, \& {Williams}}]{Mueller_2020}
{M{\"u}ller}, D., {St. Cyr}, O.~C., {Zouganelis}, I., {et~al.} 2020, AAP, 642,
  A1

\bibitem[{{M{\"u}ller-Mellin} {et~al.}(2008){M{\"u}ller-Mellin},
  {B{\"o}ttcher}, {Falenski}, {Rode}, {Duvet}, {Sanderson}, {Butler},
  {Johlander}, \& {Smit}}]{MuellerMellin.etal:08}
{M{\"u}ller-Mellin}, R., {B{\"o}ttcher}, S., {Falenski}, J., {et~al.} 2008,
  \ssr, 136, 363

\bibitem[{{Ogilvie} \& {Desch}(1997)}]{1997AdSpR..20..559O}
{Ogilvie}, K.~W., \& {Desch}, M.~D. 1997, Advances in Space Research, 20, 559

\bibitem[{{Ogilvie} {et~al.}(1995){Ogilvie}, {Chornay}, {Fritzenreiter},
  {Hunsaker}, {Keller}, {Lobell}, {Miller}, {Scudder}, {Sittler}, {Torbert},
  {Bodet}, {Needell}, {Lazarus}, {Steinberg}, {Tappan}, {Mavretic}, \&
  {Gergin}}]{Ogilvie.etal:95}
{Ogilvie}, K.~W., {Chornay}, D.~J., {Fritzenreiter}, R.~J., {et~al.} 1995,
  \ssr, 71, 55

\bibitem[{Ogilvie {et~al.}(2021)Ogilvie, Fitzenreiter, Lazarus, Kasper, \& Stevens}]{Ogilvie2021}
Ogilvie, K.~W., Fitzenreiter, R.~J., Lazarus, A.~J., Kasper, J.~C., \& Stevens, M. 2021, Wind Solar Wind Experiment (SWE) 92-sec Definitive Solar Wind Proton Data, NASA Space Physics Data Facility, DOI:10.48322/NASD-J276


\bibitem[{{Owen} {et~al.}(2020){Owen}, {Bruno}, {Livi}, {Louarn}, {Al Janabi},
  {Allegrini}, {Amoros}, {Baruah}, {Barthe}, {Berthomier}, {Bordon},
  {Brockley-Blatt}, {Brysbaert}, {Capuano}, {Collier}, {DeMarco}, {Fedorov},
  {Ford}, {Fortunato}, {Fratter}, {Galvin}, {Hancock}, {Heirtzler}, {Kataria},
  {Kistler}, {Lepri}, {Lewis}, {Loeffler}, {Marty}, {Mathon}, {Mayall}, {Mele},
  {Ogasawara}, {Orlandi}, {Pacros}, {Penou}, {Persyn}, {Petiot}, {Phillips},
  {P{\v{r}}ech}, {Raines}, {Reden}, {Rouillard}, {Rousseau}, {Rubiella},
  {Seran}, {Spencer}, {Thomas}, {Trevino}, {Verscharen}, {Wurz}, {Alapide},
  {Amoruso}, {Andr{\'e}}, {Anekallu}, {Arciuli}, {Arnett}, {Ascolese},
  {Bancroft}, {Bland}, {Brysch}, {Calvanese}, {Castronuovo},
  {{\v{C}}erm{\'a}k}, {Chornay}, {Clemens}, {Coker}, {Collinson}, {D'Amicis},
  {Dandouras}, {Darnley}, {Davies}, {Davison}, {De Los Santos}, {Devoto},
  {Dirks}, {Edlund}, {Fazakerley}, {Ferris}, {Frost}, {Fruit}, {Garat},
  {G{\'e}not}, {Gibson}, {Gilbert}, {de Giosa}, {Gradone}, {Hailey}, {Horbury},
  {Hunt}, {Jacquey}, {Johnson}, {Lavraud}, {Lawrenson}, {Leblanc}, {Lockhart},
  {Maksimovic}, {Malpus}, {Marcucci}, {Mazelle}, {Monti}, {Myers}, {Nguyen},
  {Rodriguez-Pacheco}, {Phillips}, {Popecki}, {Rees}, {Rogacki}, {Ruane},
  {Rust}, {Salatti}, {Sauvaud}, {Stakhiv}, {Stange}, {Stubbs}, {Taylor},
  {Techer}, {Terrier}, {Thibodeaux}, {Urdiales}, {Varsani}, {Walsh}, {Watson},
  {Wheeler}, {Willis}, {Wimmer-Schweingruber}, {Winter}, {Yardley}, \&
  {Zouganelis}}]{Owen.etal:20}
{Owen}, C.~J., {Bruno}, R., {Livi}, S., {et~al.} 2020, \aap, 642, A16

\bibitem[{{Papaioannou} {et~al.}(2022){Papaioannou}, {Kouloumvakos}, {Mishev},
  {Vainio}, {Usoskin}, {Herbst}, {Rouillard}, {Anastasiadis}, {Gieseler},
  {Wimmer-Schweingruber}, \& {K{\"u}hl}}]{Papaioannou2022}
{Papaioannou}, A., {Kouloumvakos}, A., {Mishev}, A., {et~al.} 2022, \aap, 660,
  L5

\bibitem[{{Pascoe} {et~al.}(2019){Pascoe}, {Smyrli}, \& {Van
  Doorsselaere}}]{Pascoe2019}
{Pascoe}, D.~J., {Smyrli}, A., \& {Van Doorsselaere}, T. 2019, \apj, 884, 43

\bibitem[{{Patsourakos} \& {Vourlidas}(2012)}]{Patsourakos2012}
{Patsourakos}, S., \& {Vourlidas}, A. 2012, \solphys, 281, 187

\bibitem[{{Pesnell} {et~al.}(2012){Pesnell}, {Thompson}, \&
  {Chamberlin}}]{2012SoPh..275....3P}
{Pesnell}, W.~D., {Thompson}, B.~J., \& {Chamberlin}, P.~C. 2012, Sol. Phys.,
  275, 3

\bibitem[{{Plowman} \& {Caspi}(2020)}]{Plowman2020}
{Plowman}, J., \& {Caspi}, A. 2020, \apj, 905, 17

\bibitem[{Poletto \& Kopp(1986)}]{Poletto1986}
Poletto, G., \& Kopp, R.~A. 1986, in The Lower Atmospheres of Solar Flares, ed.
  D.~F. Neidig (National Solar Observatory), 453--465

\bibitem[{{Qiu}(2021)}]{Qiu2021}
{Qiu}, J. 2021, \apj, 909, 99

\bibitem[{Qiu {et~al.}(2012)Qiu, Liu, \& Longcope}]{Qiu2012}
Qiu, J., Liu, W.-J., \& Longcope, D.~W. 2012, ApJ, 752, 124

\bibitem[{Qiu \& Longcope(2016)}]{Qiu2016}
Qiu, J., \& Longcope, D.~W. 2016, ApJ, 820, 14

\bibitem[{{Reeves} \& {Forbes}(2005)}]{Reeves2005}
{Reeves}, K.~K., \& {Forbes}, T.~G. 2005, \apj, 630, 1133

\bibitem[{{Reeves} {et~al.}(2017){Reeves}, {Freed}, {McKenzie}, \&
  {Savage}}]{Reeves2017}
{Reeves}, K.~K., {Freed}, M.~S., {McKenzie}, D.~E., \& {Savage}, S.~L. 2017,
  \apj, 836, 55

\bibitem[{{Reeves} {et~al.}(2019){Reeves}, {T{\"o}r{\"o}k}, {Miki{\'c}},
  {Linker}, \& {Murphy}}]{Reeves2019}
{Reeves}, K.~K., {T{\"o}r{\"o}k}, T., {Miki{\'c}}, Z., {Linker}, J., \&
  {Murphy}, N.~A. 2019, \apj, 887, 103

\bibitem[{{Reeves} \& {Warren}(2002)}]{Reeves2002}
{Reeves}, K.~K., \& {Warren}, H.~P. 2002, \apj, 578, 590

\bibitem[{{Richardson} \& {Cane}(2010)}]{Richardson2010}
{Richardson}, I.~G., \& {Cane}, H.~V. 2010, \solphys, 264, 189

\bibitem[{{Rivera} {et~al.}(2019){Rivera}, {Landi}, {Lepri}, \&
  {Gilbert}}]{Rivera2019}
{Rivera}, Y.~J., {Landi}, E., {Lepri}, S.~T., \& {Gilbert}, J.~A. 2019, \apj,
  874, 164

\bibitem[{{Rivera} {et~al.}(2023){Rivera}, {Raymond}, {Reeves}, {Lepri},
  {Lionello}, {Downs}, {Wilson}, \& {Trueba}}]{Rivera2023}
{Rivera}, Y.~J., {Raymond}, J.~C., {Reeves}, K.~K., {et~al.} 2023, \apj, 955,
  65

\bibitem[{{Rivera} {et~al.}(2024){Rivera}, {Badman}, {Stevens}, {Verniero},
  {Stawarz}, {Shi}, {Raines}, {Paulson}, {Owen}, {Niembro}, {Louarn}, {Livi},
  {Lepri}, {Kasper}, {Horbury}, {Halekas}, {Dewey}, {De Marco}, \&
  {Bale}}]{Rivera2024}
{Rivera}, Y.~J., {Badman}, S.~T., {Stevens}, M.~L., {et~al.} 2024, Science,
  385, 962

\bibitem[{{Rivera} {et~al.}(2025){Rivera}, {Badman}, {Verniero}, {Varesano},
  {Stevens}, {Stawarz}, {Reeves}, {Raines}, {Raymond}, {Owen}, {Livi}, {Lepri},
  {Landi}, {Halekas}, {Ervin}, {Dewey}, {De Marco}, {D'Amicis}, {Dakeyo},
  {Bale}, \& {Alterman}}]{Rivera2025}
{Rivera}, Y.~J., {Badman}, S.~T., {Verniero}, J.~L., {et~al.} 2025, \apj, 980,
  70

\bibitem[{{Rodr{\'\i}guez-Pacheco} {et~al.}(2020){Rodr{\'\i}guez-Pacheco},
  {Wimmer-Schweingruber}, {Mason}, {Ho}, {S{\'a}nchez-Prieto}, {Prieto},
  {Mart{\'\i}n}, {Seifert}, {Andrews}, {Kulkarni}, {Panitzsch}, {Boden},
  {B{\"o}ttcher}, {Cernuda}, {Elftmann}, {Espinosa Lara}, {G{\'o}mez-Herrero},
  {Terasa}, {Almena}, {Begley}, {B{\"o}hm}, {Blanco}, {Boogaerts}, {Carrasco},
  {Castillo}, {da Silva Fari{\~n}a}, {de Manuel Gonz{\'a}lez}, {Drews},
  {Dupont}, {Eldrum}, {Gordillo}, {Guti{\'e}rrez}, {Haggerty}, {Hayes},
  {Heber}, {Hill}, {J{\"u}ngling}, {Kerem}, {Knierim}, {K{\"o}hler}, {Kolbe},
  {Kulemzin}, {Lario}, {Lees}, {Liang}, {Mart{\'\i}nez Hell{\'\i}n}, {Meziat},
  {Montalvo}, {Nelson}, {Parra}, {Paspirgilis}, {Ravanbakhsh}, {Richards},
  {Rodr{\'\i}guez-Polo}, {Russu}, {S{\'a}nchez}, {Schlemm}, {Schuster},
  {Seimetz}, {Steinhagen}, {Tammen}, {Tyagi}, {Varela}, {Yedla}, {Yu},
  {Agueda}, {Aran}, {Horbury}, {Klecker}, {Klein}, {Kontar}, {Krucker},
  {Maksimovic}, {Maland raki}, {Owen}, {Pacheco}, {Sanahuja}, {Vainio},
  {Connell}, {Dalla}, {Dr{\"o}ge}, {Gevin}, {Gopalswamy}, {Kartavykh},
  {Kudela}, {Limousin}, {Makela}, {Mann}, {{\"O}nel}, {Posner}, {Ryan},
  {Soucek}, {Hofmeister}, {Vilmer}, {Walsh}, {Wang}, {Wiedenbeck}, {Wirth}, \&
  {Zong}}]{RodriguezPacheco2020}
{Rodr{\'\i}guez-Pacheco}, J., {Wimmer-Schweingruber}, R.~F., {Mason}, G.~M.,
  {et~al.} 2020, \aap, 642, A7

\bibitem[{Rosner {et~al.}(1978)Rosner, Tucker, \& Vaiana}]{Rosner1978}
Rosner, R., Tucker, W.~H., \& Vaiana, G.~S. 1978, ApJ, 220, 643

\bibitem[{{Roy} {et~al.}(in review){Roy}, {Musset}, {Reeves}, {Tripathi}, \&
  {Moore}}]{roy_thermal}
{Roy}, S., {Musset}, S., {Reeves}, K., {Tripathi}, D., \& {Moore}, C. in
  review, \apj, under review in ApJ

\bibitem[{{Scherrer} {et~al.}(2012){Scherrer}, {Schou}, {Bush}, {Kosovichev},
  {Bogart}, {Hoeksema}, {Liu}, {Duvall}, {Zhao}, {Title}, {Schrijver},
  {Tarbell}, \& {Tomczyk}}]{scherrer2012}
{Scherrer}, P.~H., {Schou}, J., {Bush}, R.~I., {et~al.} 2012, \solphys, 275,
  207

\bibitem[{{Seaton} {et~al.}(2023){Seaton}, {Berghmans}, {Bloomfield}, {De
  Groof}, {D'Huys}, {Nicula}, {Rachmeler}, \& {West}}]{Seaton2023}
{Seaton}, D.~B., {Berghmans}, D., {Bloomfield}, D.~S., {et~al.} 2023, \solphys,
  298, 92

\bibitem[{{Serio} {et~al.}(1991){Serio}, {Reale}, {Jakimiec}, {Sylwester}, \&
  {Sylwester}}]{Serio1991}
{Serio}, S., {Reale}, F., {Jakimiec}, J., {Sylwester}, B., \& {Sylwester}, J.
  1991, A\&A, 241, 197

\bibitem[{{Shih} {et~al.}(2009){Shih}, {Lin}, \& {Smith}}]{Shih2009}
{Shih}, A.~Y., {Lin}, R.~P., \& {Smith}, D.~M. 2009, \apjl, 698, L152

\bibitem[{{Stone} {et~al.}(1998){Stone}, {Frandsen}, {Mewaldt}, {Christian},
  {Margolies}, {Ormes}, \& {Snow}}]{1998SSRv...86....1S}
{Stone}, E.~C., {Frandsen}, A.~M., {Mewaldt}, R.~A., {et~al.} 1998, SSR, 86, 1

\bibitem[{{Sun} {et~al.}(2012){Sun}, {Hoeksema}, {Liu}, {Wiegelmann},
  {Hayashi}, {Chen}, \& {Thalmann}}]{sun2012}
{Sun}, X., {Hoeksema}, J.~T., {Liu}, Y., {et~al.} 2012, \apj, 748, 77

\bibitem[{Szabo(1994)}]{Szabo:94}
Szabo, A. 1994, Journal of Geophysical Research: Space Physics, 99, 14737

\bibitem[{{Thalmann} {et~al.}(2020){Thalmann}, {Sun}, {Moraitis}, \&
  {Gupta}}]{Thalmann2020}
{Thalmann}, J.~K., {Sun}, X., {Moraitis}, K., \& {Gupta}, M. 2020, \aap, 643,
  A153

\bibitem[{{Thomas} {et~al.}(1985){Thomas}, {Crannell}, \& {Starr}}]{Thomas1985}
{Thomas}, R.~J., {Crannell}, C.~J., \& {Starr}, R. 1985, Solar~Phys., 95, 323

\bibitem[{Trotta {et~al.}(2023)Trotta, Horbury, Lario, Vainio, Dresing,
  Dimmock, Giacalone, Hietala, Wimmer-Schweingruber, Berger, \&
  Yang}]{Trotta2023c}
Trotta, D., Horbury, T.~S., Lario, D., {et~al.} 2023, The Astrophysical Journal
  Letters, 957, L13

\bibitem[{{Valori} {et~al.}(2013){Valori}, {D{\'e}moulin}, {Pariat}, \&
  {Masson}}]{Valori2013}
{Valori}, G., {D{\'e}moulin}, P., {Pariat}, E., \& {Masson}, S. 2013, \aap,
  553, A38

\bibitem[{{{\v{C}}alogovi{\'c}} {et~al.}(2021){{\v{C}}alogovi{\'c}},
  {Dumbovi{\'c}}, {Sudar}, {Vr{\v{s}}nak}, {Martini{\'c}}, {Temmer}, \&
  {Veronig}}]{Calogovic2021}
{{\v{C}}alogovi{\'c}}, J., {Dumbovi{\'c}}, M., {Sudar}, D., {et~al.} 2021,
  \solphys, 296, 114

\bibitem[{Vi{\~n}as \& Scudder(1986)}]{Vinas.Scudder:86}
Vi{\~n}as, A.~F., \& Scudder, J.~D. 1986, Journal of Geophysical Research:
  Space Physics, 91, 39

\bibitem[{{Vourlidas} {et~al.}(2010){Vourlidas}, {Howard}, {Esfandiari},
  {Patsourakos}, {Yashiro}, \& {Michalek}}]{Vourlidas2010}
{Vourlidas}, A., {Howard}, R.~A., {Esfandiari}, E., {et~al.} 2010, \apj, 722,
  1522

\bibitem[{{Vr{\v{s}}nak} {et~al.}(2013){Vr{\v{s}}nak}, {{\v{Z}}ic}, {Vrbanec},
  {Temmer}, {Rollett}, {M{\"o}stl}, {Veronig}, {{\v{C}}alogovi{\'c}},
  {Dumbovi{\'c}}, {Luli{\'c}}, {Moon}, \& {Shanmugaraju}}]{Vrsnak2013}
{Vr{\v{s}}nak}, B., {{\v{Z}}ic}, T., {Vrbanec}, D., {et~al.} 2013, \solphys,
  285, 295

\bibitem[{{Walker} {et~al.}(2024){Walker}, {Allen}, {Ho}, {Mason}, {Li},
  {Cohen}, \& {Lee}}]{Walker.etal:24}
{Walker}, M., {Allen}, R.~C., {Ho}, G.~C., {et~al.} 2024, in AGU Fall Meeting
  Abstracts, Vol. 2024, SH43A--2871

\bibitem[{{Walker} {et~al.}(2025){Walker}, {Allen}, {Li}, {Ho}, {Mason},
  {Rodriguez-Pacheco}, {Wimmer-Schweingruber}, \&
  {Kouloumvakos}}]{Walker.etal:25}
{Walker}, M.~H., {Allen}, R.~C., {Li}, G., {et~al.} 2025, \aap, 693, A230

\bibitem[{{Wang} {et~al.}(2001){Wang}, {Richardson}, \& {Burlaga}}]{Wang2001}
{Wang}, C., {Richardson}, J.~D., \& {Burlaga}, L. 2001, \solphys, 204, 413

\bibitem[{{Warmuth} \& {Mann}(2020)}]{Warmuth2020}
{Warmuth}, A., \& {Mann}, G. 2020, \aap, 644, A172

\bibitem[{{Warren} {et~al.}(2018){Warren}, {Brooks}, {Ugarte-Urra}, {Reep},
  {Crump}, \& {Doschek}}]{Warren2018}
{Warren}, H.~P., {Brooks}, D.~H., {Ugarte-Urra}, I., {et~al.} 2018, \apj, 854,
  122

\bibitem[{{Welsch} \& {Fisher}(2016)}]{welsch2016}
{Welsch}, B.~T., \& {Fisher}, G.~H. 2016, \solphys, 291, 1681

\bibitem[{{Wiegelmann} {et~al.}(2012){Wiegelmann}, {Thalmann}, {Inhester},
  {Tadesse}, {Sun}, \& {Hoeksema}}]{wiegelmann2012}
{Wiegelmann}, T., {Thalmann}, J.~K., {Inhester}, B., {et~al.} 2012, \solphys,
  281, 37

\bibitem[{{Wimmer-Schweingruber} {et~al.}(2021){Wimmer-Schweingruber},
  {Janitzek}, {Pacheco}, {Cernuda}, {Espinosa Lara}, {G{\'o}mez-Herrero},
  {Mason}, {Allen}, {Xu}, {Carcaboso}, {Kollhoff}, {K{\"u}hl}, {Freiherr von
  Forstner}, {Berger}, {Rodriguez-Pacheco}, {Ho}, {Andrews}, {Angelini},
  {Aran}, {Boden}, {B{\"o}ttcher}, {Carrasco}, {Dresing}, {Eldrum}, {Elftmann},
  {Evans}, {Gevin}, {Hayes}, {Heber}, {Horbury}, {Kulkarni}, {Lario}, {Lees},
  {Limousin}, {Malandraki}, {Mart{\'\i}n}, {O'Brien}, {Prieto Mateo},
  {Ravanbakhsh}, {Rodriguez-Polo}, {S{\'a}nchez Prieto}, {Schlemm}, {Seifert},
  {Terasa}, {Tyagi}, {Vainio}, {Walsh}, \& {Yedla}}]{Wimmer2021}
{Wimmer-Schweingruber}, R.~F., {Janitzek}, N.~P., {Pacheco}, D., {et~al.} 2021,
  \aap, 656, A22

\bibitem[{{Woods} {et~al.}(2024){Woods}, {Eden}, {Eparvier}, {Jones},
  {Woodraska}, {Chamberlin}, \& {Machol}}]{woods24}
{Woods}, T.~N., {Eden}, T., {Eparvier}, F.~G., {et~al.} 2024, Journal of
  Geophysical Research (Space Physics), 129, 2024JA032925

\bibitem[{{Zhang} {et~al.}(2004){Zhang}, {Dere}, {Howard}, \&
  {Vourlidas}}]{Zhang2004}
{Zhang}, J., {Dere}, K.~P., {Howard}, R.~A., \& {Vourlidas}, A. 2004, \apj,
  604, 420

\bibitem[{{Zhu} {et~al.}(2020){Zhu}, {Qiu}, {Liewer}, {Vourlidas}, {Spiegel},
  \& {Hu}}]{Zhu2020}
{Zhu}, C., {Qiu}, J., {Liewer}, P., {et~al.} 2020, ApJ, 893, 141

\bibitem[{Zhu {et~al.}(2018)Zhu, Qiu, \& Longcope}]{Zhu2018}
Zhu, C., Qiu, J., \& Longcope, D.~W. 2018, ApJ, 856, 27

\bibitem[{{Zurbuchen} \& {Richardson}(2006)}]{Zurbuchen2006}
{Zurbuchen}, T.~H., \& {Richardson}, I.~G. 2006, \ssr, 123, 31

\end{thebibliography}

\end{CJK*}
\end{document}